\documentclass[twocolumn, floatfix]{aastex62}

\usepackage{scrextend}
\usepackage{booktabs}

\usepackage{savesym}
\savesymbol{tablenum}
\usepackage{siunitx}
\restoresymbol{SIX}{tablenum}
\usepackage{hyperref}

\usepackage{amsmath,bm}
\makeatletter
\newcommand{\thickhline}{%
    \noalign {\ifnum 0=`}\fi \hrule height 1pt
    \futurelet \reserved@a \@xhline
}
\makeatother

\stepcounter{secnumdepth}
\stepcounter{tocdepth}
\newcommand{ \angstrom}{\textup{\AA}}

\graphicspath{{./}{figures/}}
\usepackage[T1]{fontenc}
\usepackage[section]{placeins}
\usepackage{amssymb}
\usepackage{float}

\newcommand{\templ}{{\cal T}}
\newcommand{\losvd}{{\cal L}}
\newcommand{\additive}{{\cal A}}
\newcommand{\multiplicative}{{\cal M}}
\newcommand{\model}{{\cal S}}

\submitjournal{ApJ}

\shorttitle{Non-parametric LOSVDs in massive ETGs}
\shortauthors{Mehrgan et al.}

\begin{document}
\title{Detailed shapes of the line-of-sight velocity distributions in massive early-type galaxies from non-parametric spectral models}

\author{Kianusch Mehrgan}
\affil{Max-Planck-Institut f\"ur extraterrestrische Physik, Giessenbachstrasse, D-85748 Garching} \affil{Universit\"ats-Sternwarte M\"unchen, Scheinerstrasse 1, D-81679 M\"unchen, Germany}
\email{kmehrgan@mpe.mpg.de}
\author{Jens Thomas}
\affil{Max-Planck-Institut f\"ur extraterrestrische Physik, Giessenbachstrasse, D-85748 Garching} \affil{Universit\"ats-Sternwarte M\"unchen, Scheinerstrasse 1, D-81679 M\"unchen, Germany}

\author{Roberto Saglia}
\affil{Max-Planck-Institut f\"ur extraterrestrische Physik, Giessenbachstrasse, D-85748 Garching} \affil{Universit\"ats-Sternwarte M\"unchen, Scheinerstrasse 1, D-81679 M\"unchen, Germany}

\author{Taniya Parikh}
\affil{Max-Planck-Institut f\"ur extraterrestrische Physik, Giessenbachstrasse, D-85748 Garching} 

\author{Ralf Bender }
\affil{Max-Planck-Institut f\"ur extraterrestrische Physik, Giessenbachstrasse, D-85748 Garching} \affil{Universit\"ats-Sternwarte M\"unchen, Scheinerstrasse 1, D-81679 M\"unchen, Germany}

\begin{abstract}
We present the first systematic study of the detailed shapes of the line-of-sight velocity distributions (LOSVDs) in nine massive early-type galaxies (ETGs) using the novel non-parametric modelling code WINGFIT. High-signal spectral observations with MUSE at the VLT allow us to measure between 40 and 400 individual LOSVDs in each galaxy at a signal-to-noise level better than 100 per spectral bin and to trace the LOSVDs all the way out to the highest stellar velocities. We extensively discuss potential LOSVD distortions due to template mismatch and strategies to avoid them. Our analysis uncovers a plethora of complex, large scale kinematic structures for the shapes of the LOSVDs. Most notably, in the centers of all ETGs in our sample, we detect faint, broad LOSVD ``wings'' extending the line-of-sight velocities, $v_{los}$, well beyond $3\sigma$ to $v_{los} \sim \pm 1000 - \SI{1500}{km/s}$ on both sides of the peak of the LOSVDs. These wings likely originate from PSF effects and contain velocity information about the very central unresolved regions of the galaxies. In several galaxies, we detect wings of similar shape also towards the outer parts of the MUSE field-of-view. We propose that these wings originate from faint halos of loosely bound stars around the ETGs, similar to the cluster-bound stellar envelopes found around many brightest cluster galaxies.
\end{abstract}


\section{Introduction} 
\label{sec:intro}

Stellar kinematics from spectroscopic observations hold many interesting clues on the internal structure of galaxies. It is known since long that the angular momentum content of early-type galaxies (ETGs) varies as a function of mass. Less massive ETGs are strongly rotating while the most massive elliptical galaxies are mostly slowly rotating systems. This encodes basic information about the main aspects of the evolution of these galaxies (e.g. the dominance of dissipational vs dissipationless processes). In addition, stellar kinematics contains the fundamental information required for dynamical modeling, i.e. to uncover properties of the principle components of these galaxies, meaning the mass of the central supermassive black hole (SMBH), the stellar mass-to-light ratio, the shape of the dark matter (DM) halo and the stellar orbital anisotropy profiles, all at once. 
Various dynamical modelling methods have been developed, being either based on velocity moments, or on orbits, or even on particles \citep{Syer1996, Schwarzschild1979, Cretton1999, Siopis2000, Gebhardt2000,  Thomas2004,delorenzi2007,vanDenBosch2008,Cappellari2008,Napolitano2011,Rusli2013b, Thomas2014, Mehrgan2019, Jethwa2020,Vasilev2020,Cappellari2020, Quenneville2021, Quenneville2022}. The widely used Schwarzschild Orbit Superposition Method in particular has recently seen several important advancements (\citealt{denicola2020,Neureiter2021,Lipka2021,ThomasLipka2022, Liepold2020ApJ}, Neureither et al. (submitted to MNRAS), de Nicola et al. (submitted to NMRAS)).

The stellar kinematics of an ETG manifests in the line-of-sight velocity distribution (LOSVD) of stars in the galaxy, which is obtained from the broadening and shifting of stellar absorption features due to the Doppler shifts of stars in projected motion along the line of sight, blended together. 
A simple Gaussian parameterization of the LOSVD fails to capture the subtle difference between velocity contributions from stellar orbital anisotropy and those from steeper mass profiles \citep{Binney1982, Dejonghe1992}. For this reason, a Gauss-Hermite series expansion is often used to parameterise the LOSVDs \citep{vanDerMarel1993, Gerhard1993, Merritt1993, Bender1994}. Already the lowest-order coefficients that describe deviations from a purely Gaussian LOSVD (e.g. $h_4$) contain valuable information required to solve the mass-anisotropy degeneracy. 
Recently, also higher orders beyond $h_4$ have gained more interest \citep{Krajnovic2015, Veale2018, Quenneville2022, Thater2022}.

In principle, even more information can be obtained from stellar absorption features. E.g., one of the most
important methods, the Fourier Correlation Quotient (FCQ) method \citep{Bender1990}, produces in principle non-parametric LOSVDs and other codes for non-parametric LOSVD reconstructions have been developed as well \citep{Gebhardt2002, Pinkney2003,Houghton2006,Fabricius2014,FalconBarroso2021}. Detailed knowledge about the LOSVD shapes is currently of particular importance because recent developements in Schwarzschild dynamical modelling have shown that a surprisingly high level of accuracy and precision can be reached given adequate kinematic data (\citealt{Neureiter2021,Lipka2021}, Neureiter et al. in prep, De Nicola et al. in prep).
Pushing the limits of the kinematic information that we can extract from galactic spectra is therefore important to achieve mass measurements with a precision necessary, e.g.,  to unlock the stellar initial mass function (IMF) of ETGs and to understand its seeming variation between ETGs (and probably also within individual galaxies), which is one of the most important missing puzzle pieces of galaxy formation \citep[e.g.][]{Thomas2011, Spiniello2011, Cappellari2012, Parikh2018}.


In \citet{Mehrgan2019} we used our new spectral fitting code WINGFIT (Thomas et al. in prep) to recover non-parametric LOSVDs for the brightest cluster galaxy (BCG) of Abell~85, Holm~15A, all the way to the cut-off velocity (where the LOSVD signal goes to zero) which provides a lower limit for the escape velocity of the potential \citep{Mehrgan2019}. Our dynamical modeling of the LOSVDs showed that the radial profile of the cut-off velocity of the non-parametric LOSVDs provided an important constraint on the anisotropy profile of the galaxy - which, in turn provided constraints on its formation history from major mergers, illustrating the importance and advantages of fitting the \textit{full} shape of the LOSVD. 
However, improved knowledge about the LOSVDs in galaxies is not only interesting from the point of view of high-precision dynamical modeling. In Holm15A, for example, the measured cut-off velocities were very high, $v_{esc} \gtrsim \SI{1500}{km/s}$, extending the tails of the distribution in the form of faint ``wings''. Here, the LOSVDs indicated a faint component in velocity space that seems separated from the galaxy, which could correspond to a weakly bound, faint, large-scale stellar envelope surrounding the BCG,  similar to the kinematic evidence in the central galaxy of Abell~2199, NGC~6166 \citep[e.g.][]{Bender2015}. Indeed, our photometric decomposition of Holm15A suggested such an outer envelope as is evident for many other BCGs, illustrating the amount of detail stored in the tails of the velocity distribution -- detail, that, using a simple Gaussian or even fourth order Gauss-Hermite parameterization, would have been partially or entirely lost to us.


Therefore, we will here conduct the first systematic investigation of non-parametrically determined LOSVD shapes in a sample of ETGs. This sample consists of 9 ETGs, for which we acquired high-resolution, wide-field spectral observations from MUSE. The ETGs are part of a larger sample of galaxies which have been previously analyzed for their photometric, kinematic and dynamical properties using different instruments in previous publications 
\citep{Thomas2011,Rusli2013a, Rusli2013b, Thomas2014, Mazzalay2016, Erwin2018}. In subsequent publications, we will use the non-parametric LOSVDs from the present study to construct new Schwarzschild orbital dynamical models with stellar mass-to-light ratio gradients in order to investigate the IMF (including possible radial gradients) and the details of stellar orbital system and dark matter halos, including their triaxiality (Mehrgan et al. in prep, Neureiter et al., in prep, Parikh et al. in prep, Thomas et al. in prep).

We have structured our study as follows.
Section \ref{sec:kinData} details our observations and treatment of the MUSE spectroscopy for the sample galaxies, and also briefly introduces WINGFIT. Since we are trying to achieve previously unattained levels of accuracy for the shape of LOSVD, we performed detailed mock-tests using synthetic mock-galaxy spectra to, in Section \ref{sec:mocktesting}, explore the distorting effects of so-called template-mismatch, and, in Section \ref{sec:strategy}, to stress-test the setup and approach with which we here treat the sample galaxies. Section \ref{sec:kinResults} presents the results of our kinematic fits, which are discussed and interpreted as to their physical origin in Section \ref{sec:discussion}. Finally, we sum up our results and conclusions in Section \ref{sec:conclusion}, where we also propose further necessary investigations in order to fully understand the kinematic structures we discovered in the non-parametric shapes of the LOSVDs of the ETGs.

\section{Kinematic data and fitting code}
\label{sec:kinData}
\subsection{MUSE observations and data reduction}
\label{subsec:museReduc}

We obtained wide-field spectroscopic data of 9 early-type galaxies (ETGs) from the
Multi-Unit Spectroscopic Explorer (MUSE) at the \textit{Very Large Telescope} (VLT)
at Paranal between June and September 2015 \footnote{Observations were collected at the European
  Organisation for Astronomical Research in the Southern Hemisphere under ESO program 099.B-0193(A), P.I. J. Thomas}. The galaxies and general information on 
their observations, such as dates of observation, exposure times
and seeing conditions are listed in Tab. \ref{tab:museObsTab}.

\begin{table*}
\centering
 \begin{tabular}{l l l c c c}
 \hline
 \hline
Galaxy        & Date(s) of obs.      & $\#$ OBs                & $T_{exp} \ [s]$                & PA $[^{\circ}]$              & PSF  (FWHM) $['']$                 \\
\hline 
NGC~0307      & 07.09                & $1$                     & $1800 + 300$                   & $82$                         & $2.10$                     \\
NGC~1332      & 07.09                & $1$                     & $1800 + 300$                   & $114$                        & $2.12$                       \\
NGC~1407      & 07.09                & $1$                     & $1800 + 300$                   & $40$                         & $1.93$                     \\
NGC~4751      & 13.08                & $1$*                    & $1800 + 300$                   & $175$                        & $1.59$                       \\ 
NGC~5328      & 10, 11.08            & $2$                     & $4800 + 600$                   & $85$                         & $1.28$                             \\ 
NGC~5419      & 08.08                & $2$                     & $4800 + 600$                   & $78$                         & $1.56$                    \\ 
NGC~5516      & 11.08                & $1$                     & $2800 + 300$                   & $0$                          & $2.00$                             \\
NGC~6861      & 16.06                & $1$                     & $1800 + 300$                   & $142$                        & $0.75$                            \\ 
NGC~7619      & 18.08                &$1$                      & $1800 + 300$                   & $30$                         & $2.00$                        \\

\end{tabular}
    \caption{MUSE dataset of ETGs. All date(s) of observation relate to the year 2015. The number of Observational blocks (OBs) describes the number of object-sky-object exposure sequences, as described in Subsection \ref{subsec:museReduc}. (*) For NGC~4751 two OBs were performed, on the 08.08 and 13.08 of 2015, but for the former, both object exposures were taken with bad sky transparency conditions (thin cirrus) 
     and were therefore not used in the final analysis. Exposure times, $T_{exp}$ are listed as object plus sky exposure times, $T_{exp, total, object} + T_{exp, total, sky}$, where total exposure times are the sum of all individual object or sky exposure times from all OBs. The position angles (PA) of the observations were chosen such that they are aligned with the major axes of the galaxies as determined from previous photometric observations \citep{Rusli2013b, Rusli2013a,Mazzalay2016, Erwin2018} . Values of the FWHM of the PSF were measured directly from point-sources found in either the object or sky exposures of the galaxies, except for NGC~0307, NGC~1332, and NGC~7619 where we found no such sources and therefore used  measurements from the ESO DIMM instead.
     }
\label{tab:museObsTab}
\end{table*}

\begin{table*}
\centering
 \begin{tabular}{l c c c c c c c c c c c c}
 \hline
 \hline
Galaxy           & D [$Mpc$]  & $M_V$ [$mag$]   & Cored      & $r_{e}$ $['']$   & $\sigma_{e/2}$ $[km/s]$  & $\sigma_{e}$ $[km/s]$  & $\sigma_{0}$ $[km/s]$ & $\lambda_{e/2}$ & $\lambda_{e}$     &  $\epsilon_{e/2}$   &  $\epsilon_{e}$   & Rotator\\
\hline
NGC~0307         & $52.8$       & $-20.8^*$       &  no      & $4.8$            & $213.5$                  & $190.5$                  & $218.2$               & $0.33$         & $0.43$           & $0.36$              & $0.37$            & fast  \\
NGC~1332         & $22.3$       & $-21.5^*$       &  no      & $28.0$           & $263.1$                  & $-$                      & $334.2$               & $0.37$         & $-$               & $0.31$              & $-$               & fast \\ 
NGC~1407         & $28.1$       & $-22.7$       &  yes     & $70.33$          & $258.9$                  & $-$                      & $300.8$               & $0.09$         & $-$               & $0.05$              & $-$               & interm.\\
NGC~4751         & $26.9$       & $-20.8$       &  no      & $22.76$          & $249.0$                  & $219.3$                  & $382.8$               & $0.64$         & $0.69$           & $0.51$              & $0.53$            & fast \\
NGC~5328         & $64.1$       & $-22.8$       &  yes     & $22.2$           & $306.7$                  & $291.3$                  & $333.5$               & $0.06$         & $0.15$           & $0.31$              & $0.31$            & slow \\ 
NGC~5419         & $56.2$       & $-23.1$       &  yes     & $43.4$           & $309.6$                  & $-$                      & $347.7$               & $0.04$         & $-$               & $0.20$              & $-$               & slow  \\ 
NGC~5516         & $58.4$       & $-22.9$       &  yes     & $22.1$           & $283.8$                  & $274.2$                  & $315.8$               & $0.08$         & $0.07$           & $0.16$              & $0.17$            & slow  \\ 
NGC~6861         & $27.3$       & $-21.4$       &  no      & $17.7$           & $294.2$                  & $275.5$                  & $414.6$               & $0.55$         & $0.57$           & $0.42$              & $0.43$            & fast  \\ 
NGC~7619         & $51.5$       &  $-22.9$      &  yes     & $36.9$           & $270.8$                  & $-$                      & $333.0$               & $0.14$         & $-$               & $0.26$              & $-$               & interm. \\ 
\end{tabular}
\caption{General, kinematic and morphological properties of the sample galaxies. We have adopted the  distance, absolute $V$-Band magnitude, presence/absence of a core, effective radius $r_{e}$ and ellipticity $\epsilon$ from previous publications: from \citet{Rusli2011} in the case of NGC~1332, \citet{Mazzalay2016} for NGC~5419, \citet{Erwin2018} for NGC~0307 and \citet{Rusli2013a, Rusli2013b, Thomas2014} for the rest. All distances except NGC~1332 and NGC~5328 are determined from values which were each determined from the respective radial velocity (HyperLEDA), corrected for the in-fall velocity of the local group into the Virgo cluster. Distance values for NGC~1332  and NGC~5328 were taken from the SBF Survey of galaxy distances \citep{Tonry2001}, after a Cepheid zeropoint correction of $\SI{-0.06}{mag}$\citep{Mei2005}. Based on the new stellar kinematics presented here, the stellar velocity dispersion and angular momentum per unit mass \citep{Emsellem2007} within one $r_e$, $\sigma_{e}$ and $\lambda_e$, and within half of $r_e$, $\sigma_{e/2}$ and $\lambda_{e,2}$, are luminosity-weighted averages over elliptical apertures following the guidelines laid out by \citet{Emsellem2007, Emsellem2011}.  Luminosity-weighted averages of the ellipticity, $\epsilon_{e}$ and $\epsilon_{e/2}$ also follow the same procedure but are based on photometry originating from the studies listed above. The central velocity dispersion, $\sigma_{0}$ is a luminosity weighted average over a circular region within a radius equal to the FWHM of the PSF.  (*) For NGC~0307 and NGC~1332 we used $B$-band magnitude and B-V color of RC3.}
\label{tab:museKinTab}

\end{table*}

Observations were carried out in one or two observational blocks (OBs), each consisting of two dithered object exposures plus one sky exposure in between. Individual object exposures were adapted to each object ($900$ - \SI{1200}{s} long), the sky exposures were always \SI{300}{s} long. The MUSE field-of-view (FOV) covers approximately $\SI{1}{\arcmin} \times \SI{1}{\arcmin}$ on the sky, which encompasses
the effective radius, $r_{e}$, of each galaxy in our sample (see Tab. \ref{tab:museKinTab}).

We performed the reduction of the MUSE data for all 9 galaxies 
using version 2.9.1 of the standard Esoreflex MUSE pipeline
supplied by ESO \citep{Esoreflex2013}. The pipeline runs several
recipes on all exposures such as flat-field and wavelength
calibrations and returns combined data cubes, covering the optical
domain from about $\SI{4800}{\angstrom}$ to
$\SI{9400}{\angstrom}$ with a spectral resolution of $\SI{1.25}{\angstrom}$. 
At the redshifts of the galaxies in our sample, $z \sim 0.005 - 0.015$, this wavelength range covers several  stellar absorption features important for the recovery of stellar kinematics such as H$\beta$,
the Mgb region and several Fe lines, particularly those at $\SI{5270}{\angstrom}$ and $\SI{5335}{\angstrom}$.

Sky emissions were removed separately from all galaxy exposures using
the sky-field from offset sky-exposures, taking into account the instrumental line spread function for each of the 24 IFUs that MUSE consists of. While the correction of telluric absorption features is fully covered by the standard Esoreflex pipeline, we noticed that for all galaxies the telluric spectrum, which is derived from observations of a telluric standard star,
was scaled by a too large factor resulting in a correction where the division by the telluric spectrum left strong residuals on the galaxies' spectra. This was particularly significant in the telluric absorption region between $\sim 7590 - \SI{7710}{\angstrom}$. To obtain a better telluric correction, we decided to optimise the scale factor for the telluric standard star spectrum manually such as to minimize the residuals in the strong telluric absorption-region between $\sim 7590 - \SI{7710}{\angstrom}$ of the corrected spectrum with respect to the continuum on both sides of this region. For two galaxies (NGC~7619 and NGC~5328) we also used \textit{Molecfit} \citep{Molecfit2015a, Molecfit2015b}, a software tool for the correction of telluric absorption features developed by ESO, to do the telluric correction. We found that \textit{Molecfit} produced comparable results and therefore continued to manually optimise the scaling factors for the ESO-provided telluric standard stars.

We resampled
all data cubes to a spaxel size of $0\farcs4 \times 0\farcs4$, which at the respective redshifts of the galaxies corresponds to spatial resolutions
of approximately $40$-\SI{120}{pc}. In cases where we found point sources such as active galactic nuclei or stars in the object or sky exposures we estimated the full width at half maximum (FWHM) of the point spread function for our observations directly from their broadening, otherwise we used the seeing measurement of the ESO Differential Image Motion Monitor \citep[ESO DIMM, ][]{ESODIMM1990} at the VLT. The FWHM of the PSF ranges between $\sim0.8\arcsec - 2.2\arcsec$ for our sample.

\subsection{Spatial binning}

We spatially binned the MUSE data cubes of the nine galaxies using the Voronoi tessellation method of \citet{Voronoi2003} for a target SNR of 150
per spectral bin between  $\sim 4800 - \SI{5500}{\angstrom}$ In this way we also achieved a SNR $> 100$ in every spectral bin over the entire MUSE wavelength
range (except for spectral bins where strong skyline-residuals persisted in the data reduction steps of Sect.~\ref{subsec:museReduc}).
For the calculation of the SNR of the spectral bins for the spatial binning, we used the variance $\sigma^2 (x, y, \lambda)$ which is generated alongside the spectroscopic data by the Esoreflex pipeline as a secondary slice of the product data cube. Pixels belonging to foreground sources such as other galaxies, AGN or stars were removed from the data before binning.  
This resulted in $60 - 400$ spatial bins for each galaxy.

\subsection{Morphological properties of the sample galaxies}
\label{subsec:MuseWINGFIT}

\citet{KormendyBender1997} and \citet{Faber1997} introduced a sequencing of ETGs into two types as a refinement of the historical Hubble classification scheme: 1) luminous/massive ETGs with shallow central surface brightness cores, boxy isophotes, and little rotation. 2) less luminous/massive ETGs with steep central power-law surface brightness profiles, disky isophotes, and significant rotational support. There are more properties and studies of these two types, for which we refer to the review of \citet{Lauer2012}. To give a first brief overview of our sample galaxies based just on our spectroscopic MUSE data, we have classified our ETGs into the slow/fast rotator categories of \citet{Emsellem2007, Emsellem2011} in Tab.~\ref{tab:museKinTab}.

\begin{figure}
\centering
 \includegraphics[width=0.9\columnwidth]{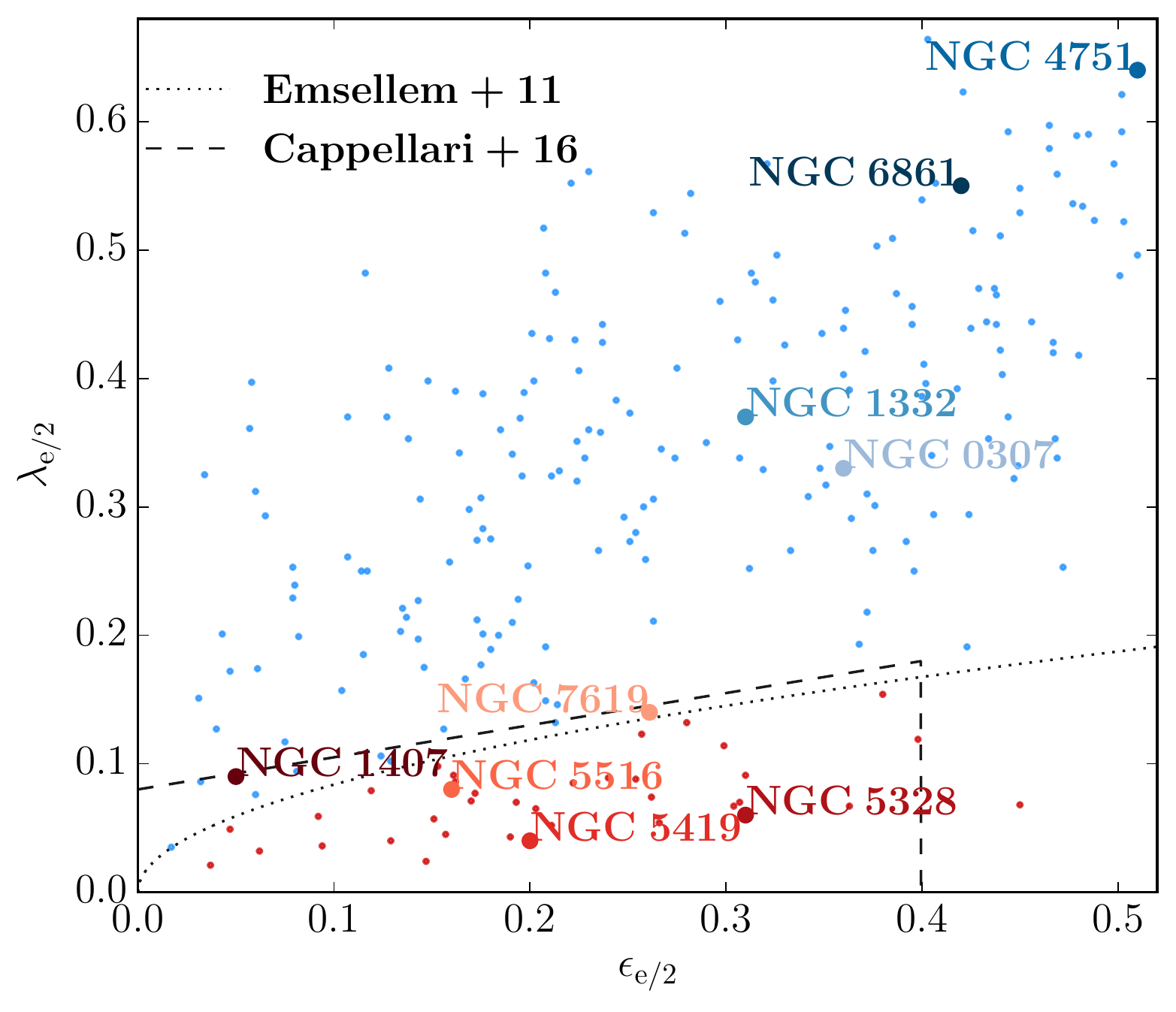}
 \caption{Apparent angular momentum $\lambda$ per unit mass and ellitpicity $\epsilon$ measured at half the effective radius for the 9 sample-galaxies (large points) compared to ETGs from \citet{Emsellem2011} (small points), including the old and revised dividing line (\citealp{Emsellem2011} and \citealp{Cappellari2016}, respectively) between slow rotators (red points) and fast rotators (blue points). NGC~1407 and NGC~7619, which we refer to as ``intermediate'' rotators in the text are here grouped in with slow rotators.}
   \label{fig:lamvell}
\end{figure}

We calculate the $\lambda$-parameter for the slow/fast rotator classification using the spectroscopic data out to $r_e/2$ (or out to $r_e$ if the data allow). In total our sample includes 4 fast rotators, 3 slow rotators and 2  ``intermediate'' cases. The latter are very close to the dividing line between slow and fast rotators from \citet{Emsellem2011} and to the revised line from \citet{Cappellari2016} (see Fig. \ref{fig:lamvell}). As we will argue later, these two galaxies are dynamically much more similar to slow rotators than to fast rotators.
All slow and intermediate rotators in our sample have depleted stellar cores in their light profiles \citep{Rusli2013a}. At least two of these galaxies seem to have kinematical misalignements and/or kinematically decoupled cores. We will provide a more detailed discussion of the kinematical structure of the galaxies below (Sec.~\ref{sec:discussion}).

\subsection{Non-parameteric LOSVD fitting}
We will perform the main part of our kinematic analysis using WINGFIT, a new spectral fitting code that allows to derive LOSVDs in a non-parametric fashion (Thomas et al, in prep.). In WINGFIT, the spectral model $\model$ is composed as 
\begin{equation}
\model = \left[\sum_{j=1}^{N_c}  \left( \sum_{i=1}^{N_j} w_{ij} \templ_{ij} \right) \ast \losvd_j \right] \times \left[ \sum_{l=1}^{N_m} d_l \multiplicative_l \right] + \sum_{k=0}^{N_a} b_k \additive_k.    
\end{equation}
$\templ_{ij}$ are template spectra which are superimposed to construct $N_c$ kinematical components with different LOSVDs $\losvd_j$. For elliptical galaxies a typical model will have a single stellar kinematical component composed of one or several template stellar spectra convolved with the LOSVD of the stars $\losvd$. If emission lines are present, one or more additional kinematical components can be added, e.g. composed of template spectra with the respective emission lines and associated to separate $\losvd$ to derive the kinematics of the respective gas component. In more complex galaxies with several distinct stellar populations that have different kinematics one can also use more than one stellar kinematical component. $\additive_l$ and $\multiplicative_l$ are Legendre Polynomials of order $l$ that can be included in the model in order to account for additive spectral components or flux calibration uncertainties. In this paper we will use WINGFIT only in the non-parametric mode, where the LOSVDs $\losvd_j$ are modelled in a non-parametric fashion. To this end, each LOSVD is represented by its values at $N_\mathrm{vel}$ line-of-sight velocities, equally spaced between a minimum and maximum velocity. The width of the velocity bins is determined by the resolution of the data spectrum. \\In this study we sample the LOSVDs over $N_\mathrm{vel} = 87$ velocity bins within $-\SI{2900}{km/s} < v_\mathrm{los} < \SI{2900}{km/s}$ which covers more than $\pm 6 \sigma_\mathrm{max}$ even in the galaxies where the maximum velocity dispersion is $\sigma_\mathrm{max} \sim \SI{400}{km/s}$. The sole exception is the galaxy NGC~0307, where a sampling between $-\SI{1500}{km/s} < v_\mathrm{los} < \SI{1500}{km/s}$ or $N_\mathrm{vel} = 47$, respectively, is sufficient due to its low velocity dispersion.

The free parameters of the galaxy model are the template weights $w_{ij}$, the LOSVDs $\losvd_j$ and the polynomial coefficients $d_l$ and $b_k$.
The code utilizes a Levenberg-Marquardt algorithm to minimize the $\chi^2$ over all spectral pixel of a model stellar spectrum relative to the data. In the non-parametric mode -- which we will consider in this paper -- the code uses penalty function that minimises the second derivative of the LOSVD to reject physically implausible solutions. The code uses a data-driven optimisation method that is based on a generalisation of the classical Akaike Information Criterion \citet{ThomasLipka2022}. The smoothing penalty and the method to determine its optimal strength individually for each spectrum in each Voronoi bin are exactly the same as described in eq. 3 of \citep{ThomasLipka2022}. A full description of the code will be given in Thomas et al. (in prep).

Since this is the first systematic investigation of non-parametrically determined LOSVD shapes in a sample of massive elliptical galaxies, it is important to carefully investigate how well the LOSVDs of the stars can be recovered and which potential systematic issues could affect the results.
For the data in our sample (e.g. assuming the MUSE spectral coverage and the SNR used in our binned spectra) we show in Sec.~\ref{sub:idealrecov} that the non-parametric LOSVD recovery in principle works without bias {\it if} the exact template spectrum is used. In this case, the LOSVD recovery is largely independent of the fitting setup. This means, the recovery is independent of the particular wavelength region used in the fit, it is independent of one or more spectral regions being masked and excluded from the fit or not and it is independent of whether or not additive or multiplicative polynomials have been used in the fit. 

In general, however, it is unrealistic to assume that the exact stellar mix is available as a template spectrum for the fit. This is particularly true for the fits in this study, since we use the MILES library of \textit{empirical} stellar spectra from stars in the Milky Way -- i.e. from an environment very different from our target ETGs. In the next two sections, we discuss results of a number of mock-tests aimed at investigating how template mismatch can affect the shapes of LOSVDs in observed galaxies. We put particular emphasis on the question which setup (wavelength region, masks, template selection, usage of polynomials) is the best given the fact that some template mismatch is almost unavoidable. The analysis of the galaxies will be presented in Sec.~\ref{sec:kinResults}.

\section{LOSVD distortions induced by template mismatch}
\label{sec:mocktesting}

The ``naive'' approach to spectral fitting would be to simply minimize the residuals of the fit to the spectrum.
We here leave aside the question of how to find the optimal trade-off between smoothing and goodness-of-fit (see \citealt{ThomasLipka2022} and Thomas et al. in prep). Beyond this, the fit can often be improved by tweaking the fitting setup: The number of template stellar spectra used, the orders of additive and multiplicative polynomials, and spectral regions which are masked or excluded from the fit. Typically, the fit to the spectrum will remain very similar for most setups, but the recovered LOSVD, particularly at large projected velocities may differ. While for many purposes the detailed shapes of the high-velocity tails of galaxy LOSVDs may not be important,
they certainly are for accurate and precise dynamical modelling (Neureiter et al., de Nicola et al. in prep).

\begin{figure*}
\centering
 \includegraphics[width=2\columnwidth]{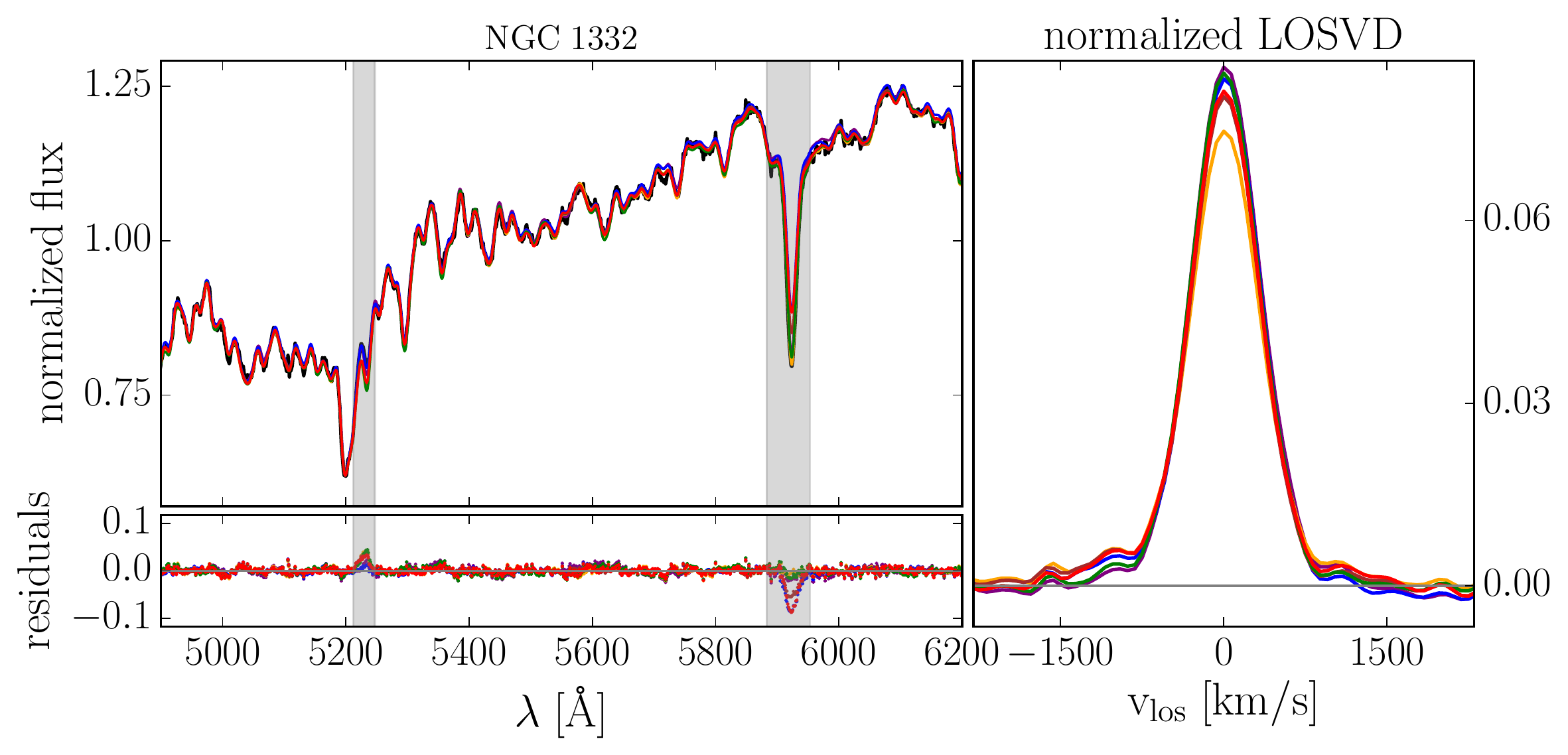}

 \caption{Top-left: a spectrum measured near the centre of NGC~1332 (black) fitted with different models (colored lines), with different orders of additive and multiplicative polynomials, as well as spectral masking with WINGFIT, using a template set of around 30 (mostly K-type) MILES stars. The spectral models are very similar, hence in spectral intervals where only one color is visible, the models overlap. The residuals are shown in the bottom-left panel, in the same colors. Grey shaded areas indicate spectral regions which were masked during the fit for some models - One mask concerning potential[N I]-emission around \SI{5250}{\angstrom} and the other, NaD, around \SI{5950}{\angstrom}. All models shown in the figure are detailed in Tab. \ref{tab:naiveFits}.
 The right panel shows the non-parametric LOSVDs determined from the fits (same colors as fits to the spectrum). The line-of-sight velocities $v_{los}$ are relative to the systemic velocity of the galaxy.}
   \label{fig:examplenNaiveFit}
\end{figure*}

Fig. \ref{fig:examplenNaiveFit} and the complementary Tab. \ref{tab:naiveFits} show several example fits with different setups for the same central bin of NGC~1332. For many setups, we encounter LOSVDs which consist of a relatively narrow central component and a weaker but very broad additional component that extends to large line-of-sight velocities $v_{los}$ and which is most strongly affected by variations of the fitting setup.  We will henceforth refer to these broad components as ``wings'': a faint, high-velocity structure, on \textit{both} sides of the peak of the LOSVD that extends well beyond $3\sigma$ (i.e. typically to $v_{los} \sim \pm 1000 - \SI{1500}{km/s}$).

Increasing the orders of the additive and multiplicative polynomials, as well as the spectral masking applied to the spectrum, improves the RMS to the spectrum. Therefore, setups with higher orders of polynomials and more spectral masking seem favorable. However, given the strong impact on the wing-component, the question is: Which setup allows for the most robust recovery of LOSVDs, avoiding both over- and underprediction of the stellar light in various velocity regions, in particular at the tails of the distribution?


\begin{table*}
\centering
 \begin{tabular}{l l l l c}
 \hline
 \hline
color & add. polynomial        & mult. polynomial    & spectral masks                & rms $[10^{-3}]$ \\
\hline 
brown     &   $8$      & $8$               &[N I], NaD                    &    $5.72$                      \\
red     &   $1$      & $12$              &[N I], NaD                    & $5.85$ \\
blue        &$4$      & $8$               &[N I]                         & $6.04$   \\
orange     &   $8$      & $4$               &[N I]                         &   $6.22$                     \\
green    &    $0$      & $4$               &[N I], NaD                    & $7.3$2                        \\
purple       & $0$      & $1$               &                            & $8.06$                        \\

\end{tabular}
    \caption{Setup parameters and root-mean-square (RMS) of the fits to the data of the models shown in Fig. \ref{fig:examplenNaiveFit}. Models are identified by their color-coding in the figure. The spectral models are very similar, hence in spectral intervals where only one color is visible in the figure, the models overlap. The setup parameters include the orders of the additive and multiplicative polynomials used in the fit as well as the spectral features which were masked.}
\label{tab:naiveFits}
\end{table*}
\subsection{LOSVD distortions due to line-strength mismatch in a single isolated line}
\label{subsec:mismatch_general}

\begin{figure*}
\centering
 \includegraphics[width=2\columnwidth]{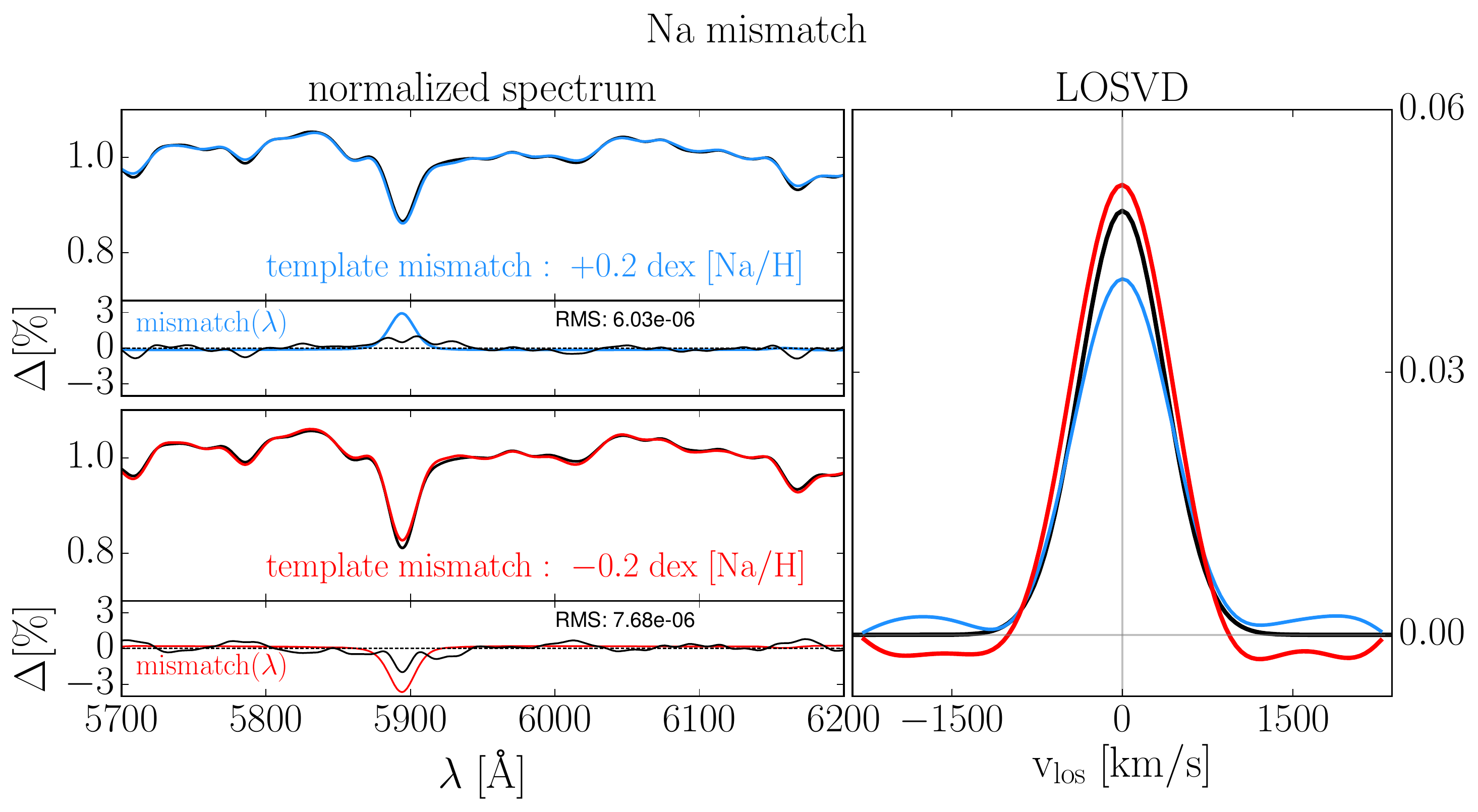}
 \caption{Fits to a mock spectrum with artificially induced template mismatch, with respect to the elemental abundance of Na. The left row shows the mock (black) which has been Na-reduced (top sub-panel) or Na-enhanced (bottom sub-panel) relative to the original mock (dashed black). Fits to the modified mocks using the unmodified template, are show in blue and red for reduced and enhanced elemental abundances, respectively. In the right hand panel, we show, in the same color-coding, the LOSVD-recoveries which have been distorted by template mismatch compared to the true LOSVD (dashed black). Below either spectrum we indicate the template mismatch of different absorption features via the residuals between the modified and unmodified mock. No additive or multiplicative polynomials were used during the fit.}
   \label{fig:NaDmismatch}
\end{figure*}

To illustrate the basic mechanism by which LOSVD shapes and template mismatch can interfere, we start with fits around the strong and relatively isolated NaD feature.
We use $\textit{alf}$ (\href{https://github.com/cconroy20/alf}{v2.1}) to create a synthetic template-star spectrum with solar abundances. We also create a mock galaxy spectrum by convolving this stellar spectrum with a simple Gaussian LOSVD\footnote{The specifics of how we create mocks and templates in this section are detailed in Appendix \ref{ap:MockCreation}.}. Then, we simulate template mismatch by creating two additional mock galaxy spectra, based on the same Gaussian LOSVD, but modifying the abundance of [Na/H] by $\pm \SI{0.2}{dex}$ relative to solar. The fits of these two mocks with the unmodified, solar abundance template star and the resulting distortions of the recovered LOSVDs are shown in Fig. \ref{fig:NaDmismatch}. To keep things simple at first, we use no multiplicative or additive polynomials during the fits and fit noise-free spectra.

The Gaussian input LOSVD, per definition, has no wings. However, fitting the NaD feature with a template that is overabundant in [Na/H] (achieved by using the solar template to fit the the mock spectrum with subsolar [Na/H]), produces a wingy LOSVD (blue). The opposite is true for the fit where the template is underabundant in [Na/H] and the recovered LOSVD has ``negative'' wings. This result is intuitive:
After convolution with a wingy LOSVD, the absorption feature of the template star appears weaker. After convolution with a LOSVD with "negative" wings, instead, the absorption feature appears stronger than in the unconvolved template.

When simply fitting an isolated feature, template mismatch can only occur in form of this mismatch in the line strength. 
As we will show in the following, various combinations of over- or underabundances in different elements can lead to more complicated net distortions of the recovered LOSVDs.

\begin{table*}
\centering
 \begin{tabular}{l | c c| c c |c c | c c | c c}
 mismatch & Na$\blacktriangle$   &  Na$\blacktriangledown$  &  Mg$\blacktriangle$  &    Mg$\blacktriangledown$ &  Fe$\blacktriangle$  &    Fe$\blacktriangledown$  &  Mg/Fe$\blacktriangle$ &  Mg/Fe$\blacktriangledown$ & Mg/Fe$^{mpoly}\blacktriangle$  & Mg/Fe$^{mpoly}\blacktriangledown$ \\
 \hline
 \thickhline
 
$\Delta h_3$  &  $0.008$ &   $-0.010$&                  $0.007$ &   $-0.016$ &        $-0.004$ & $0.001$ &        $0.015$ &  $0.002$     &      $-0.003$ & $-0.003$   \\
$\Delta h_5$ &  $0.003$ & $-0.005$ &                  $0.032$ &  $-0.032$ &           $-0.014$ &  $0.016$ &       $0.038$ & $0.001$   &  $0.008$ & $-0.001$        \\
$\Delta h_7$ &  $0.000$ &  $0.001$ &                  $0.033$ &  $-0.033$ &           $-0.018$ & $0.022$ &        $0.034$ &  $0.002$ &       $0.009$ & $0.000$  \\ 
\thickhline 
$\Delta h_4$  &  $-0.003$ &  $-0.067$         &       $-0.023$ &  $-0.007$ &         $0.014$ &  $-0.042$ &      $0.019$ &  $0.012$ &           $0.036$  & $0.069$  \\
$\Delta h_6$ &  $0.038$ & $-0.04$ &                       $0.003$ & $-0.008$ &          $0.017$ &  $-0.023$ &        $0.037$ &  $0.019$ &     $0.038$ &  $0.055$  \\
$\Delta h_8$ & $0.050$ & $-0.039$ &                    $0.025$ &   $-0.032$ &         $-0.002$ & $-0.017$ &       $0.044$ &  $0.017$ &     $0.036$ & $0.045$        \\

\end{tabular}
    \caption{
Gauss-Hermite coefficients of LOSVDs recovered simulating different forms of template-mismatch. While the true input LOSVD is purely Gaussian, the recovered LOSVDs are distorted depending on the template mismatch. We parameterise the distortion by fitting the recovered LOSVDs with 8th order Gauss-Hermite polynomials. We group the Hermite moments into even and odd order moments. Since the input LOSVD is Gaussian in every case ($h_i = 0; \ i = 3,4,...$), the Hermite parameter $h_i$ of this fit is a measure of the recovery mismatch and we denote it by $\Delta h_i$.
    Upward triangles indicate that the associated elemental abundance of the template spectrum is too low relative to the mock spectrum. Downward triangles indicate template mismatch in the opposite direction. Mg/Fe$^{mpoly}$ refers to the mismatch tests presented in Fig. \ref{fig:PolyProblem}, wherein multiplicative polynomials were used in the recovery of the LOSVD. We here only list the distorting effects of the fit with multiplicative polynomials as those of the fit with additive polynomials are virtually identical. 
    }

\label{tab:parameterizeMismatch}
\end{table*}

\subsection{The origin of asymmetric and symmetric LOSVD distortions}
\label{subsec:kindsofmimsatch}
We performed a number of similar mock-tests with controlled template mismatch in particular for Mg and Fe.
These are the elements associated with the most predominant absorption features (besides NaD) within the wavelength-interval $4800 - \SI{6200}{\angstrom}$ that we finally decided to use for our stellar kinematic analysis (Sec.~\ref{sec:kinResults}). Gauss-Hermite parameterizations of the resulting distorted LOSVDs (including the NaD mock-test described above) are listed in Tab. \ref{tab:parameterizeMismatch}. We fit up to the 8th order, which is sufficient for the LOSVD shapes we deal with here.



\begin{figure*}
\centering
\includegraphics[width=1.5\columnwidth]{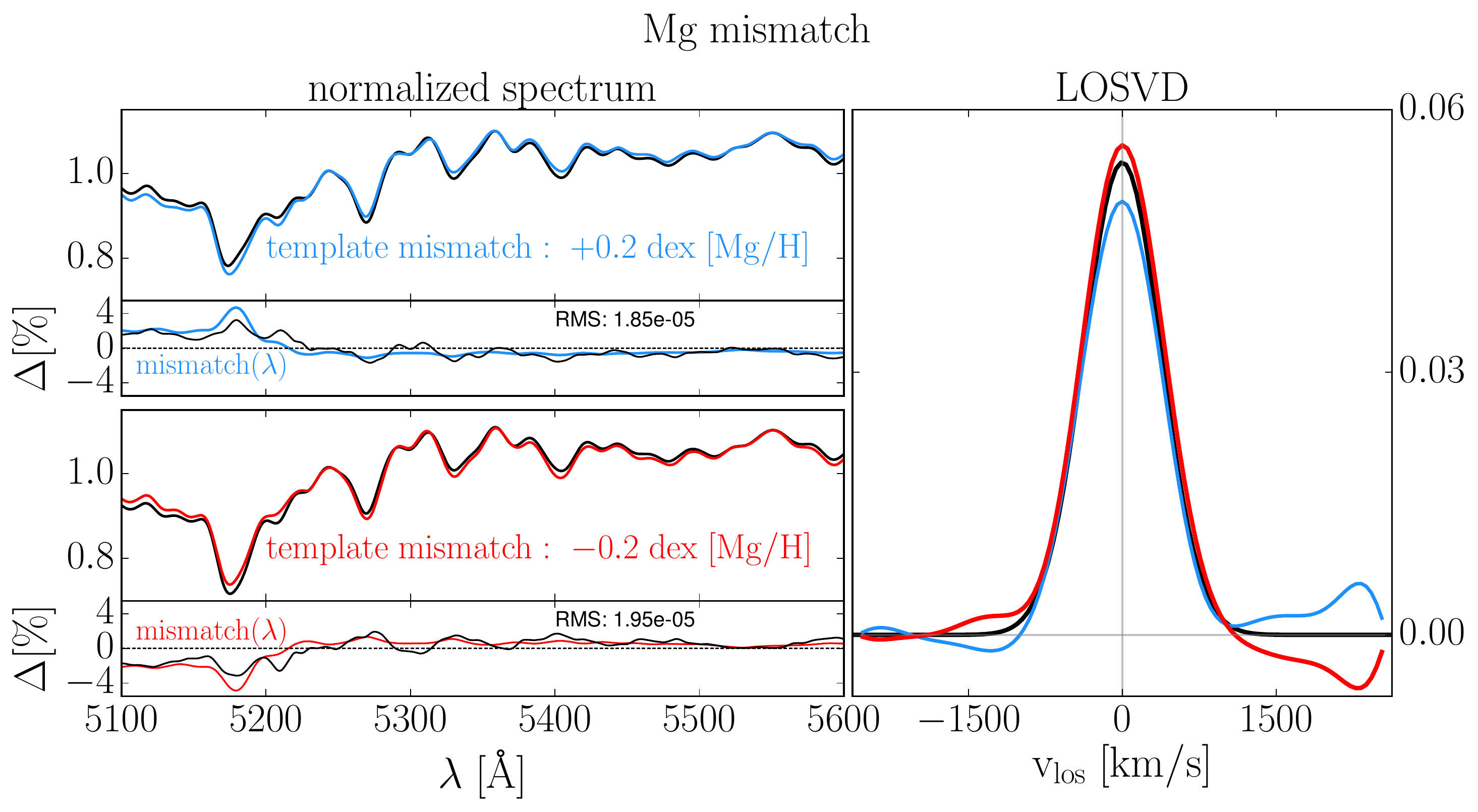}
\includegraphics[width=1.5\columnwidth]{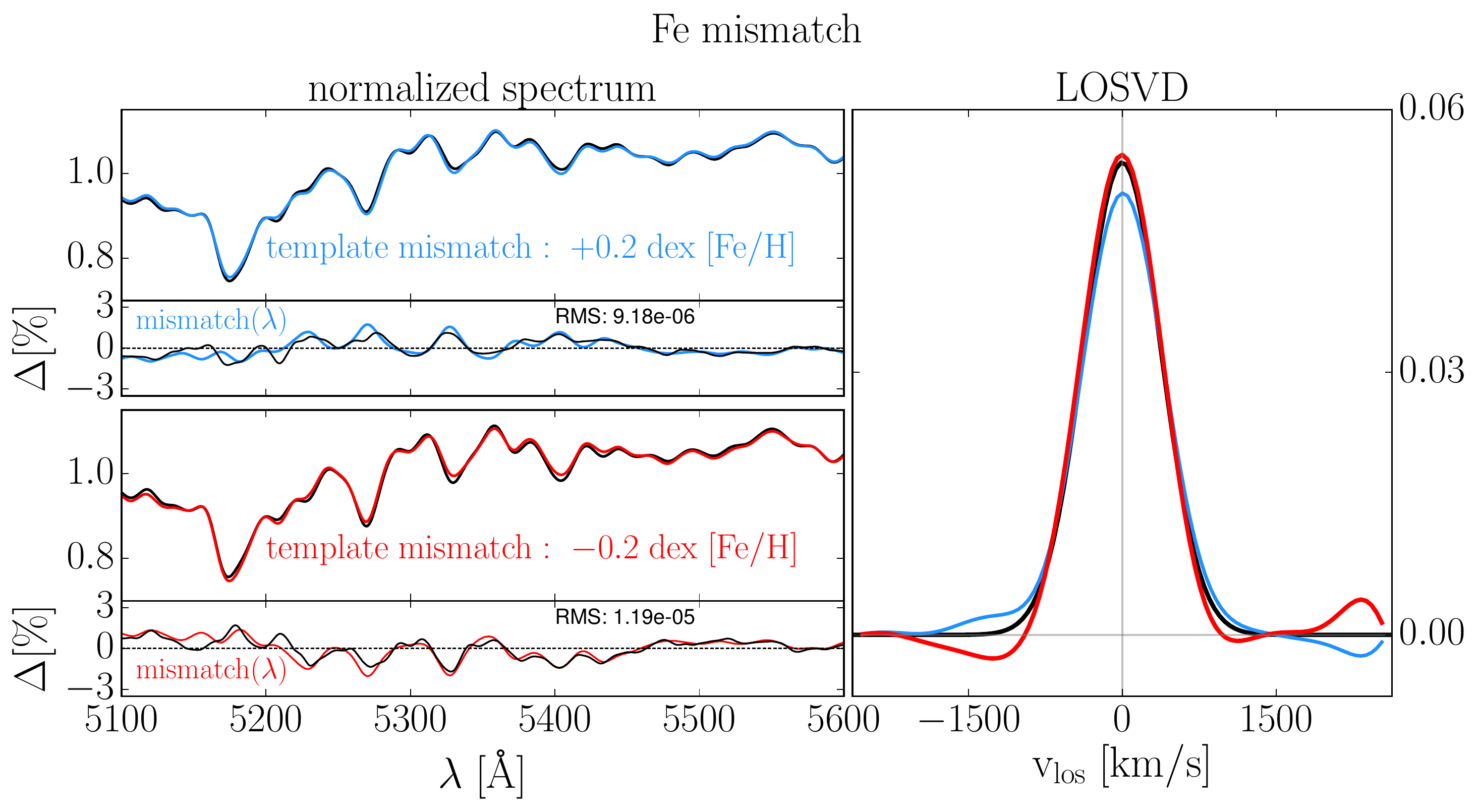}
\includegraphics[width=1.5\columnwidth]{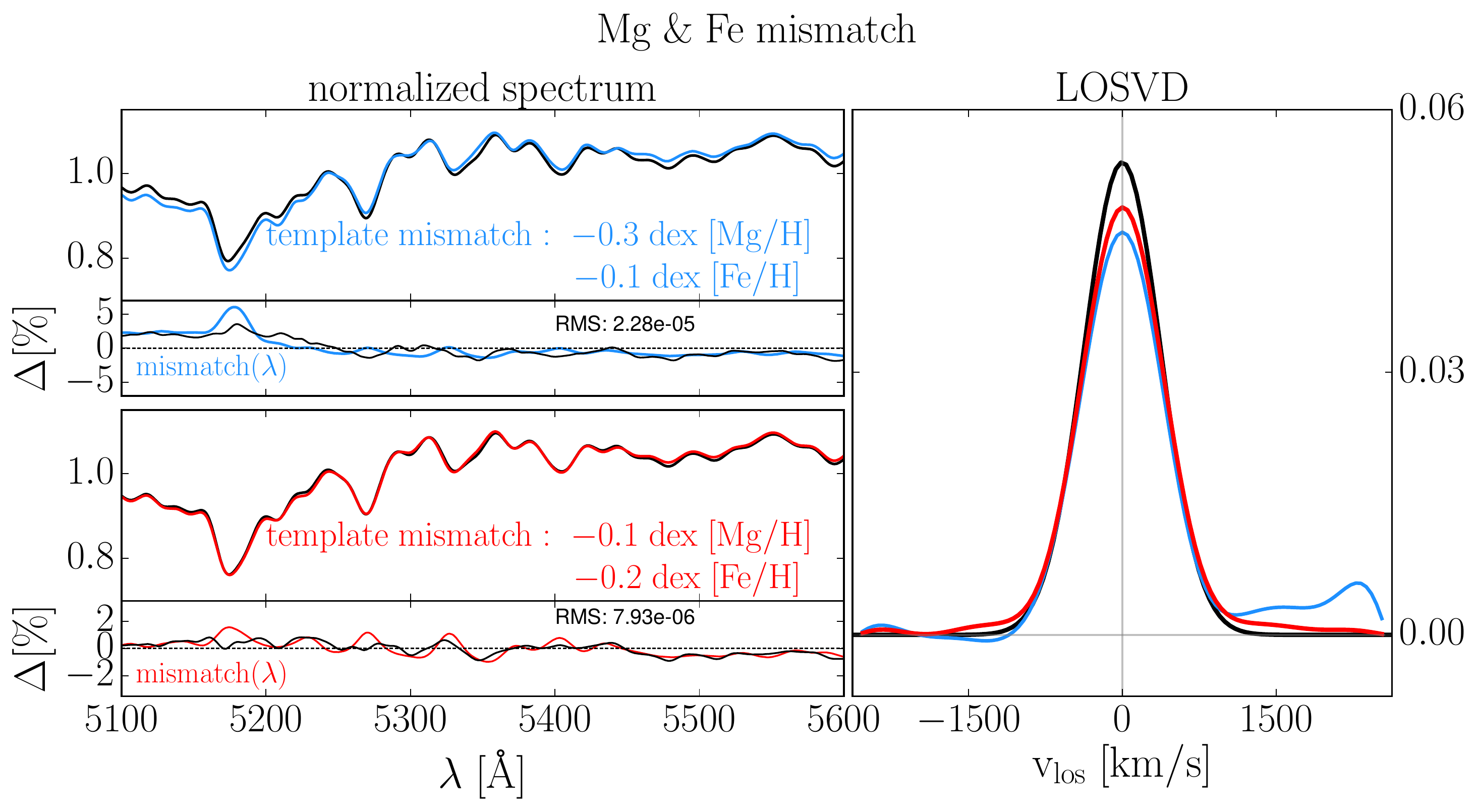}
 \caption{Test for artificially induced template mismatch, with respect to the elemental abundance of Mg, Fe and Mg \& Fe simultaneously, in analogy to the Na-mismatch tests from Fig. \ref{fig:NaDmismatch}.}
   \label{fig:Wingmechanism}
\end{figure*}


To start with, we generated four mocks by increasing and decreasing the abundances of [Mg/H] and [Fe/H] individually, each by $\pm \SI{0.2}{dex}$. As with the Na-test, we performed the first set of tests without the use of additive or multiplicative polynomials. 
It should be noted that some of the related fits, particularly those for Mg-mismatch, are so poor, that we would not accept them for any observed spectrum. However, we here intentionally probe exaggerated template mismatch in order to highlight certain trends of the recovery of the LOSVD. Later on, in Section \ref{sec:strategy}, we will describe tests which involve more realistic amounts of template mismatch.

The distortions of the exact Gaussian input LOSVD are primarily \textit{asymmetric} in the case of Mg: the fitted LOSVDs have a (positive) wing on one side of the peak and a dip (or "negative" wing) on the other  (Fig.~\ref{fig:Wingmechanism} top). This type of asymmetric template mismatch is known since long and manifests itself in a bias of the third order Gauss-Hermite coefficient $h_3$ \citep{Bender1994}. Note, however, that for the fit with too strong Mg (see Tab. \ref{tab:parameterizeMismatch}) the $h_3$ of the resulting distorted LOSVD -- or equivalently the recovery mismatch $\Delta h_3$ -- is quite small, $\Delta h_3 = 0.007$. It is actually similar to the value for the near symmetric LOSVDs from the equivalent Na-mismatch test,
$\Delta h_3 = 0.008$. The difference between the cases becomes only apparent when looking at the sequence of higher-order odd Hermite coefficients. In the Mg test (asymmetric LOSVD) also the next higher-order moments are non-zero: $\Delta h_3 = 0.007$, $\Delta h_5 = 0.032$, $\Delta h_7 = 0.033$. Instead, for the Na case (more symmetric LOSVD) the odd higher-order moments add little beyond the third order: $\Delta h_5 = 0.003$, $\Delta h_7 = 0$. Relying solely on $h_3$ and $h_4$ to describe the shape of the LOSVD is sometimes not sufficient and
an unambiguous characterization of the LOSVD distortions requires a sequence of odd Hermite coefficients (which often have the same sign).


Compared to Mg, distortions caused by template mismatch in Fe are more \textit{symmetric} (as can also be seen from a comparison of the sequence of higher order odd moments in Tab. \ref{tab:parameterizeMismatch}). Indeed, the LOSVDs recovered from fits with a template that is over- or underabundant in [Fe/H] relative to the mock galaxy spectrum show positive as well as negative distortions on the {\it same} side ($v_\mathrm{los} > 0 $;  Fig.~\ref{fig:Wingmechanism} middle).

Next, we modified both [Fe/H] and [Mg/H] at the same time. To mimic a template that is $\alpha$-deficient (which in this limited wavelength range translates to a too low [Mg/Fe]), we set [Mg/H] to $\SI{-0.1}{dex}$ and [Fe/H] to $\SI{-0.2}{dex}$. When doing so, the fit with this template yields a LOSVD distorted in such a way that even though the LOSVD is almost perfectly symmetric ($\Delta h_{3,5,7} \sim 0 $) it displays smooth artificial wings. When we try to change the ratio in the opposite direction, to mimic a template that is $\alpha$ enhanced relative to the galaxy ([Mg/H] $= - 0.3$, [Fe/H] $= -0.1$), the fitted LOSVD becomes strongly asymmetric again, with the largest $|\Delta h_{3,5,7}|$ distortions of all tests, producing only one wing on one side (Fig.~\ref{fig:Wingmechanism} bottom).

This variety of distortion patterns in the recovered LOSVDs (e.g. asymmetric vs symmetric) can be understood by comparing the degree to which the template mismatch is either {\it localized} or {\it spread out} across the fitted wavelength interval. After all, the LOSVD, by design, affects every absorption feature within the wavelength range \textit{in the same way}. Meaning that a mismatch pattern that repeats over the whole fitted wavelength region in a similar fashion can be compensated for by modifications of the LOSVD shape.

For example, when we change [Fe/H], the depths of multiple absorption features between $\sim 5200$ - $\SI{5450}{\AA}$ are modified to a roughly similar degree, leading to the many regular ups and downs in the mismatch pattern shown the little panels below each panel with a spectrum in Fig.~\ref{fig:Wingmechanism}. For the test where we additionally enhanced [Mg/H], this similarity between the line strength differences of the template and the mock spectrum even extends to Mgb, such that almost all absorption features in the fitted wavelength interval are too shallow by a similar amount (Fig.~\ref{fig:Wingmechanism} bottom). In these cases, the recovered LOSVDs show relatively symmetric distortions. The most symmetric case is the one we described last: here the LOSVD shows very smooth and symmetric wings which make the fit to the spectrum appear quite good even though both the template and the LOSVD are wrong. By raising or lowering the signal on both sides of the inner, main part of the LOSVD all the absorption features of the fitted model collectively become shallower or deeper, respectively (by the mechanism described in Sec.~\ref{subsec:mismatch_general}).

Such symmetric template mismatch, scarcely ever mentioned, has already been noted in observations by \citet{Bender1994}, who found that template mismatch not only biased their measurements of $h_3$, \textit{but also $h_4$.}. We can expand on this and say that symmetric template mismatch biases the sequence of even order moments, $h_{4,6,8,...}$ in the same direction (as with asymmetric template mismatch for the odd order moments).

In the remaining cases shown in Fig.~\ref{fig:Wingmechanism} the mismatch pattern is very inhomogeneous over the fitted wavelength region and, in fact, mostly \textit{localised} (around Mgb). In these cases, the recovered LOSVD gets \textit{asymmetrically} distorted. The fitting-optimization process has to seek a solution compensating for the local template mismatch, e.g. by distorting the LOSVD. However, at the same time, the freedom in compromising the LOSVD shape to improve the fit is limited by the fact that all other features in the fit region do not need a compensation (or need a different one, for example in the opposite direction). If a single feature is particularly dominant in the spectrum (and the continuum on both sides of the feature particularly smooth and featureless -- e.g. as discussed above with NaD) this compromise can exceptionally result in symmetric LOSVD-distortions as described above. Typically, however, the LOSVD settles for a shape that is asymmetrically distorted. Moreover, in comparison to non-localised template mismatch, the improvements in the fit achieved by deforming the LOSVD are very limited when the mismatch is localised and the residuals remain relatively large. 

The predominant distortion effect of template-mismatch appears to be positive or negative wing signals. Nonetheless, the dispersion $\sigma$ for all template mismatch mock-tests conducted in this study is biased \textit{high}. For the tests in this section the average $\sigma$ mismatch is $\sim 10 \%$. For tests with more realistic amounts of template mismatch which we encounter in Section \ref{sec:strategy} this bias is $\sim 1-5 \%$.

\FloatBarrier

\begin{figure*}
\centering
\includegraphics[width=2\columnwidth]{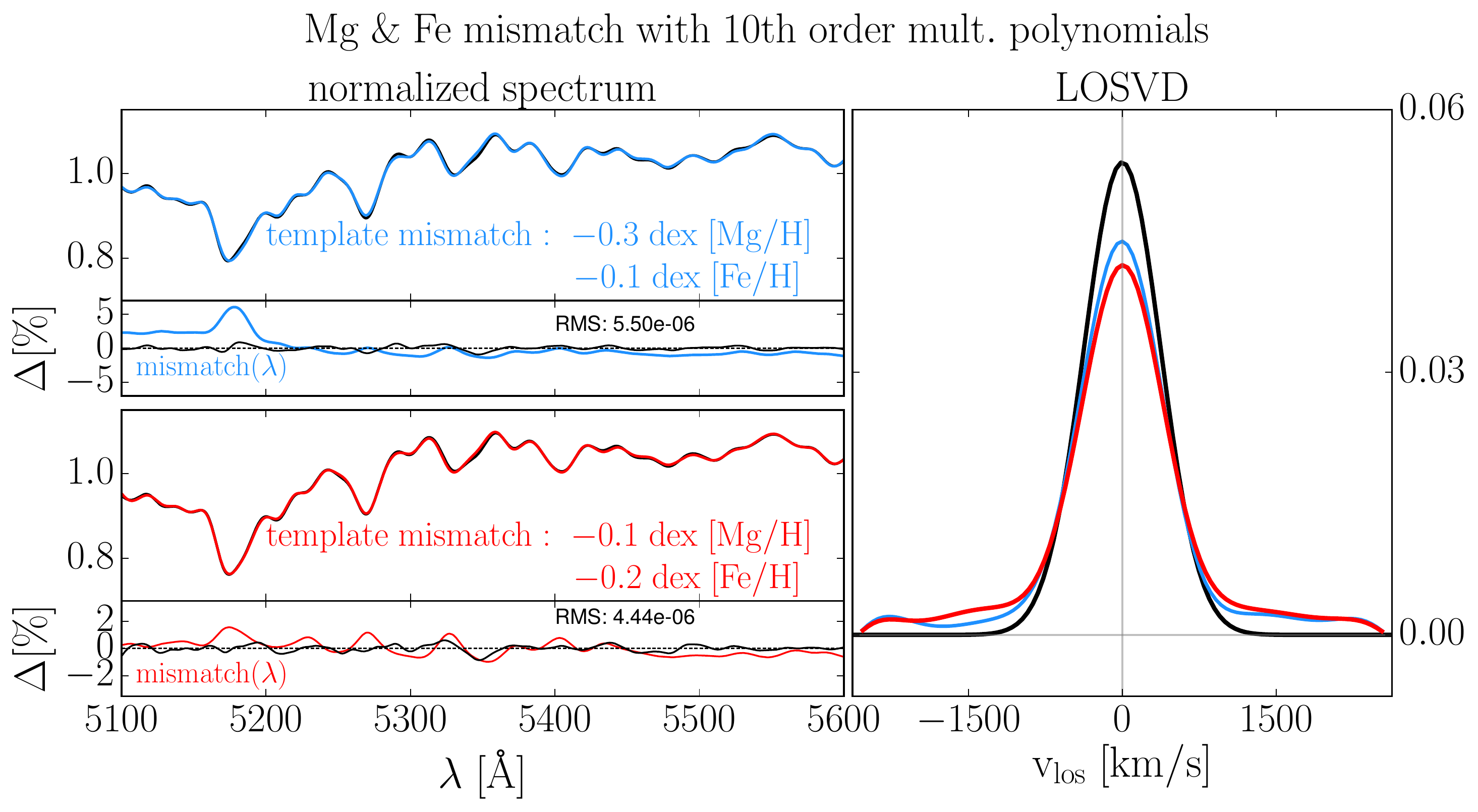}
\includegraphics[width=2\columnwidth]{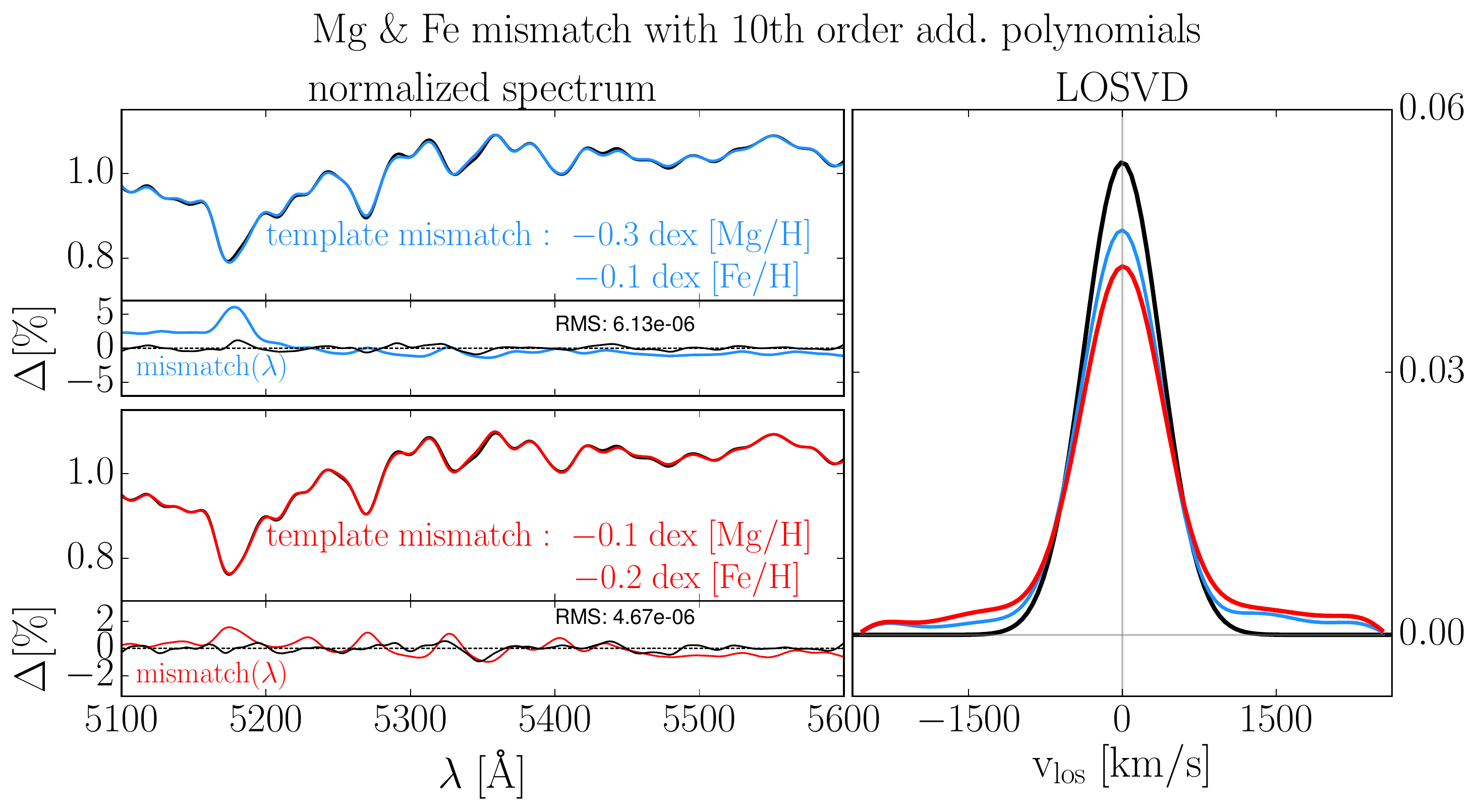}
 \caption{Test for artificially induced template mismatch, with respect to the elemental abundance of Mg \& Fe simultaneously, in analogy to the Na-mismatch tests from Fig. \ref{fig:NaDmismatch} and Fig. \ref{fig:Wingmechanism}, but with the use of additive/multiplicative polynomials.The ways in which these polynomials change the shape of the effective template during the fit is indicated in black in the mismatch-panels: here we compare the spectral difference between the modified and unmodified mock \textit{after} the use of the polynomials, to highlight how the polynomials interfere with template mismatch to produce even further distorted LOSVDs.}
   \label{fig:PolyProblem}
\end{figure*}

\subsection{Polynomials can amplify the effect of template mismatch}
\label{subsec:polytests}

The quality of the fits in the previous subsection is often lacking, particularly for those mock-tests with more spectrally localized template mismatch. These fits can be improved substantially by the use of polynomials, which is standard practice for stellar kinematics. This however, turns out to be problematic for the recovery of the LOSVD shape. Fig.~\ref{fig:PolyProblem} shows again fits with mismatched [Mg/Fe] like the ones in the bottom panel of Fig.~\ref{fig:Wingmechanism}. However, this time we also use additive and multiplicative polynomials in the fit. To clearly highlight the trend in the LOSVD recovery we use polynomials with an order up to $10$.

The result for the positive [Mg/Fe] mismatch-mock (blue) is particularly concerning: The use of either type of polynomials increased the quality of the fit from unacceptable (RMS $= 2.3 \times 10^{-5}$) to a level that would be considered more than sufficient in case of an observed spectrum (RMS $\sim 0.6 \times 10^{-5}$). At the same time, however, while removing almost any asymmetry from the recovered LOSVDs (compare with Fig. \ref{fig:Wingmechanism}), the usage of the polynomials raised strong wings. This is also evident in the Gauss-Hermite parameterization where $|\Delta h_{3,5,7}|$ decrease, while $\Delta h_{4,6,8}$ increase.

What happens here is that the freedom provided by the polynomials is used by the code to homogenise the template mismatch in different spectral regions. If this is possible, then symmetric LOSVD distortions can be used to collectively compensate for the mismatch (as described above). Together, this leads to a significant improvement of the fit. However, the LOSVD shape develops a bias towards extended wings. 
This can be seen for the much more uniform template mismatch mock-test after the application of polynomials.

Furthermore, even for the mismatch-test where the template is underabundant in [Mg/Fe] (red) the already symmetric distortions of the LOSVD become more pronounced through the use of polynomials. The additive/multiplicative polynomials modify the effective template such that template mismatch in Mgb and the Fe features is actually \textit{increased}. However it is increased in such a way that at the same time the mismatch is homogenized over the wavelength region such that in combination with a respectively distorted LOSVD the fit to the spectrum becomes overall much better (RMS $= 0.8 \times 10^{-5}$ vs $0.5 \times 10^{-5}$). Again, the LOSVD distortion consists of strong artificial wings (compare with Fig. \ref{fig:Wingmechanism}), producing the largest values of $h_{4,6,8}$ so far.

These tests demonstrate that polynomials should be used with care. For both tests, the fit to the spectrum became significantly better, but at the expense of strong LOSVD distortions. Hence, when template mismatch could be an issue, a too liberal use of polynomials can lead to biased LOSVD shapes. Without template mismatch, polynomials do not lead to biases (Sec.~\ref{sub:idealrecov}).

As we will explain at the start of the next Section, this tendency of polynomials to exacerbate template mismatch in such a way that symmetric wings are overproduced in the fitted LOSVDs is potentially dangerous. Not the least because it is harder to identify than template mismatch that leads to asymmetric LOSVD distortions.



\section{Fitting strategy to minimise the effects of template mismatch}
\label{sec:strategy}
As we have demonstrated in the previous section, template mismatch can induce asymmetric and symmetric distortions of the recovered LOSVDs. 

Asymmetric distortions are less problematic. 
For galaxies that are in dynamical equilibrium,
\textit{in the absence of template mismatch}, stars on the same orbits, but on opposite sides of the center of the galaxy will be seen moving along the line-of-sight with velocities $\pm v_{los}$, i.e. with the same absolute value but opposite signs. Hence, an asymetric bias in the shape of the LOSVD over any spatial region that is axi- or point-symmetric about the center of the galaxy can be recognised as template mismatch
(e.g. \citealt{Bender1994}).

Similarly, LOSVD distortions that produce symmetric, negative wings 
can always easily be identified as template mismatch because a negative LOSVD signal is unphysical.

More problematic are symmetric distortions that produce positive-signal wings, because they lead to a kind of ``hidden'' template mismatch. ``Hidden'', because the fit to the spectrum in these cases is usually quite good (e.g. in the polynomial tests in Sec.~\ref{subsec:polytests}), while at the same time, the resulting LOSVD shapes are consistent with LOSVD shapes that \textit{can} occur in real dynamical systems (via e.g. radial anisotropy or variations in the circular velocity curve, see Section \ref{sec:discussion}). Hence, in this case, the form of the LOSVD itself does not allow one to judge whether it is distorted by template mismatch or not. This type of template mismatch is therefore potentially dangerous. 

We now turn to the question if one can modify the setup of the fits in a way such that any influence of template mismatch is at least minimised.


\subsection{Strategy to reduce template mismatch}
\label{subsec:TemplateSelection}

Assuming that the galaxies in our sample are in dynamic equilibrium,
we argue that asymmetric template mismatch can be efficiently suppressed by the selection of appropriate template stars.

To this end, we section the MUSE FOV into half a dozen to a dozen elliptical annuli and add together spaxels that fall within each annulus, with spaxels near the boundary of the FOV added to the outermost annulus.
Pixels which are contaminated by foreground sources such as other galaxies, AGN or stars are removed before this process. For reasons of symmetry we also removed all pixels from the FOV which were point-symmetrical to these contaminated pixels. Then, we individually fit each of the resulting spectra with all $\sim 1000$ stars of the MILES library simultaneously, \textit{with an LOSVD which we constrain to be symmetric around $v_{los} = 0$ during the fit} and then select those templates which are assigned a non-zero weight as the final template (sub-)set for all Voronoi bins whose centers lie within the annulus in question for the final kinematic analysis. Typically, this approach to template optimization yields optimized template-sets of roughly 20-30 MILES templates per elliptical annulus or mock. These empirical templates serve as an approximation of the underlying stellar mix in the spectra encompassed by the corresponding annulus. In the actual fits to the individual spectra in the annulus the weights of the empirical templates are allowed to vary freely. During the selection, we use the same wavelength interval and spectral masks as in our final fits.



\subsection{Strategy to reduce LOSVD distortions}
\label{subsec:realTemp}
Since we use MILES stars to fit the massive ETGs in our sample, we cannot exclude some residual template mismatch, even after the careful selection of templates. 
The above tests have demonstrated that in such a situation one should minimise the use of polynomials  (Sec.~\ref{subsec:polytests}).
%

In principle, additive polynomials are only required in a fit if there is an actual additive component in the spectrum, which typically means an AGN and would thus only apply to the central few arc seconds of a galaxy. In that case, for those central spectra, the lowest possible order of additive polynomials should be used. Fortunately, in the case of our galaxies, there was no evidence for significant AGN activity \citep{Thomas2011,Rusli2013a, Rusli2013b, Thomas2014, Mazzalay2016, Erwin2018} and all central spectra could be fit sufficiently well without the use of additive polynomials (see Section \ref{sec:kinResults}). Therefore, we disabled the use of additive polynomials entirely.

By contrast, fitting observed galaxy spectra entirely without the use of multiplicative polynomials as well is unrealistic given residual imperfections in the data reduction process, e.g. in the flux calibration of the observations. Typically, even very extensively flux calibrated spectra, such as our MUSE data, cannot be fitted properly without multiplicative polynomials. 
To optimise the usage of the polynomials, we tried to determine the largest multiplicative polynomial 
 order that still allows for an unbiased recovery of the true LOSVD structure. The mock tests presented later in this section indicate that a multiplicative polynomial of fourth order is best for our data and fit range (see Fig. \ref{fig:Mocksummary}). It is likely that the optimal polynomial order depends on the specific data at hand, the fitted wavelength interval, the type and mass of galaxy etc.

Concerning the wavelength interval that we fit, MILES in principle allows to model the entire region $4700 - \SI{7000}{\angstrom}$. In the mock-tests above, we only focused on isolated or few features, but including more absorption features in the fit increases the constraints on the shape of the LOSVD.
Unfortunately, the MUSE range includes multiple spectral regions contaminated by under- or over-subtracted skylines, regions in which the detector is affected by other instrumental issues, 
and sometimes strong emission lines from ionized gas, such as $H\alpha$.
All these systematics can distort the recovered LOSVD shapes. We tried a variety of different setups, before settling for the wavelength region between $\sim 4800-\SI{6200}{\angstrom}$. This interval includes H$\beta$, the Mgb-triplet, several strong Fe-absorption lines and the NaD absorption feature. Our approach involves minimal spectral masking. The specific spectral masks and wavelength limits vary from galaxy to galaxy and are determined according to a strategy that is detailed in appendix \ref{ap:masking}. For the setup-test in this section the spectral masking was chosen to recover the shape of the LOSVD without additional distortions from different model spectra after adding to them mock-emission lines for $H\beta$, [OIII] $\SI{5007}{\angstrom}$ and [NI] $\SI{5199}{\angstrom}$ (with $v_{rot} = \SI{200}{km/s}, \sigma = \SI{150}{km/s}$). \footnote{Historically, emission lines were often removed from galaxy spectra, instead of spectrally masked. This, however, can bias the shape of the LOSVD if the emission is not removed correctly. Nonetheless, we also tested this approach by fitting different models with added emission lines using the in-built multi-component fitting capabilities of WINGFIT, using a simple Gaussian for the emission-line component,  and 8th order Gauss Hermite Polynomials for the stellar component. Then, in the next step we re-fitted the mocks again with only a stellar component and two different treatments of the emission lines: spectral masking and subtraction of the emission-line models from the two-component fits. We found that the recoveries were of similar quality, though always slightly better in the case of spectral masking, the latter typically having a factor $1.1 -1.2$ advantage in terms of RMS to the input LOSVD. Therefore, we here only use spectral masking.}

Notably, we did not mask the NaD feature for our best setup. Masking the NaD absorption feature  often resulted in worse recoveries of the LOSVD (Sec.~\ref{sec:NaD}).

%

\subsection{Verification of the setup}
\label{subsec:verif}
In order to verify our fitting approach, we finally describe a set of targeted mock-tests where we explicitly try to mimic the conditions under which we fit real massive ETGs and try to answer the question of how much we can trust the shape of the recovered LOSVDs. For these final tests we again created mocks based on synthetic stellar template spectra using $\textit{alf}$ and different LOSVD shapes. We fitted these with exactly the setup that we used for the observed galaxies to obtain an estimate of the expected scatter (and potential biases) in the measurements.



%

\begin{table*}[hbt!]
\centering
 \begin{tabular}{l c c c c c c c c c}
  \hline
 \hline
Mock & \multicolumn{4}{c}{Template}  & \multicolumn{5}{c}{LOSVD}  \\ 
 \cmidrule(r){2-5} \cmidrule(l){6-10} 
                &  [Z/H]     & [Mg/H] & [Fe/H]     & [Na/H] & $v_{1}$ $[km/s]$& $\sigma_{1}$ $[km/s]$  & $w_{2}$  & $v_{2}$ $[km/s]$  & $\sigma_{2}$ $[km/s]$ \\
                
                & (dex)     & (dex)  & (dex)      & (dex)    &                &                        &          &                   &  \\ 
                 \hline
Deep features + Wings                 & $0.233$    & $0.231$  & $-0.064$  & $0.523$    & $120$ & $370$  & $0.3$ & $0$ & $900$\\
Shallow features + Gaussian       & $0.206$    & $0.190$  & $-0.139$  & $0.456$    & $123$ & $374$  & $0.0$ & $-$ & $-$\\
Solar abundance + Gaussian     & $0.0$      & $0.0$    & $0.0$     & $0.0$      & $123$ & $374$  & $0.0$ & $-$ & $-$\\
\end{tabular}
\caption{Parameters of mock spectra: stellar population parameters of the $\textit{alf}$-generated synthetic templates and Gaussian parameters of the LOSVD.
For the latter, $v_{1}$, $\sigma_1$ are parameters of the dominant kinematic component while $v_{2}$, $\sigma_2$ belong to a secondary component with relative weight $w_2$. The total LOSVD is the weighted sum of the two components, with $w_1 \equiv 1 - w_2$
}
 \label{tab:mocks}

\end{table*}

{\bf{``Deep features + Wings''}}: 
First, we set out to create a mock spectrum that is representative of the type of galactic spectra in the centers of our ETGs where we measured the most significant LOSVD wings. To this end, we first derived a stellar population model from a fit with a fourth order Gauss-Hermite LOSVD to the central regions of NGC~1332 using $\textit{alf}$. A detailed description of our stellar population fitting procedure for the sample galaxies will be given in a different publication (Parikh et al. in press.). Next, we generated a galaxy mock spectrum based on these stellar population parameters and an LOSVD with typical wings extending to $v_{los} \sim \pm \SI{1700}{km/s}$. The latter was constructed as the weighted sum of two Gaussians, one narrower component for carrying most of the signal of the distribution and one broader Gaussian component for the wings (see the parameters in Tab. \ref{tab:mocks}). 

In the following, we will refer to this mock by the short-hand DEEP.

{\bf{``Shallow features + Gaussian''}}: 
In order to investigate how well we can recover the LOSVD shape despite the danger of ``hidden'' template mismatch we attempted to find a combination of a {\it wingless} LOSVD and a different template stellar spectrum that - together - combine to a spectrum that resembles the DEEP mock as closely as possible.  Therefore, we derived an alternative stellar population model by fitting the DEEP mock with \textit{alf}, constraining the LOSVD to have a Gaussian shape. In this way, we forced \textit{alf} to implicitly construct a stellar template spectrum that fits the mock under the constraint of a wingless, Gaussian LOSVD.
The resulting model served as the second mock for our mock-tests, which we will refer to by the short-hand SHALLOW.

\begin{figure}
\centering
 \includegraphics[width=0.9\columnwidth]{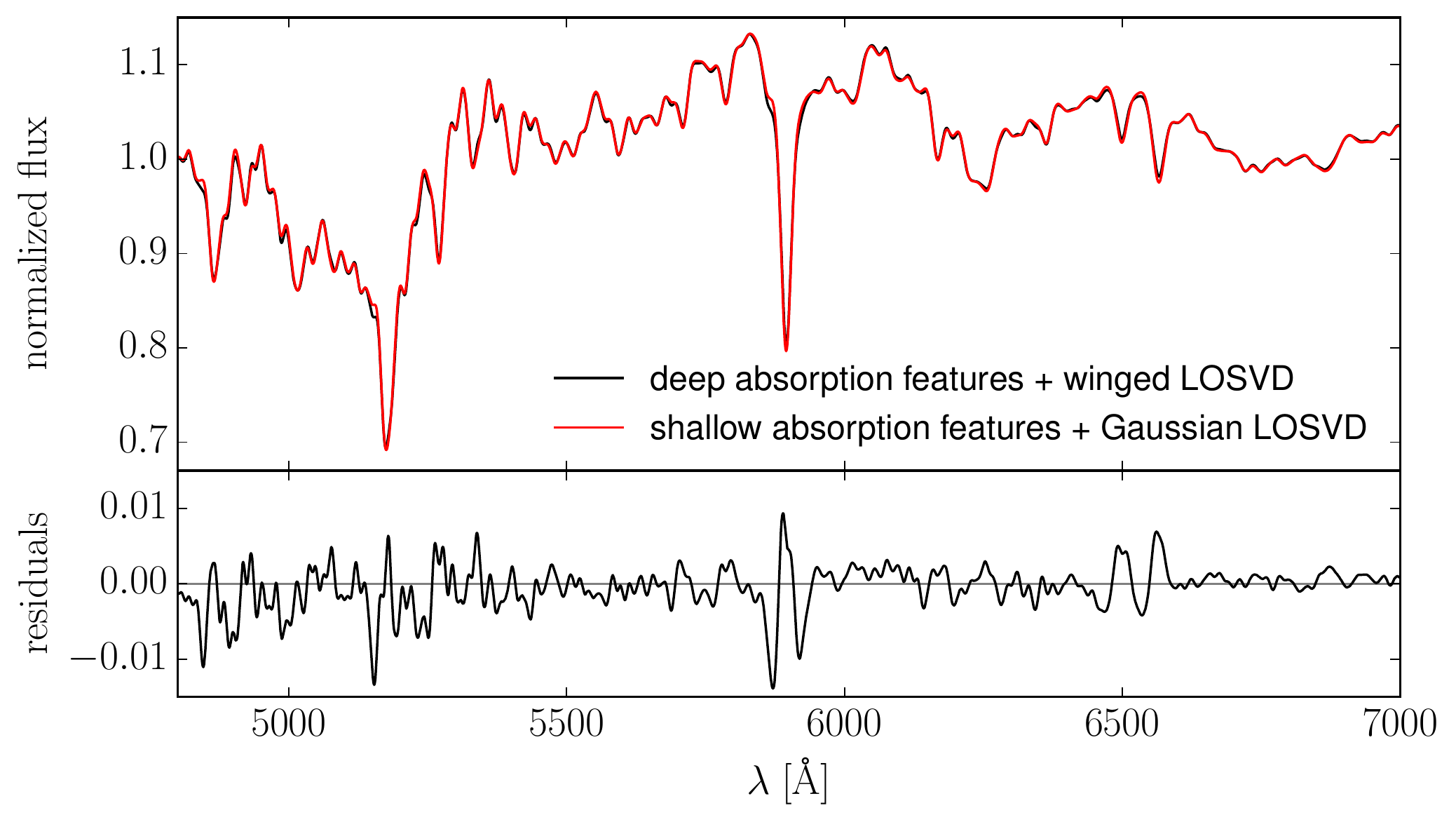}
 \caption{Deep features + Wings mock (black) compared to the Shallow features + Gaussian mock (red). While the mocks are based on different LOSVDs, the templates were adapted to make the resulting mock spectra as similar as possible. Differences in individual pixels are typically smaller than one percent.}
    \label{fig:gaussbiasModell}
\end{figure}

It is shown in Fig. \ref{fig:gaussbiasModell} in comparison with the DEEP mock. As can be ascertained from Tab. \ref{tab:mocks}, the LOSVD of the mock is essentially just the primary component of the DEEP mock, without the secondary wing component. The main difference to the latter mock in terms of the chemical abundances is the reduced [Fe/H] and [Mg/H] (as expected from the discussion in Sec.~\ref{sec:mocktesting}) resulting in \textit{intrinsically} shallower features that nonetheless, after both underlying spectra are convolved with their respective LOSVDs, appear to be almost equally deep. Indeed, we tried to provoke as much degeneracy between the models as possible: 
While the underlying population of the DEEP mock was \SI{12.8}{Gyrs}, a still plausible age, we allowed the age of the underlying population of the SHALLOW mock to assume even unphysical values, for the sake of imitating the DEEP mock as closely as possible, resulting in a stellar population age of \SI{15}{Gyrs}.

{\bf{``Solar abundance + Gaussian''}}: 
Finally, we created a reference-model whose stellar populations were purposefully easy to approximate with our empirical MILES templates, by convolving a solar-abundance template (generated by setting all abundance ratios to zero, but keeping the age of the underlying population of the DEEP mock) with the same Gaussian LOSVD as for the SHALLOW mock. The short-hand for this mock will in the following be SOLAR.

We generated 10 noisy realisations of each mock assuming a signal-to-noise ratio $\mathrm{SNR}=150$ (as in our MUSE data). We fitted all mock data sets with WINGFIT and averaged the resulting LOSVDs in each velocity bin and determined the scatter.

\subsubsection{LOSVD recovery in the absence of template mismatch}
\label{sub:idealrecov}
As a preliminary first test, we fitted the three mocks with non-parametric LOSVDs using an `ideal'' template set consisting of the three synthetic templates underlying the mocks. In this way, for each of the three mocks, one out of the three candidate templates in the template set was always the ``true'', exact template of that mock. 

\begin{figure*}
\centering
 \includegraphics[width=1.9\columnwidth]{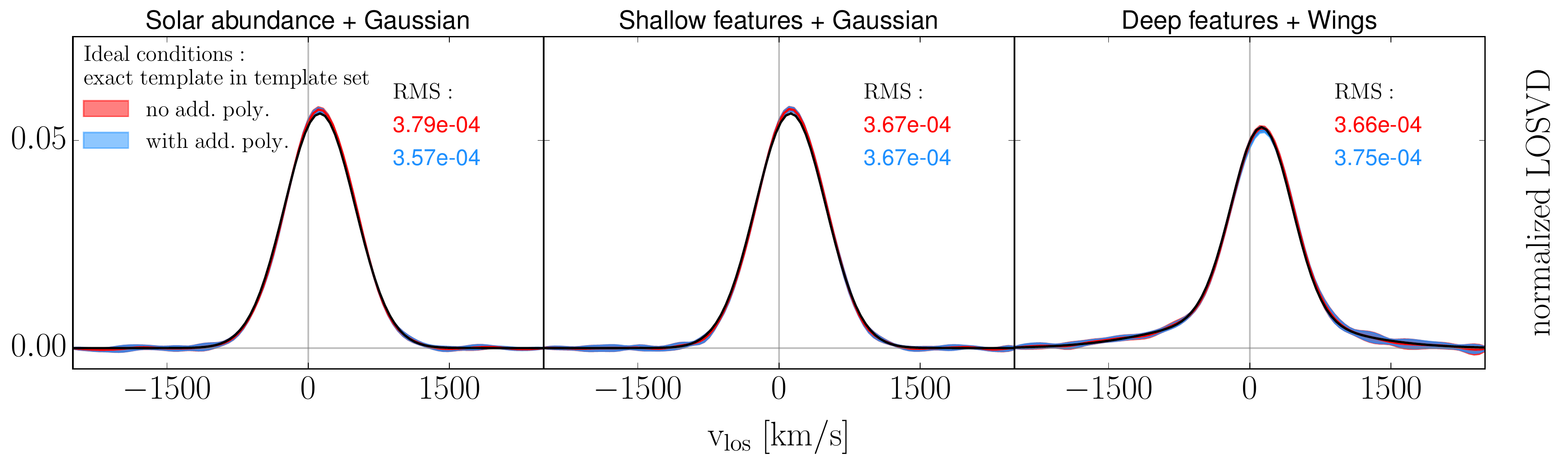}
 \caption{Non-parametric WINGFIT-recoveries of the LOSVDs of the SOLAR mock (left column), the DEEP mock (middle column) and the the SHALLOW mock (right column). All input LOSVDs are shown in black. As an example for setup-invariance of the recovery we show recoveries performed with- and without the use of additive polynomials (blue and red shaded, respectively). The barely visible width of the lines indicates the spread of the recoveries from 10 different noise realisations.} 
   \label{fig:idealMock}
\end{figure*}

The LOSVD recoveries are shown in Fig. \ref{fig:idealMock}. We tried a variety of different combinations of spectral regions for the fit, spectral masking, use/order of multiplicative and/or additive polynomials during the fit and found that the results remained invariant under these alterations:  in all cases, the recovery of the LOSVD was successful, irrespective of the LOSVD-shape, and in each fit the correct template was picked from the template set, whilst the templates belonging to the other two mocks were ignored. 

\begin{figure*}
\centering
 \includegraphics[width=1.9\columnwidth]{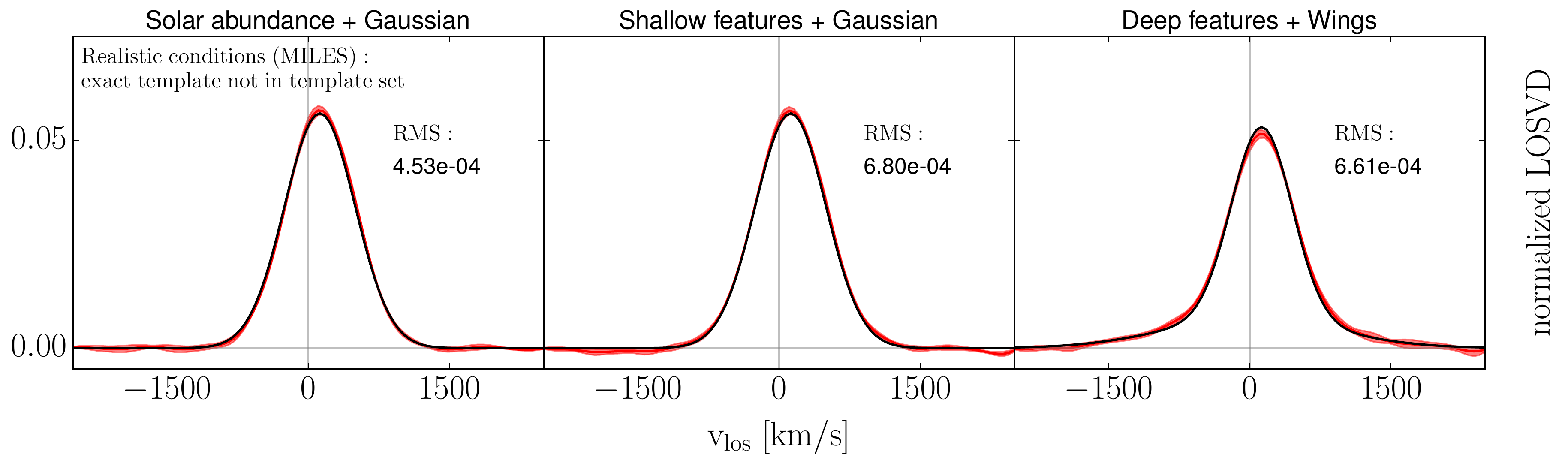}
 \caption{Summary of mock-tests: Non-parametric WINGFIT-recoveries of the LOSVDs of the SOLAR mock (left column), DEEP mock (middle column) and the the SHALLOW mock (right column) using our final fitting setup. All input LOSVDs are shown in solid black. The extend of the shaded areas indicate the spread of the recoveries from different noisy realizations of the corresponding mock-spectrum. We used a forth order multiplicative polynomial and no additive polynomials in the fit.}  
   \label{fig:Mocksummary}
\end{figure*}

\subsubsection{LOSVD recovery under realistic conditions}
\label{subsub:realrecov}

We now turn to the results under realistic conditions, i.e. when fitting with the MILES library.



{\bf{SOLAR}}: as expected, the recovery of the solar-abundance mock with the MILES stars does not pose any problems or biases. The LOSVD recovery is  essentially as good as under ideal conditions with the correct template stellar spectrum (Fig.~\ref{fig:Mocksummary} left). And, like the ideal fits which used the ``real'' template, the fit result is largely setup-invariant, i.e. it does not depend on the selected wavelength range, spectral masking or use of polynomials.


{\bf{DEEP and SHALLOW}}: The most interesting question is whether we can discriminate between the LOSVD shapes underlying the mocks with the intrinsically shallow and deep features, respectively, even though the mocks were created in such a way as to make them appear as similar as possible. As Fig.~\ref{fig:Mocksummary} shows, the recovery is indeed surprisingly good. The fits clearly reveal the different intrinsic LOSVD shapes underlying the two mocks. While the RMS between the true LOSVD and the reconstructed one did increase compared to the ideal situation with known template spectra, it is for both mocks still very low. In terms of an 8th order Gauss-Hermite parameterization, 
both the asymmetric and symmetric distortions of the LOSVD of the SHALLOW mock are negligible, $\Delta h_{3,5,7} < 0.01$, $\Delta h_{4,6,8} < 0.005$. Fitting the two component DEEP LOSVD with an 8th order Gauss-Hermite polynomial and comparing the sequence of even order Hermite moments as a representation of the strong wings, $h_4 = 0.067, h_6 = 0.033, h_8 = 0.024$, with those of the recovery, $h_4 = 0.063, h_6 = 0.017, h_8 = 0.006$, we can see that the recovery overall is quite successful, but becomes less accurate with increasing order. In terms of the actual non-parametric shape of the LOSVD, this amounts to a very slight overprediction of the LOSVD signal between $v_{los} = \pm \SI{500}{km/s}$ and $\pm \SI{1500}{km/s}$ and an even smaller underprediction of the LOSVD signal beyond $v_{los} = \pm \SI{1500}{km/s}$. There is also a small bias in $\sigma$ of $\Delta \sigma \sim 1\%$ for the SHALLOW mock and $\Delta \sigma \sim 5\%$ for the WINGS mock. Overall, however, the recovered LOSVD is consistent with the input model within the statistical uncertainties.



The results of these tests are representative
for the line strengths observed in the center of NGC~1332. The strengths of Mgb, Fe and NaD in the centers of the other galaxies of our sample are similar (e.g. Mgb $\sim 5.5-\SI{6}{\angstrom}$, <Fe>  $\sim 3.5-\SI{3.7}{\angstrom}$ and NaD $\sim 6-\SI{7}{\angstrom}$; Parikh et al., submitted). The only exception here being NGC~0307, the most lightweight ETG in our sample which is markedly closer to solar (Mgb $\sim \SI{4.7}{\angstrom}$, <Fe>  $\sim {3.2}{\angstrom}$ and NaD $\sim \SI{4.5}{\angstrom}$). Furthermore, all galaxies, including NGC~1332 and NGC~0307, follow the same radial trends, moving closer to solar with increasing radius (in the outer parts of the galaxies Mgb $\sim 4.2-\SI{4}{\angstrom}$, <Fe>  $\sim 2.5-\SI{3.2}{\angstrom}$ and NaD $\sim 3-\SI{4.7}{\angstrom}$; Parikh et al., submitted). This reduces the amount of template mismatch with MILES templates and consequently the LOSVD distortions, as the SOLAR test shows. Hence, our tests are representative for our whole sample and probe the strongest template mismatch we expect to find in the galaxies.
The lack of LOSVD distortions for these tests is particularly relevant given the fact that for both mocks there \textit{is} actually some residual template mismatch between the effective MILES template and the true one, namely such that the effective template is $\alpha$-deficient, relative to the $\alpha$-enhanced DEEP and SHALLOW spectra. In Fig. \ref{fig:Wingmechanism} we had simulated the effects of this template mismatch on the shape of the LOSVD in a more isolated fashion, i.e. using a smaller wavelength region and less template stars. 
One particular difference is that our fiducial setup for real galaxies that we use here includes the NaD feature. We note that NaD is often very strong in 
ETGs \citep[e.g.][]{vanDokkum2012, Conroy&vanDokkum2012, Spiniello2012}. One could have expected that our effective template underpredicts NaD, resulting in a suppression of wing-signal where genuine wings could exist. However, we do not find a mismatch between
the [Na/H] abundance in our effective MILES template with respect to the true one. Therefore, the inclusion of the NaD line did not bias the LOSVD recoveries in such a way that the distortion effects of the [Mg/Fe] mismatch were merely compensated in the opposite direction. In this case we would have only accidentally recovered the correct LOSVD shape. Instead, it is the strength of NaD that provides a vital constraint on the shape of the LOSVD which made the fit more robust against the mismatch present in the other features.

We note that when we used the setup with minimum RMS according to Tab.~\ref{tab:naiveFits} which made extensive use of higher-order polynomials and where we masked NaD, then the (symmetric) distortions of the LOSVDs were so strong that the recovered LOSVDs for both SHALLOW and DEEP were excessively winged and basically became indistinguishable from each other. Masking the second strongest feature in our wavelength range, Mgb, proved less critical, but still significantly over predicted the wings of the DEEP mock (raising the even order Hermite moments, particularly $\Delta h_4 \sim 0.03$). Masking Mgb also increased $\sigma$ (by $\sim 3\%$), but the effect of masking any feature also depends on the other features included in the fit and the particular mismatch of the templates that are used \citep[e.g.][]{Barth2002}.


Instead, using our best setup from this section, we conclude that even though the centers of ETGs are often $\alpha$-enriched by type II supernovae as a consequence of rapid early star formation \citep[e.g.][]{DThomas1999, DThomas2005, Conroy2014} our tests demonstrate that this is not an issue for an unbiased recovery of their LOSVD shapes.

We repeated these tests for SNRs of 20, 50 and 80 without encountering any additional bias in the LOSVD recovery\footnote{This was the case even for the lowest SNR = 20, despite the noise in this case is nominally larger than the differences between the mocks seen in Fig. \ref{fig:gaussbiasModell}. This appears to be an effect of the large number of pixels used in the fit, which still provide enough constraints for the LOSVD.}. At a SNR = 50, the recovery of the prograde wing of the DEEP-mock LOSVD looses its statistical significance (by contrast, the more extended retrograde wing can be recovered with significance even at a SNR = 20 up to $\sim \SI{1500}{km/s}$).


\section{Results}
\label{sec:kinResults}
\begin{figure*}[!t]
\centering
 \includegraphics[width=1.8\columnwidth]{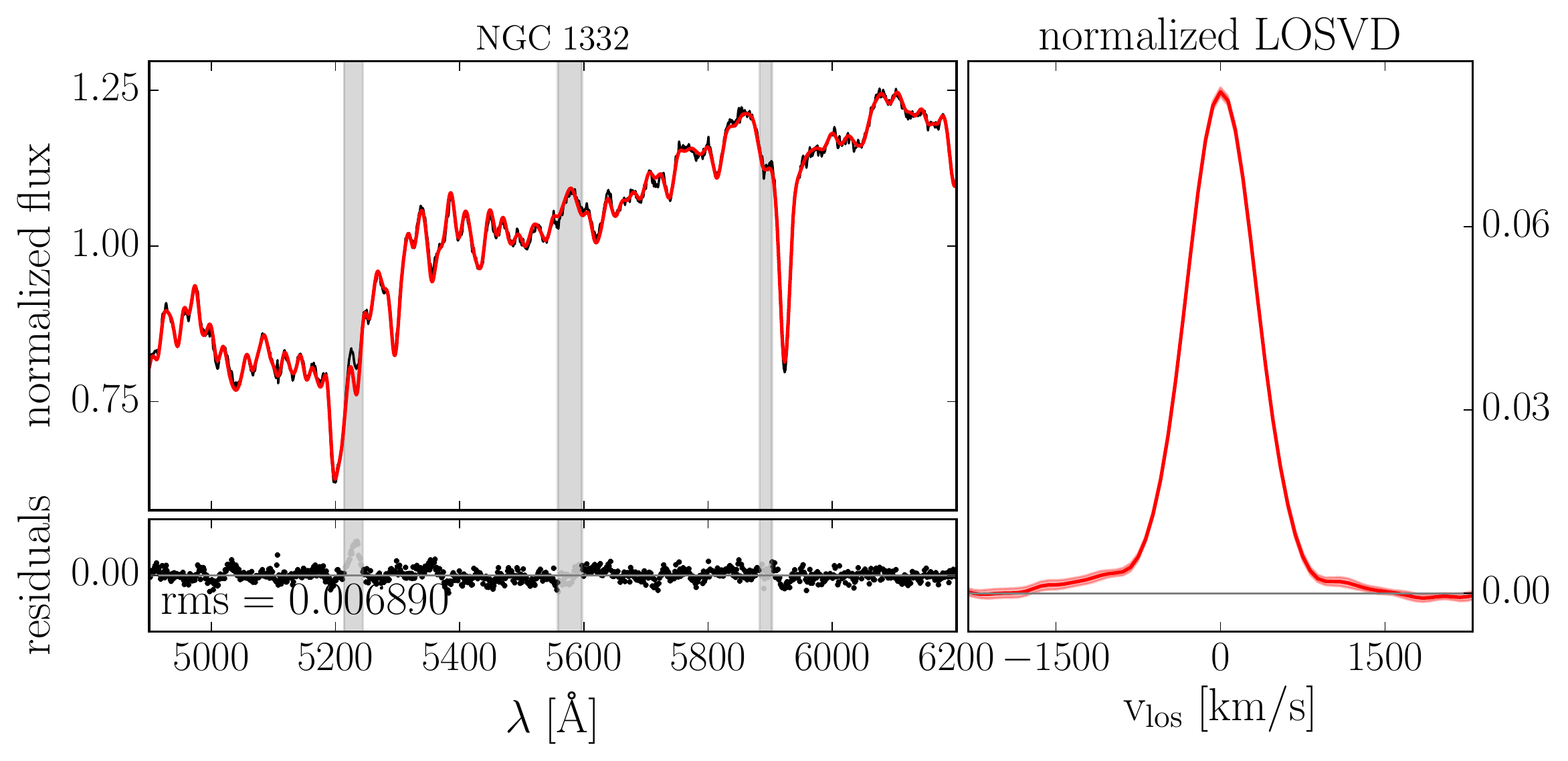}
 \caption{Left: Kinematic fit (red) to the spectrum (black) of the central Voronoi-bin of NGC~1332 using WINGFIT. Grey shaded areas indicate spectral regions which were masked during the fit. Right: Non-parametric LOSVDs (solid red) recovered from these fits. The shaded envelope indicates the statistical uncertainties of the LOSVD from 100 Monte-Carlo simulations. The line-of-sight velocities $v_{los}$ are relative to the systemic velocity of the galaxy.}
   \label{fig:exampleSpec}
\end{figure*}

In the previous Sections we have demonstrated that the non-parametric shapes of LOSVDs
can be measured with high accuracy and precision. We fitted non-parametric LOSVDs with WINGFIT for all spatial bins of our sample galaxies using the setup that turned out best in the previous Sections. As an example of these fits, Fig. \ref{fig:exampleSpec} shows a kinematic fit to a Voronoi-Bin from the central regions of NGC~1332. 



\subsection{Non-parametric LOSVDs} 
\label{sec:nonparaResults}

\begin{figure*}[t!] 
\centering
 \includegraphics[width=0.9\columnwidth]{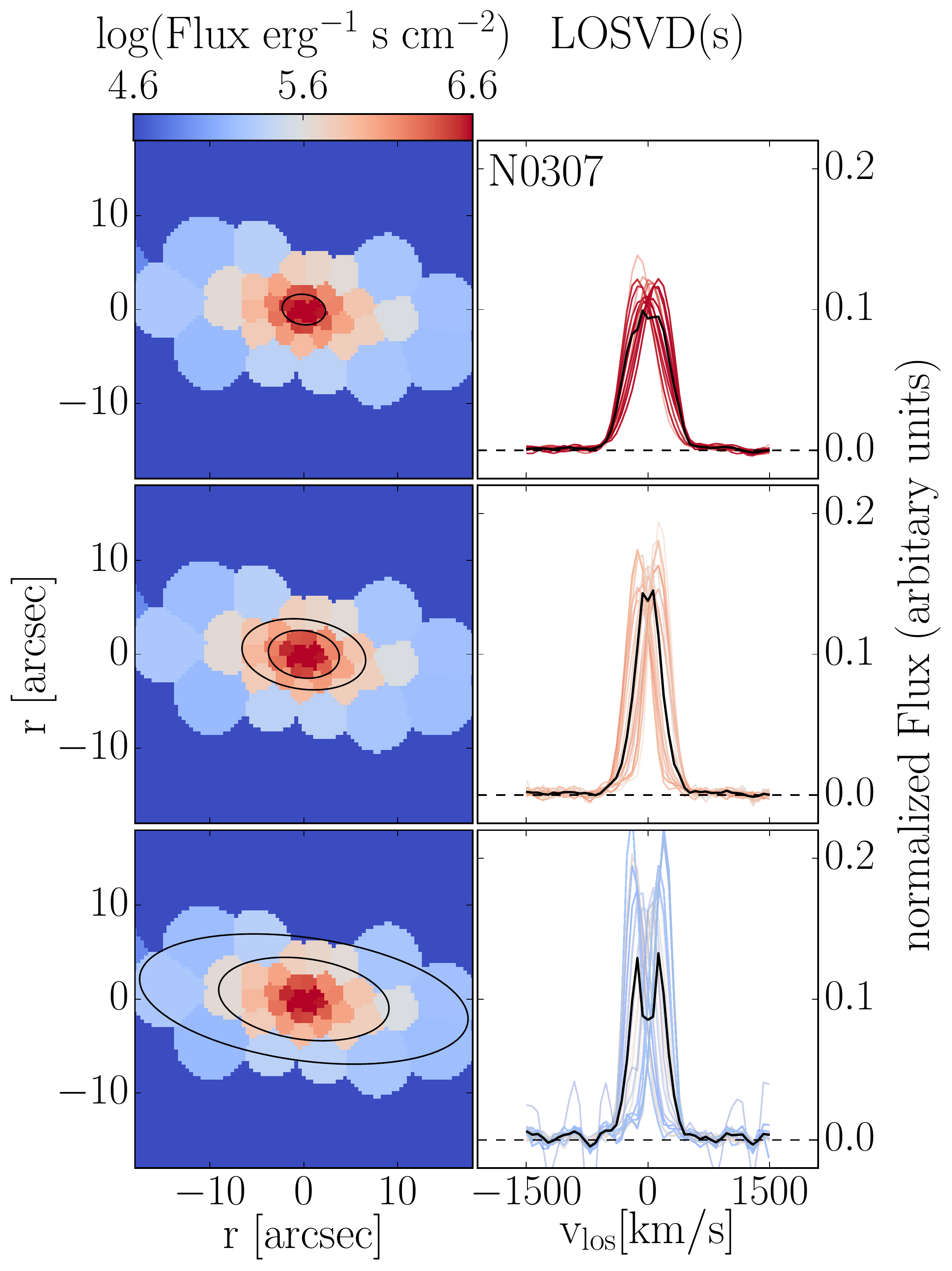}
  \includegraphics[width=0.9\columnwidth]{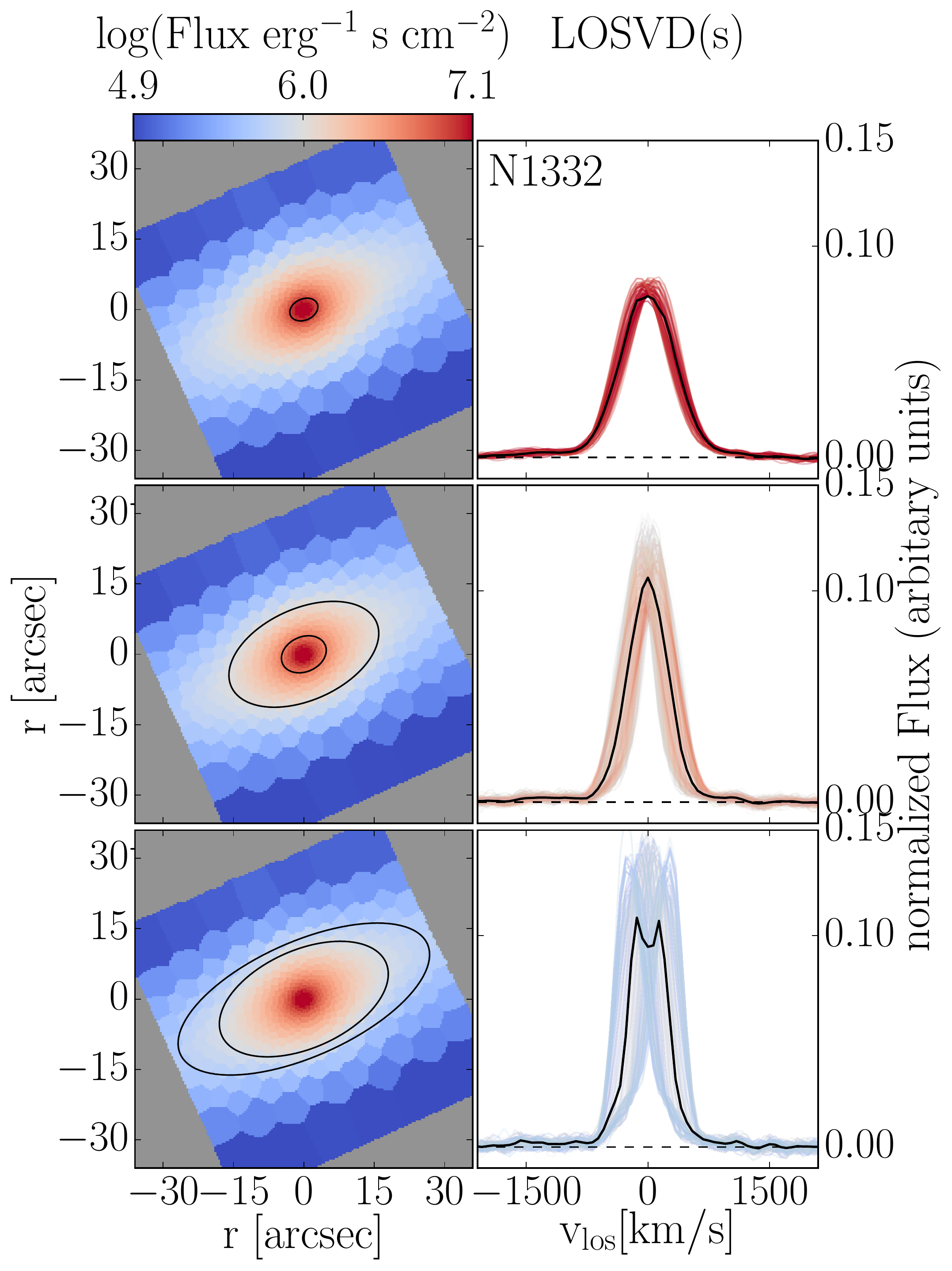}
   \includegraphics[width=0.9\columnwidth]{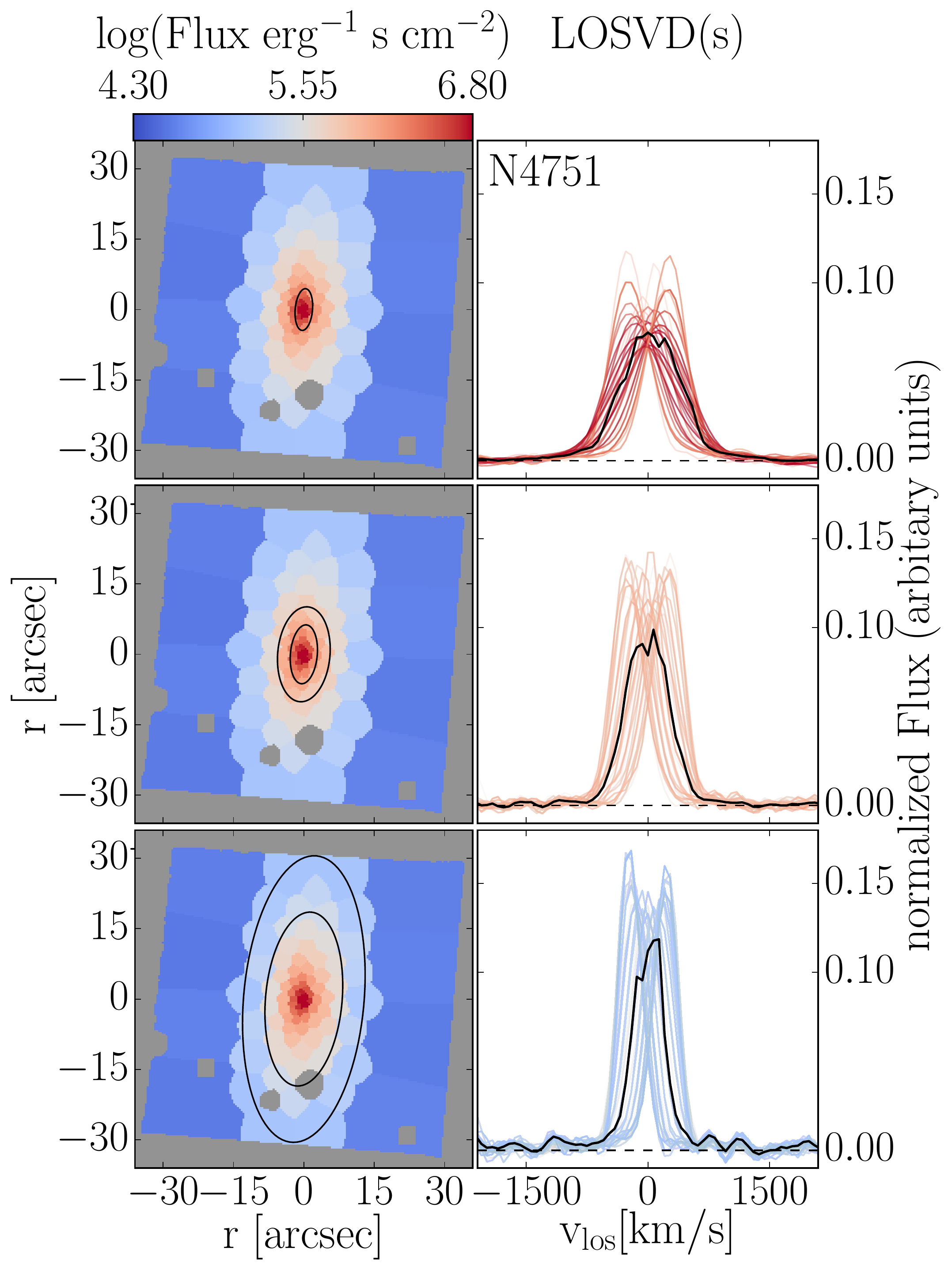}
    \includegraphics[width=0.9\columnwidth]{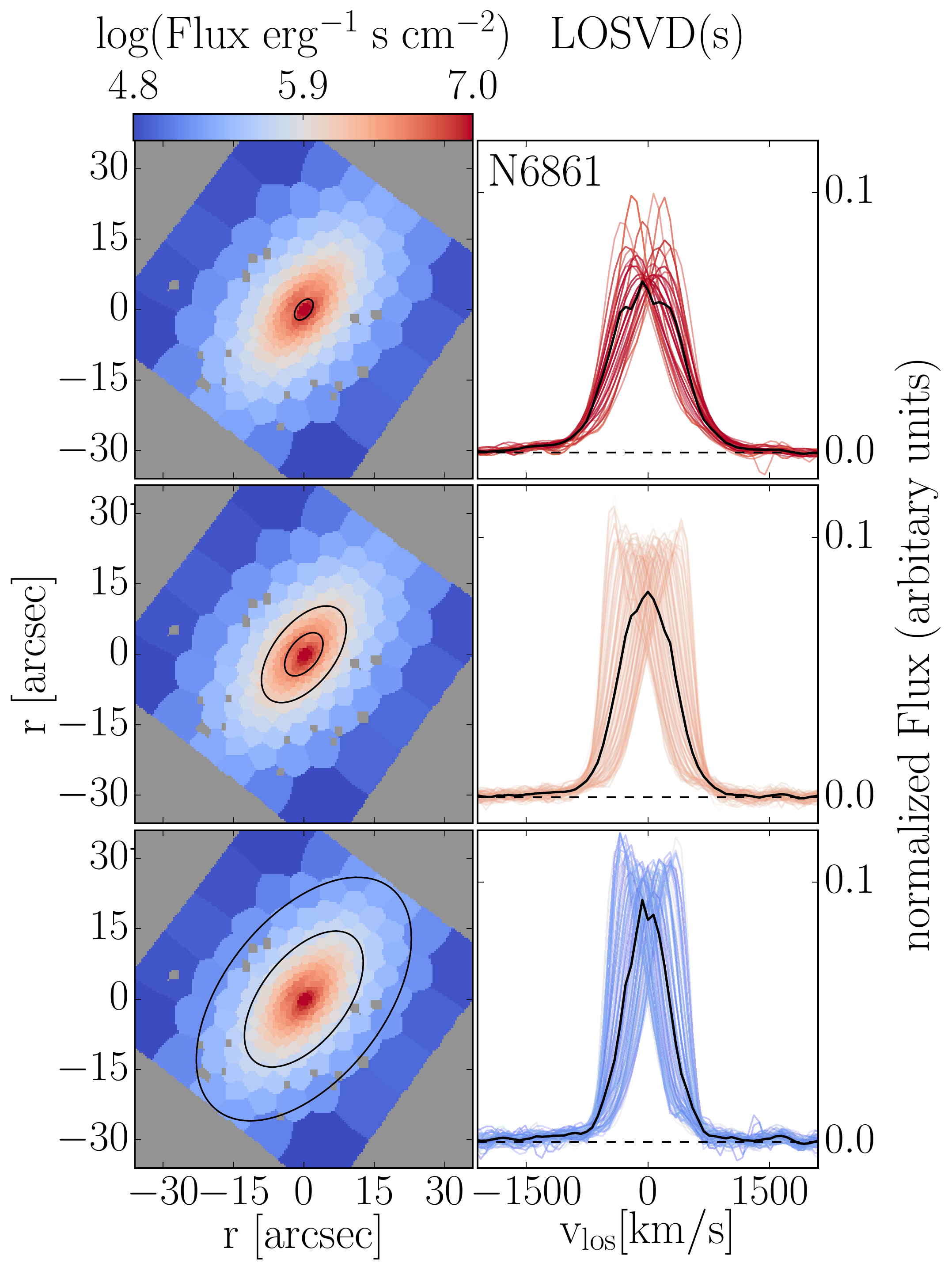} 
 \caption{Non-parametric LOSVDs for fast rotating ETGs NGC~0307, NGC~1332, NGC~4751 and NGC~6861. For each sub-figure, LOSVDs are shown across three different spatial regions. Each region, in turn, is represented by one row and defined by all bins between two isophotes (black lines) shown in the left panels of the sub-figures. The respective non-parametric LOSVDs from our WINGFIT analysis are shown in the right panels of the sub-figures. Each LOSVD is colored according to the flux in the respective Voronoi bin. LOSVDs in solid black show the light-weighted average LOSVD in the respective area. }
   \label{fig:FRellipses}
\end{figure*}

In Figs.~\ref{fig:FRellipses} and \ref{fig:SRellipses},  we show all non-parametric LOSVDs from our WINGFIT-analysis grouped into fast and slow + intermediate rotators, respectively. We show individual LOSVDs in different spatial regions, together with their light-weighted average over these regions.  While individual LOSVDs sometimes show oscillations larger than the statistical uncertainties from Monte Carlos simulations, the process of averaging LOSVDs over larger spatial regions helps to identify the most robust structures (in particular in the high-velocity tails of the LOSVDs). All galaxies, to a lesser or greater extent, show LOSVDs with wings. These are well defined in different parts of the galaxies but typically decrease from the center outwards. 

\begin{figure*}[!htbp]
\centering
 \includegraphics[width=0.9\columnwidth]{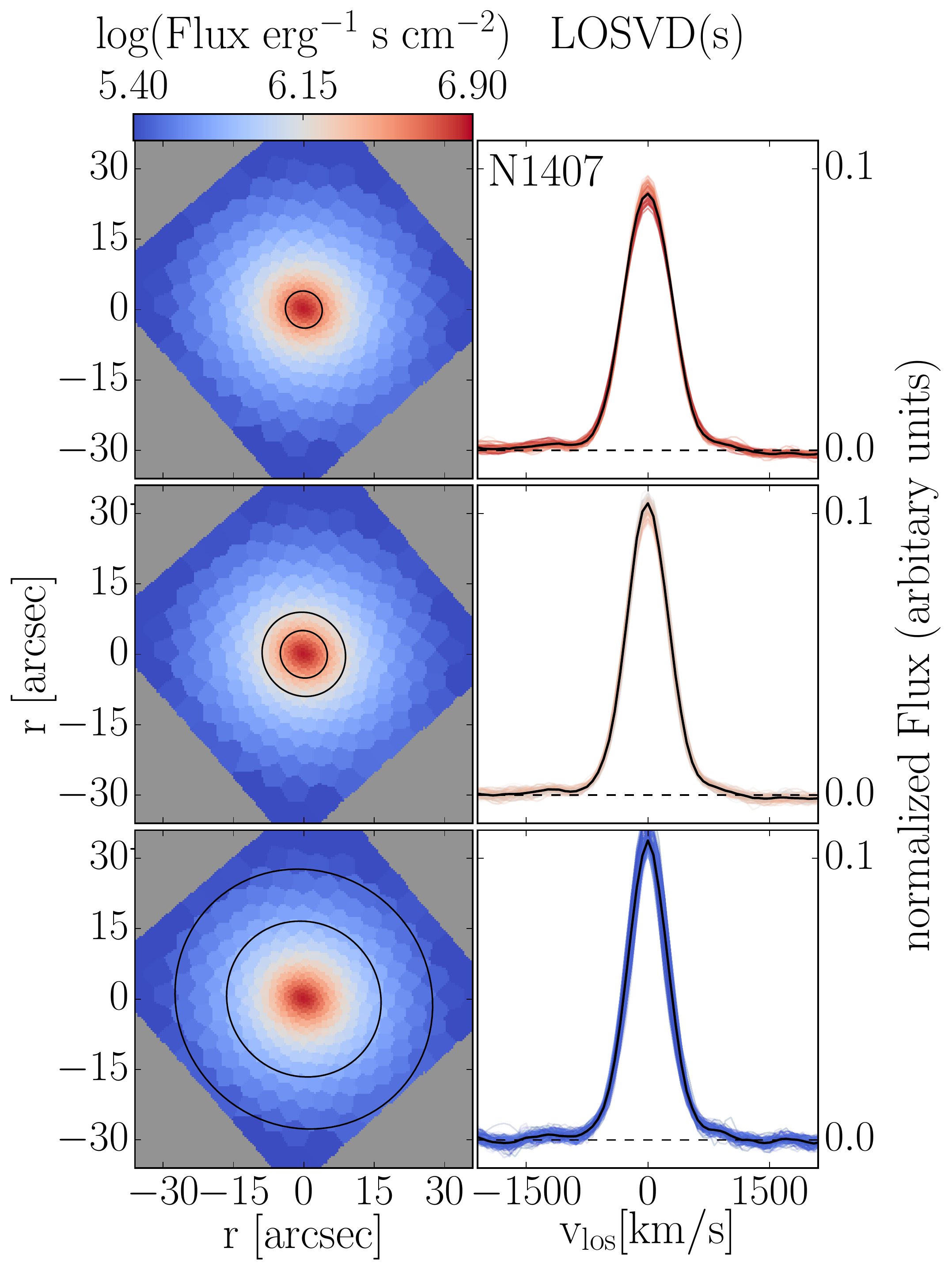}
  \includegraphics[width=0.9\columnwidth]{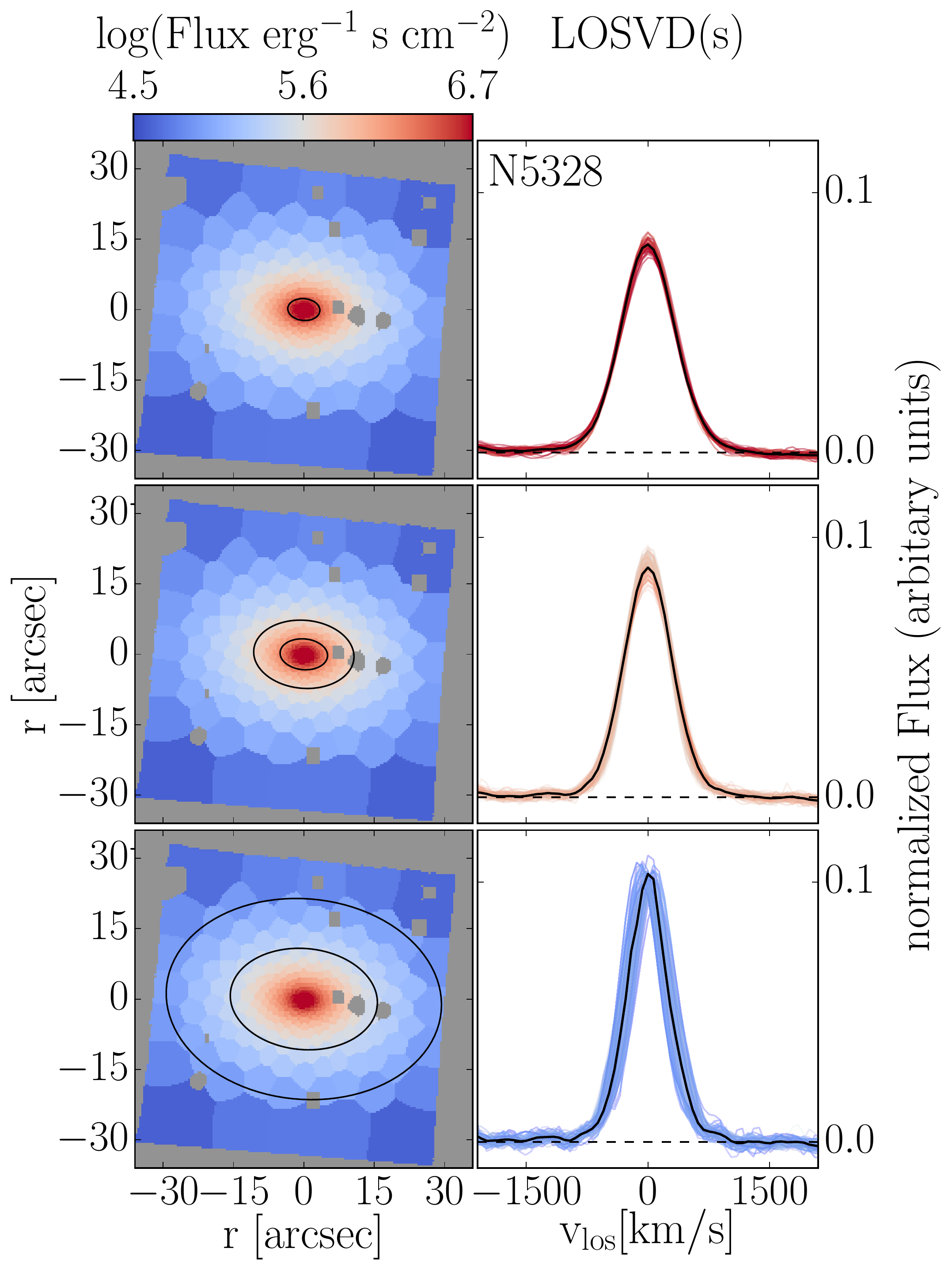}
   \includegraphics[width=0.9\columnwidth]{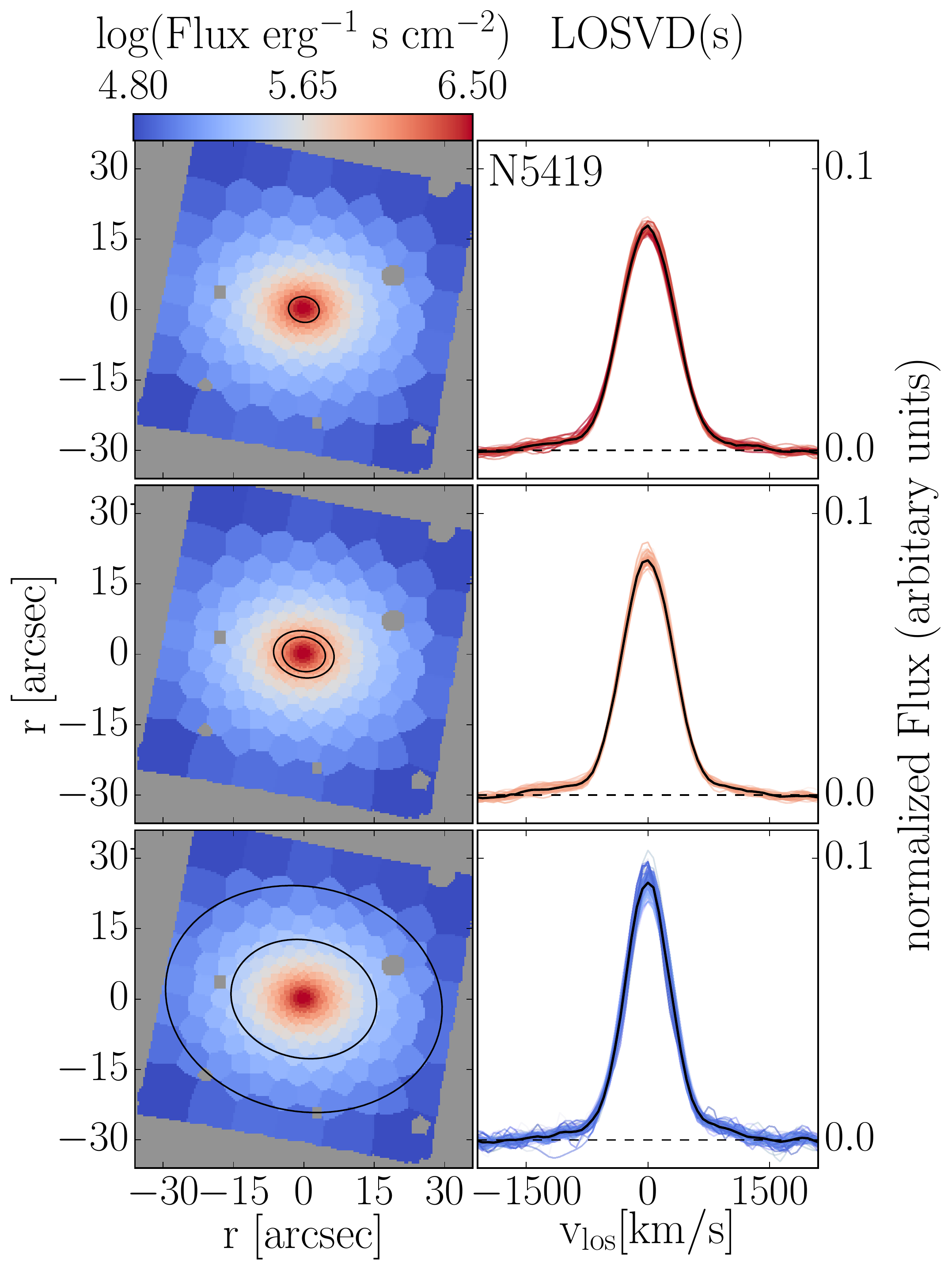}
    \includegraphics[width=0.9\columnwidth]{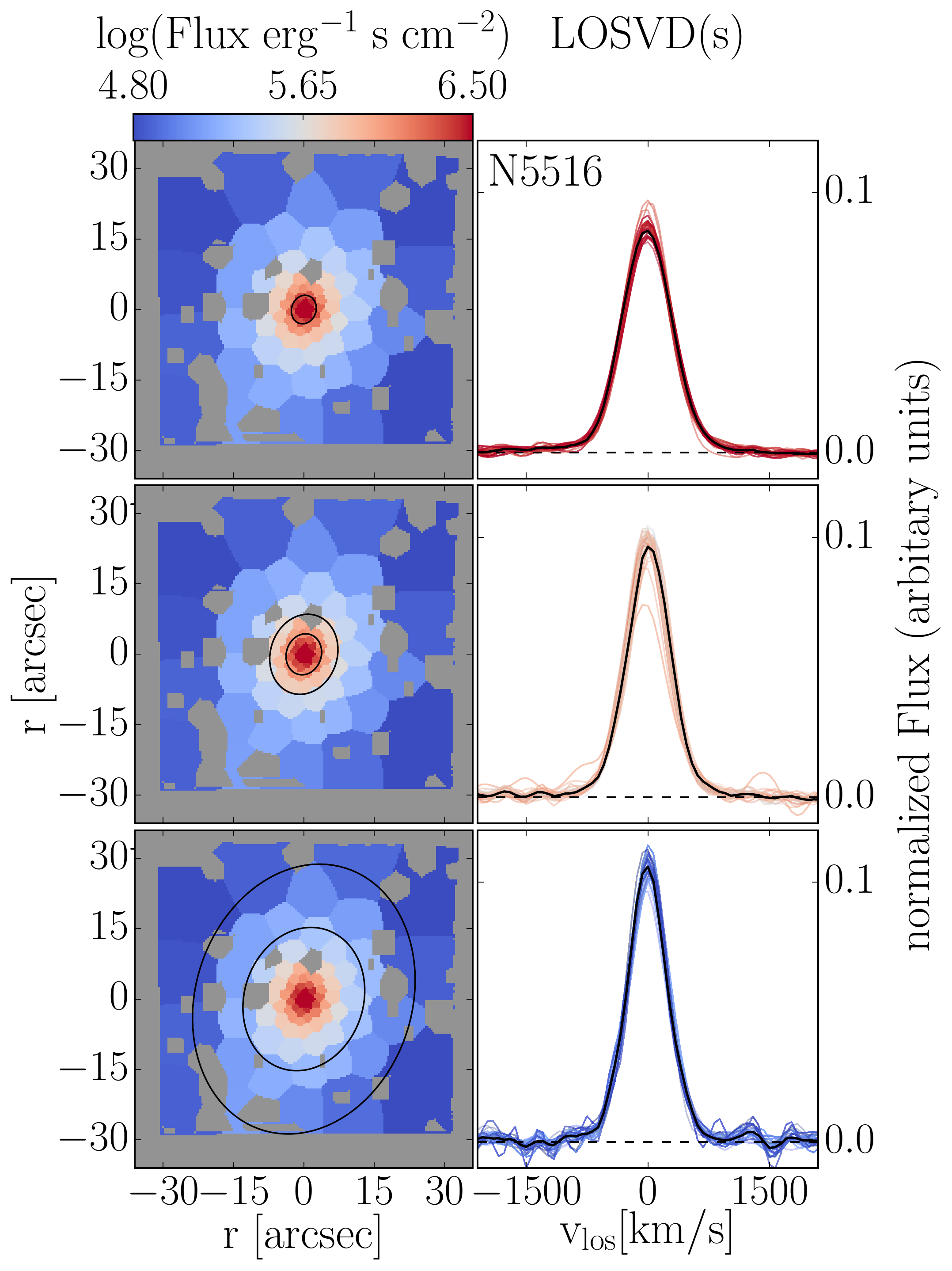} 
 \caption{Non-parametric LOSVDs for slow rotating ETGs NGC~1407, NGC~5328, NGC~5419, NGC~5516 and NGC~7619}
   \label{fig:SRellipses}
\end{figure*}

 \addtocounter{figure}{-1}
\begin{figure}[!htbp]
\centering
 \includegraphics[width=0.97\columnwidth]{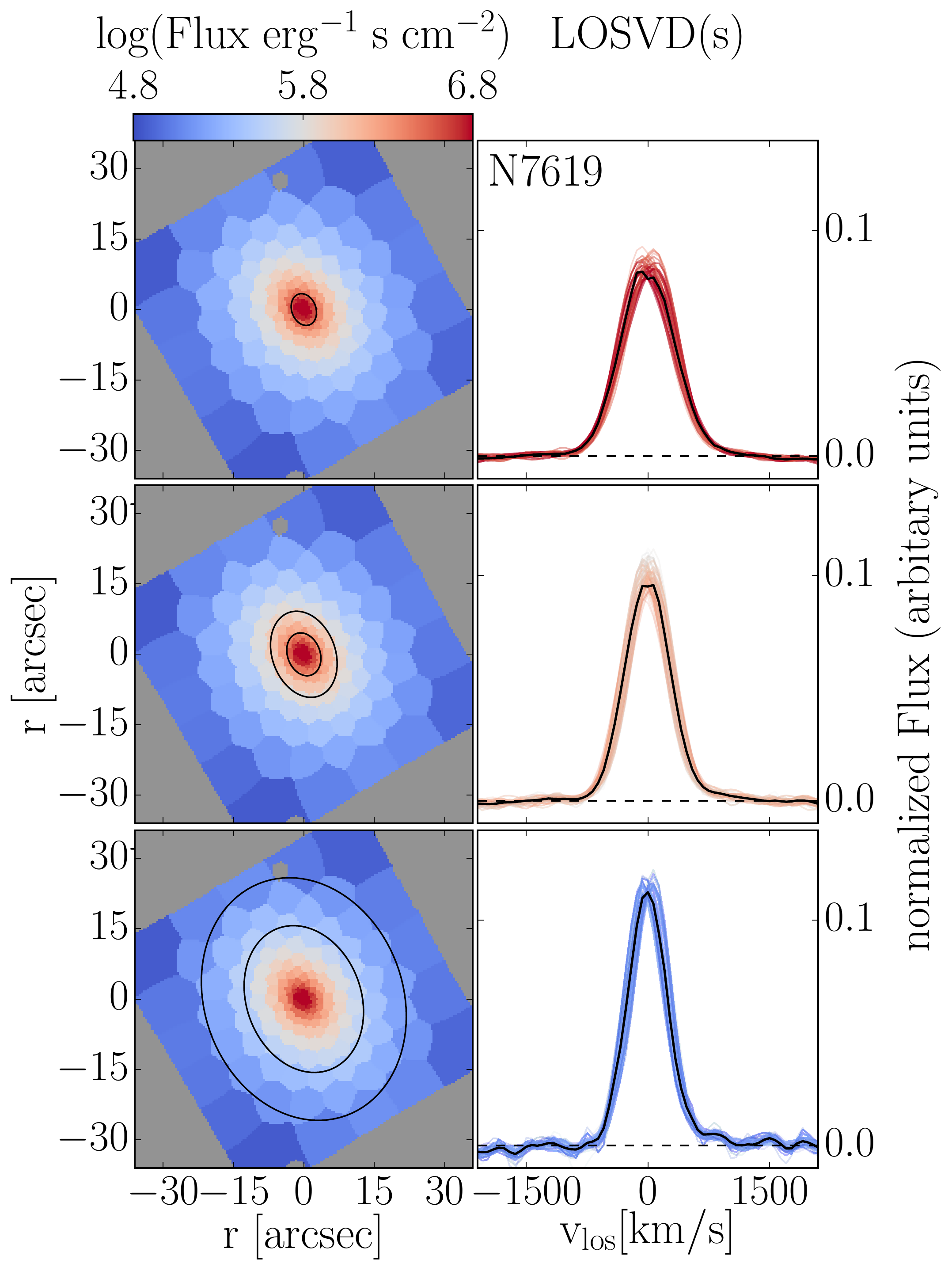}
 \caption{(continued)}
\end{figure}
\subsection{Parameterization of the LOSVDs:  higher-order Gauss-Hermite moments and large-scale trends of stellar kinematics}
Fitting the non-parametric LOSVDs from the previous section aposteriori with Gauss-Hermite polynomials, we also present the results of our stellar kinematic analysis in the form of 2D kinematic maps of the rotational velocity $v_{rot}$, velocity dispersion $\sigma$ and the higher-order Gauss-Hermite coefficients $h_3, h_4, h_5, ...$: Maps for the fast and slow + intermediate rotators are shown in Figs. \ref{fig:FRkinmap} and \ref{fig:SRkinmap}, respectively.  The maps help to highlight coherent stellar kinematic structures and patterns across the MUSE FOV. The highest order of Gauss-Hermite coefficients at which we can still visually identify coherent structures in the kinematic maps, is larger for the fast rotators in our sample, than for the slow + intermediate rotators. 
Therefore, we here present the kinematic maps of the fast rotators in our sample up to $h_{8}$ and the ones for the slow rotators up to $h_{6}$. These orders should be regarded as the highest \textit{common} order for either group. Individual galaxies show structures up to even higher-order moments.

In general, increasing orders of Gauss-Hermite moments permit an increase in the amount of signal at higher line-of-sight velocities, $v_\mathrm{los}$ in units of $\sigma$, which can be generated for a parametric LOSVD. As a result, higher orders of coefficients are necessary to produce LOSVD-wings. This is particularly important for power-law galaxies as rotation increases the distance of the peak of the LOSVD from the terminal velocity of the tails/wings on the opposite site of the rotation. Consequently, higher orders and larger values of Hermite coefficients are needed to represent the full shape of their LOSVDs. Higher orders also add complexity to the shape of the distribution, which is, again, particularly relevant for the power-law galaxies in our sample, as they have LOSVDs with a stark contrast between dynamically cold, low-dispersion, disk- or disk-like kinematic components and a broader, high-dispersion, bulge- and wings- components (Thomas et al., in prep).

In NGC~6861, the kinematic maps show such spatially coherent patterns along its major axis up to even higher moments than shown here. We will present a detailed discussion of the kinematics of this galaxy in a separate paper (Thomas et al. in prep). 

\subsubsection{Odd order Hermite moments}

\begin{figure*}
 \centering
 \includegraphics[width=1.7\columnwidth]{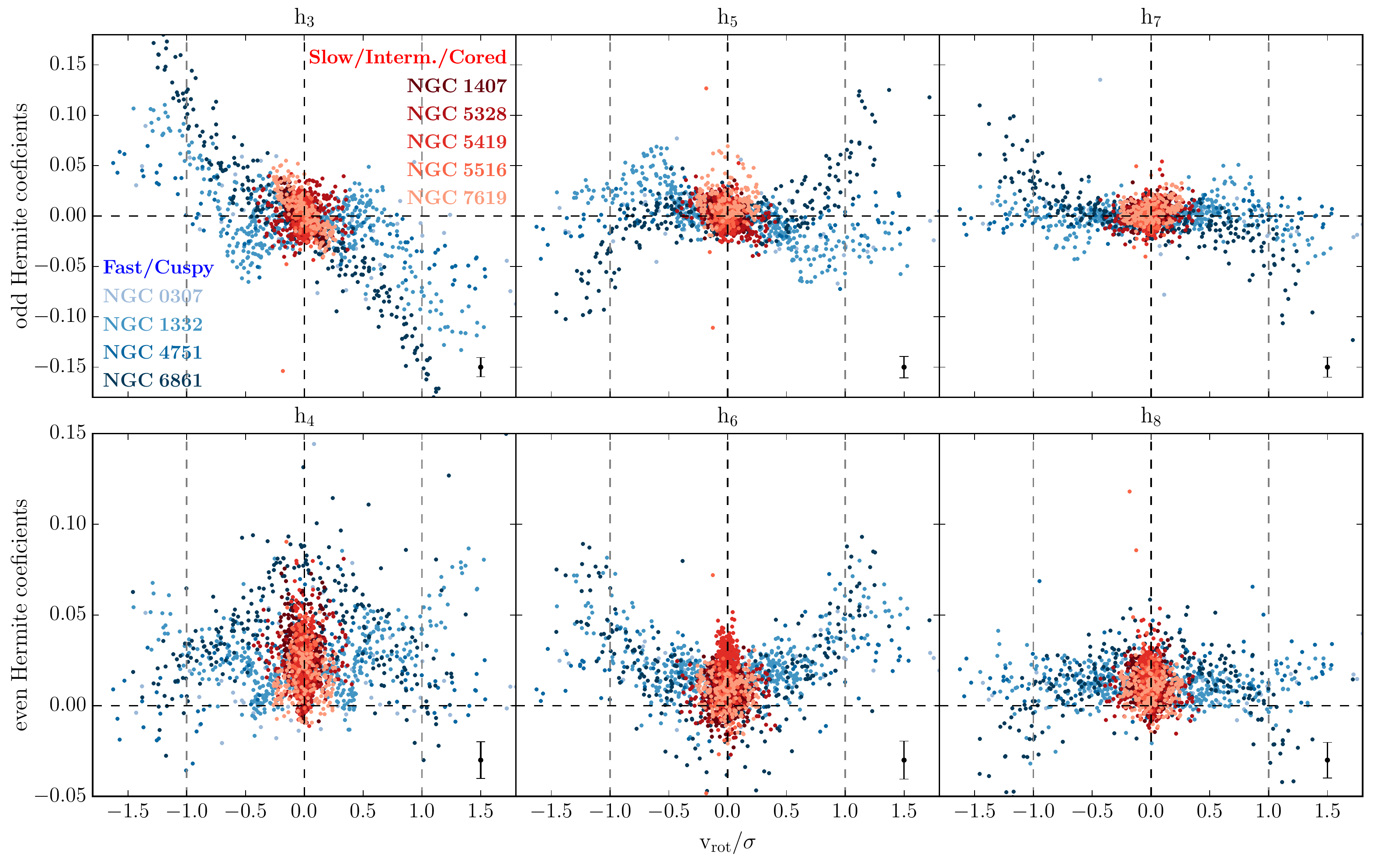}
 \caption{Higher order Hermite moments $h_3, ..., h_8$ against $v_{rot}/ \sigma$, separated into odd (first row) and even order moments (second row). 
The average error bar of each Hermite moment over all galaxies is indicated in black in each panel.} The Hermite moments of fast rotating, power-law ETGs (shades of blue) cover on average a larger dynamic range than those of slow \& intermediate rotating, cored ETGs (shades of red). The even order moments of both galaxy types are on average positive, while the odd moments are centered on zero. Lines at $v_{rot}/ \sigma = \pm 1$ indicate transition points between dynamically hot and cold regions.
   \label{fig:RotationProfiles}
\end{figure*}

Fig.~\ref{fig:RotationProfiles} shows the higher order Gauss-Hermite moments of the ETGs against $v_{rot}/\sigma$, separated into fast and slow+intermediate rotators by color.

Odd order Hermite moments add asymmetries to the LOSVDs.
In the first row of the figure, the $h_{3,5,6}$ profiles of fast rotating power-laws are spread out wide over $v_{rot}/\sigma$, with the profiles developing pronounced, curved, sometimes spiral-arm like $v_{rot}$ - $h_{2n+1}$ (anti-) correlation patterns. Correlations or anti-correlations often reverse their order from one odd order moment to the next, resulting in the oscillation of coefficients around zero between these moments seen in the kinematic maps. In general, odd order Hermite moments are largely influenced by projection effects related to rotation, e.g. the famous
anti-correlation between $v_\mathrm{rot}$ and $h_3$ in most axisymmetric galaxies. These typical hallmarks extend beyond $h_{3}$ to higher odd Hermite coefficients.

The spiral-arm shapes originate from the slope changes in the diagrams, most notably around $|v_{rot}/\sigma| \sim 1$ and $|v_{rot}/\sigma| \sim 0.5$. Typically, such slope changes are associated with transitions between different dynamical components, such as disks and bulges \citep[e.g.][]{Erwin2015, Veale2018}. This is particularly visible for NGC~6861, were the $h_{2n+1}$ (anti-) correlations with $v_{rot}$ become much steeper for the transition from the dynamically hot to the dynamically cold regime, $|v_{rot}/\sigma| > 1$. This is also clearly visible in the kinematic maps (see e.g. NGC~6861 in Fig. \ref{fig:FRkinmap}), where spatial regions with the strongest odd order Hermite coefficients trace out the major axis of the galaxy. 

NGC~1332 is a particularly interesting case:  here, slope changes around $|v_{rot}/\sigma| \sim 0.5$ result in a transition from $v_{rot}-h_3$ correlation to anti-correlation, which then - via the oscillation of coefficients - reverberates to $h_5$ and to a lesser extend, $h_7$. The region of $v_{rot}, h_3$ correlation within $|v_{rot}/\sigma| \sim 0.5$ is also very noticeable in the  $h_3$-map for the galaxy (see Fig. \ref{fig:FRkinmap}), forming a ``butterfly-shape'', that, as we will argue later on, could be indicative of the presence of a bar component.

By contrast, slow + intermediate rotating cores cluster around $v_{rot}/\sigma = 0$ and $h_{3,5,7} = 0$. There is only very minimal $v_{rot}-h_{2n + 1}$ (anti-) correlation and odd moment-to-moment oscillations, with the notable exception being the intermediate rotator NGC~7619, which shows steep $v_{rot}-h_3$ anti-correlation, which also features prominently in the galaxy's kinematic maps (see Fig. \ref{fig:SRkinmap}). None of the slow + intermediate rotating galaxies have dynamically cold components, $|v_{rot}/\sigma| > 1$.

Overall, for all galaxies, the odd order Hermite moments are largely unbiased, averaging out to zero over the FOV, which is also clearly visible in the kinematic maps, suggesting that any residual asymmetric bias in the shape of the LOSVDs caused by template mismatch is likely very small.

\subsubsection{Even order Hermite moments}
Even order Hermite moments add symmetric components to the LOSVDs. They are shown in the second row of Fig.~\ref{fig:RotationProfiles}. The most notable trend is that the coefficients are overall offset from zero towards $h_{2n+2} > 0$ for all types of galaxies -- in contrast to the odd moments.
The fast rotating power-laws show mostly linear relations between $|v_{rot}|$ and $h_{4,6,8}$, which are, unsurprisingly, symmetric around $v_{los} = 0$.  Just as with the odd order coefficients, we find oscillations of the values of the coefficients from one even order to the next, albeit around values $h_{2n+2} > 0$ for most $v_{rot}/\sigma$. 

For the slow + intermediate galaxies, there appears to be little or no correlation with rotation for the even order moments. Instead, as can be seen in the kinematic maps, the even order moments are more correlated with radius than with rotation, in the sense that the even order coefficients typically increase/decrease at larger radii, while odd order coefficients are generally strongest wherever the rotational velocity is strongest. Typically, the sign of the even moments does not change from one order to the next in case of the slow rotators. These even moments are probably related to the contrasts between the widths of the narrower main part of the LOSVD and the broader wings. 

However, neither $h_4$ nor other higher-order, even Gauss-Hermite moments can be used to parameterize the wings in a straight-forward manner. This is because even order moments not only govern the tail-light and cut-off velocity, but also the narrowing or flat-topping of the trunk of the LOSVD. Consequently, adding e.g. wings to a LOSVD with a flat-topped or bimodal shape,
can result in a LOSVD with a net-$h_4$ smaller than that of a LOSVD with a narrower peak but smaller wings. 

%

We note that NGC~6861 seems to have particularly large values of $h_4$ within $|v_{rot}/\sigma| \sim 0.5$, compared to other fast rotators. This could be indicative of a larger, dynamically hot component in the galaxy contrasting against a narrower, more flattened component by way of $h_4$. However, this is likely also partly due to dust, as the $h_4$ map of the galaxy (see Fig. \ref{fig:FRkinmap}) shows its largest $h_4$ values in spatial regions associated with patchy dust bands within the galaxy's disk. We discuss the effect of dust on the measurement of the stellar kinematics in the next section.

\subsection{Radially resolved Angular momentum}
\label{subsec:angmom}
Using our new kinematics, we create profiles of the angular momentum parameter $\lambda$ of \citet{Emsellem2007, Emsellem2011} against radius, shown in Fig.~\ref{fig:LambdaProfiles}. As expected, the profiles are consistent with the typical fast/slow rotator dichotomy. However, three galaxies show trends that are worth singling out: 
NGC~7619, one of the intermediate rotating galaxies, has a $\lambda$-profile perfectly consistent with that of the fast rotators in our sample for small $r/r_e$, before shallowing out to a profile that is more similar to the slow rotators, but with overall larger $\lambda$ at the same $r/r_e$. Here, the ``intermediate'' classification seems to be particularly apt. Both NGC~5328 and NGC~5419 have local maxima of $\lambda$ around $r/r_e \sim 0.1$. This is due to the fact, that both galaxies exhibit counter rotation in their central regions (cf. the $v_{rot}$-maps in Fig.~\ref{fig:SRkinmap}). An inner maximum of the angular momentum -- although weaker -- is also present in NGC~1407. While the galaxy shows some kinematic twist in its center, there is no sign of counter rotation. All three galaxies appear to have double velocity dispersion peaks in their centers. 

\begin{figure*}
\centering
 \includegraphics[width=1.6\columnwidth]{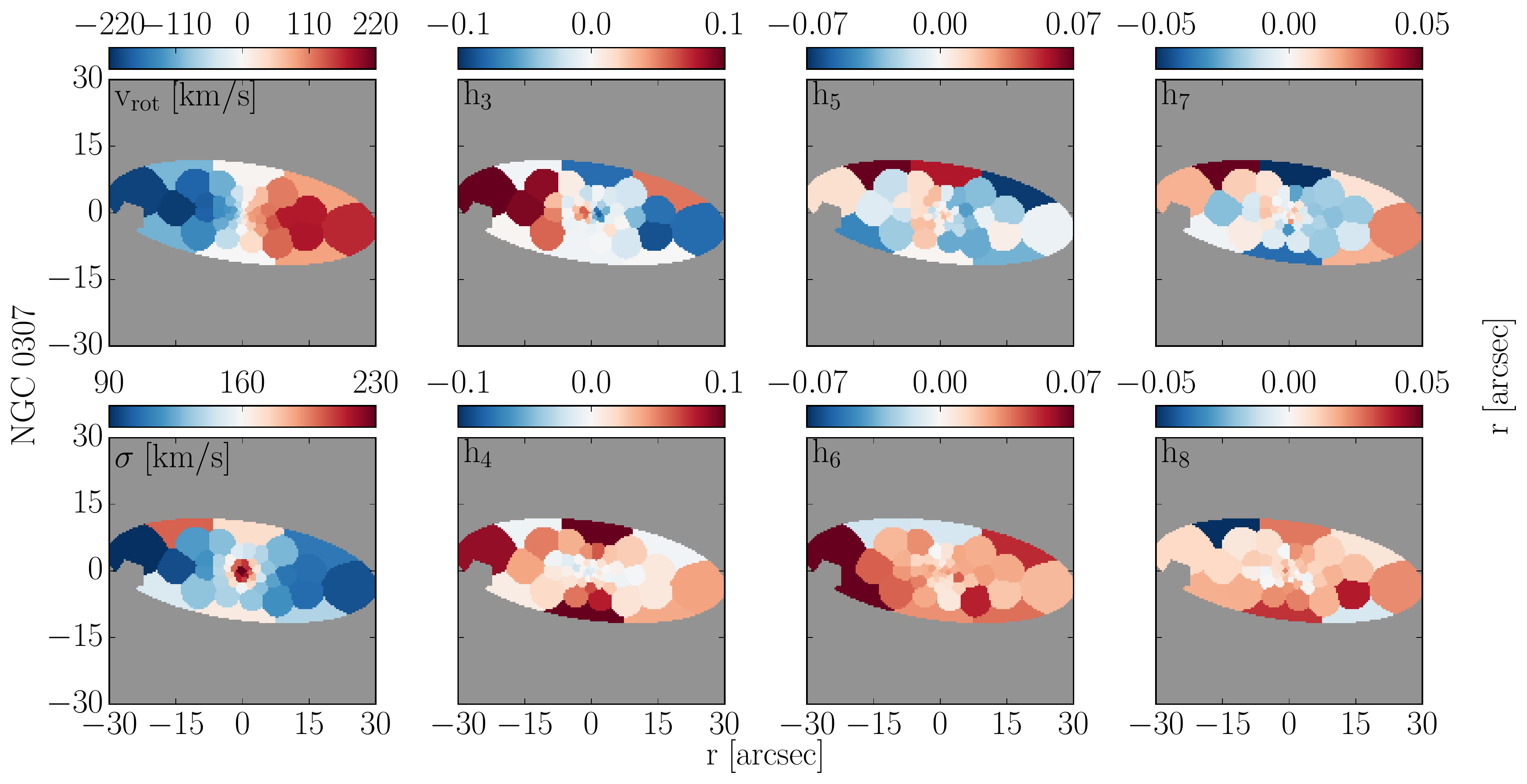}
 \includegraphics[width=1.6\columnwidth]{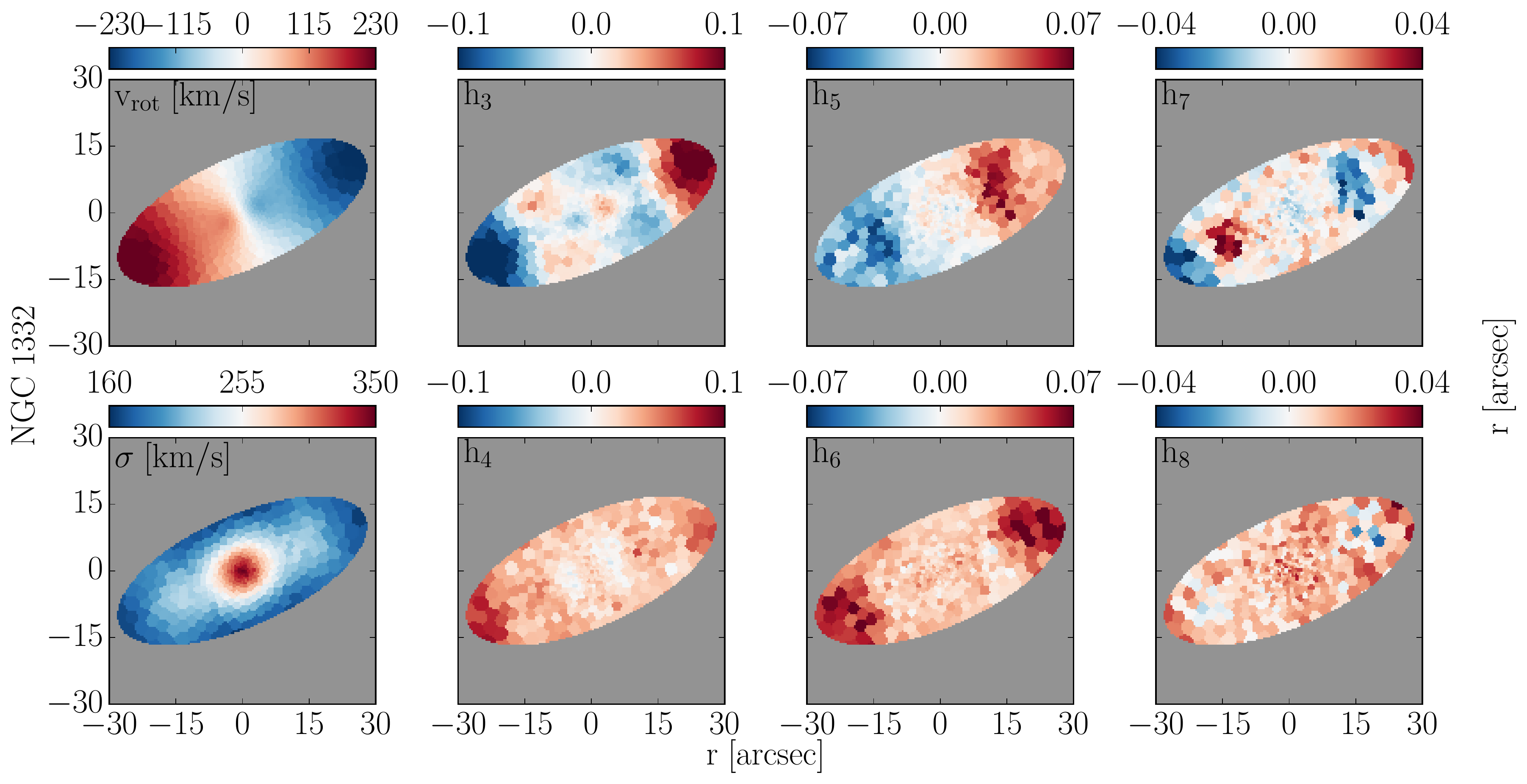}
  \includegraphics[width=1.6\columnwidth]{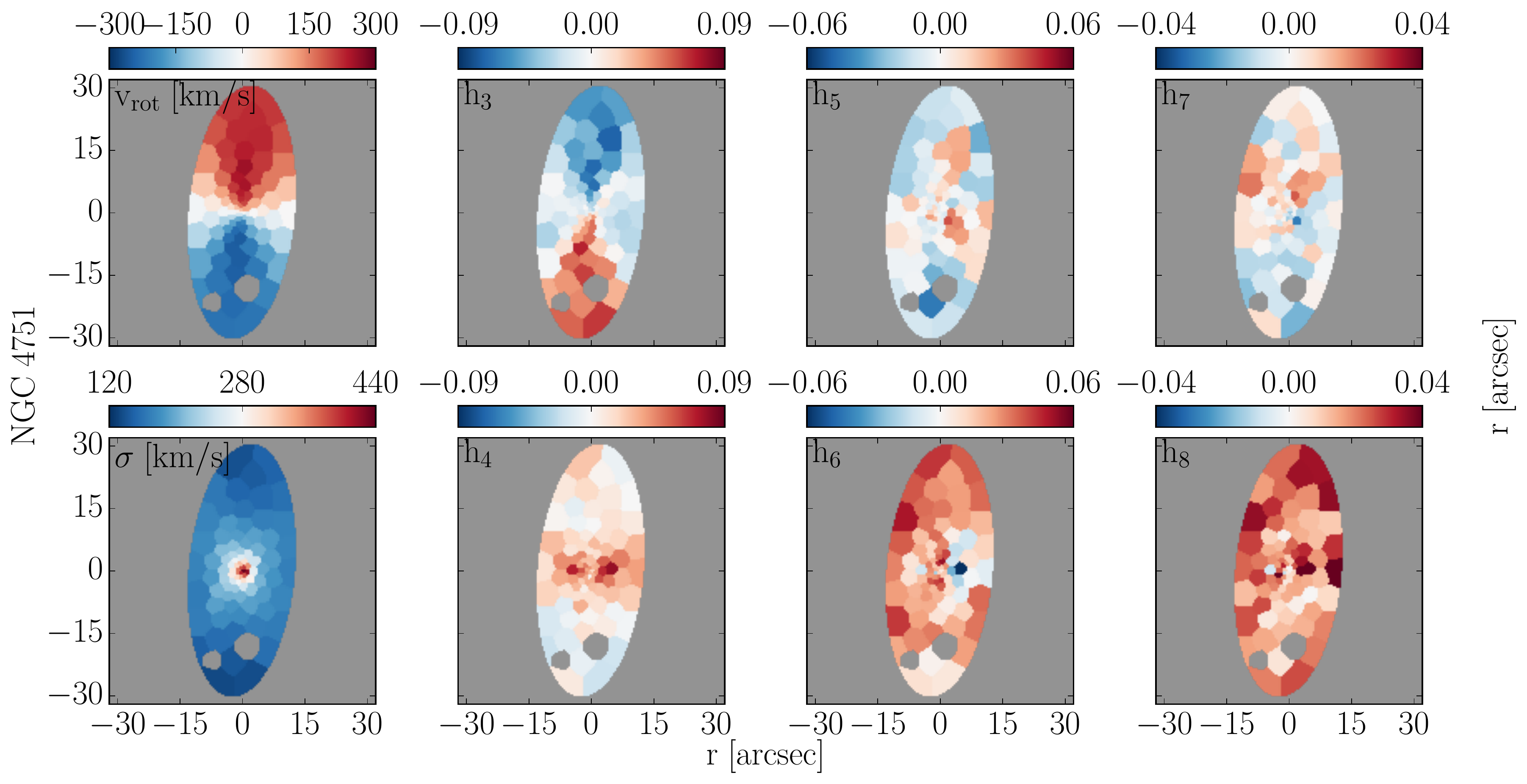}
 \caption{From left to right: MUSE-based 2D stellar kinematics of fast rotating ETGs NGC~0307, NGC~1332, NGC~4751 and NGC~6861.
   In each sub-figure, from top-left to bottom-right: maps of the
   rotational velocity $v_{rot}$, velocity dispersion $\sigma$ and 
   higher-order Gauss-Hermite coefficients $h_3$ - $h_{8}$. We only show pixels inside the largest isophote that fits wholesale into the MUSE FOV (gray areas inside this region have been excluded from the binning). This implicitly indicates the positions of the major and minor axes and gives a general idea of each galaxy's morphology.}
   \label{fig:FRkinmap}
\end{figure*}
 \addtocounter{figure}{-1}
\begin{figure*}[t]
\centering
 \includegraphics[width=1.6\columnwidth]{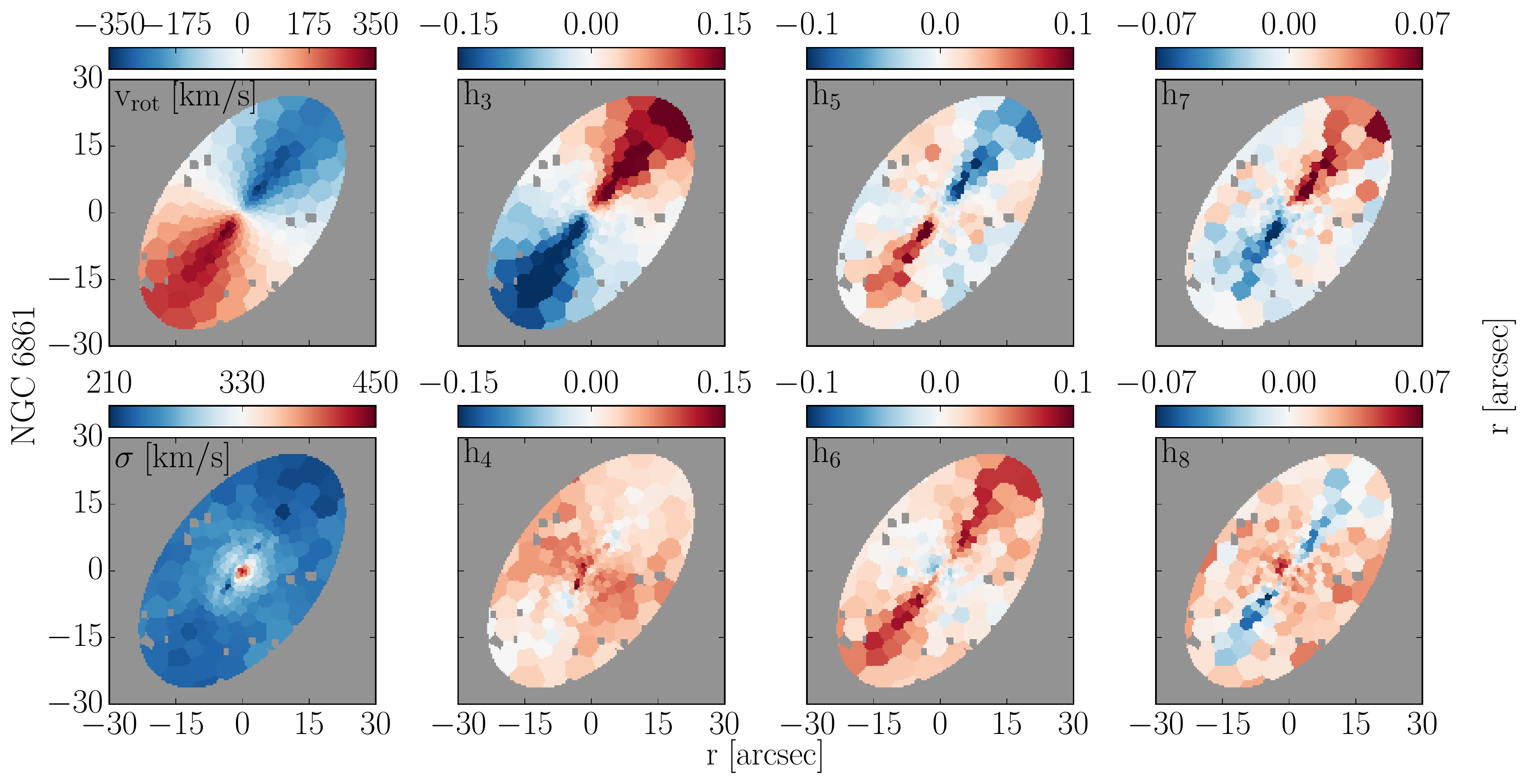}
 \caption{(continued)}
 \end{figure*}

 \begin{figure*}[!htbp]
\centering
 \includegraphics[width=1.7\columnwidth]{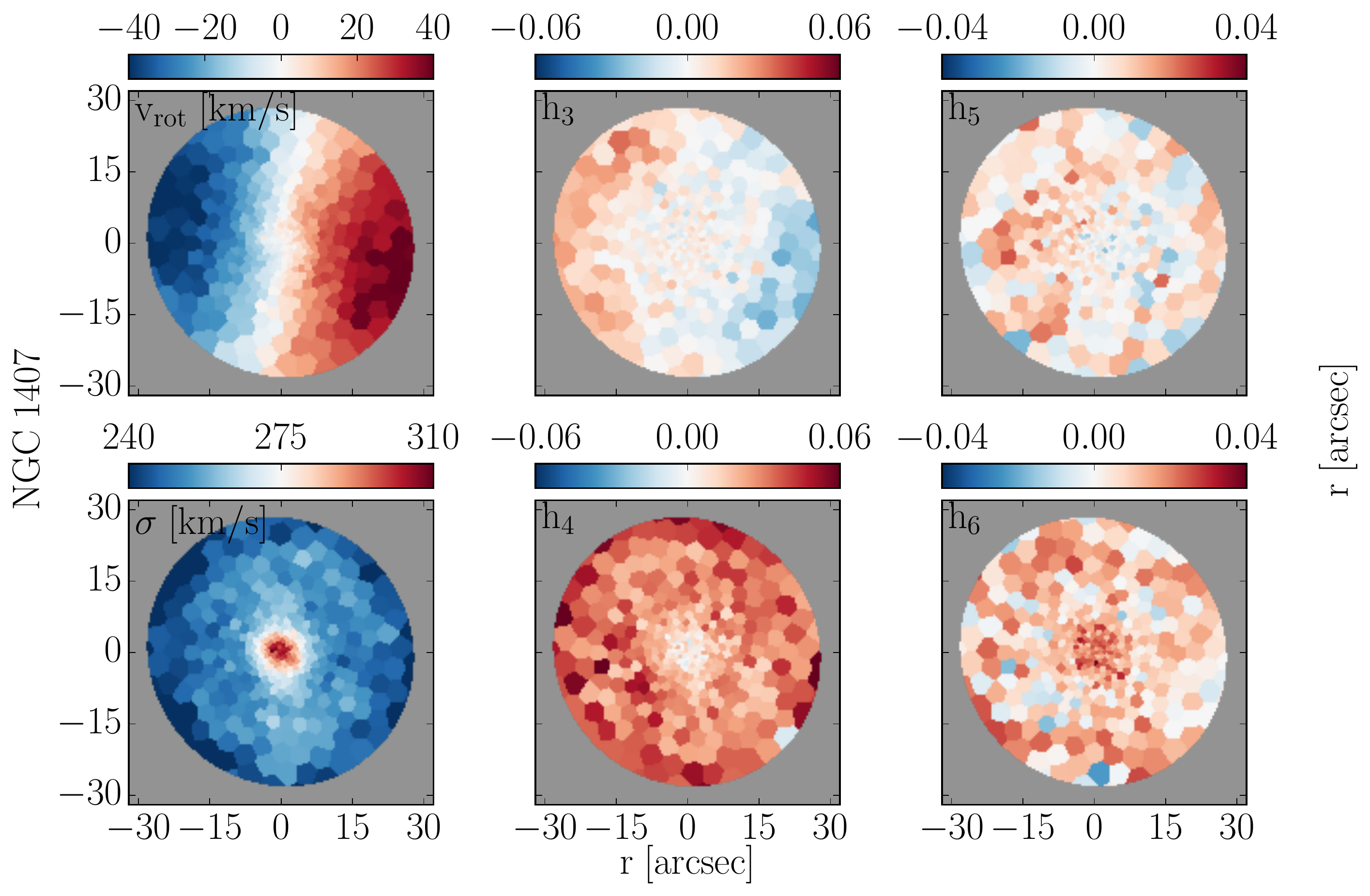}
 \caption{From left to right: MUSE-based 2D stellar kinematics of slow rotating ETGs NGC~1407, NGC~5328, NGC~5419, NGC~5516 and NGC~7619.
   In each sub-figure, from top-left to bottom-right: maps of the
   rotational velocity $v_{rot}$, velocity dispersion $\sigma$ and 
   higher-order Gauss-Hermite coefficients $h_3$ - $h_{8}$. We only show pixels inside the largest isophote that fits wholesale into the MUSE FOV (gray areas inside this region have been excluded from the binning). This implicitly indicates the positions of the major and minor axes and gives a general idea of each galaxy's morphology.}
   \label{fig:SRkinmap}
\end{figure*}

 \addtocounter{figure}{-1}
\begin{figure*}[!htbp]
\centering
 \includegraphics[width=1.7\columnwidth]{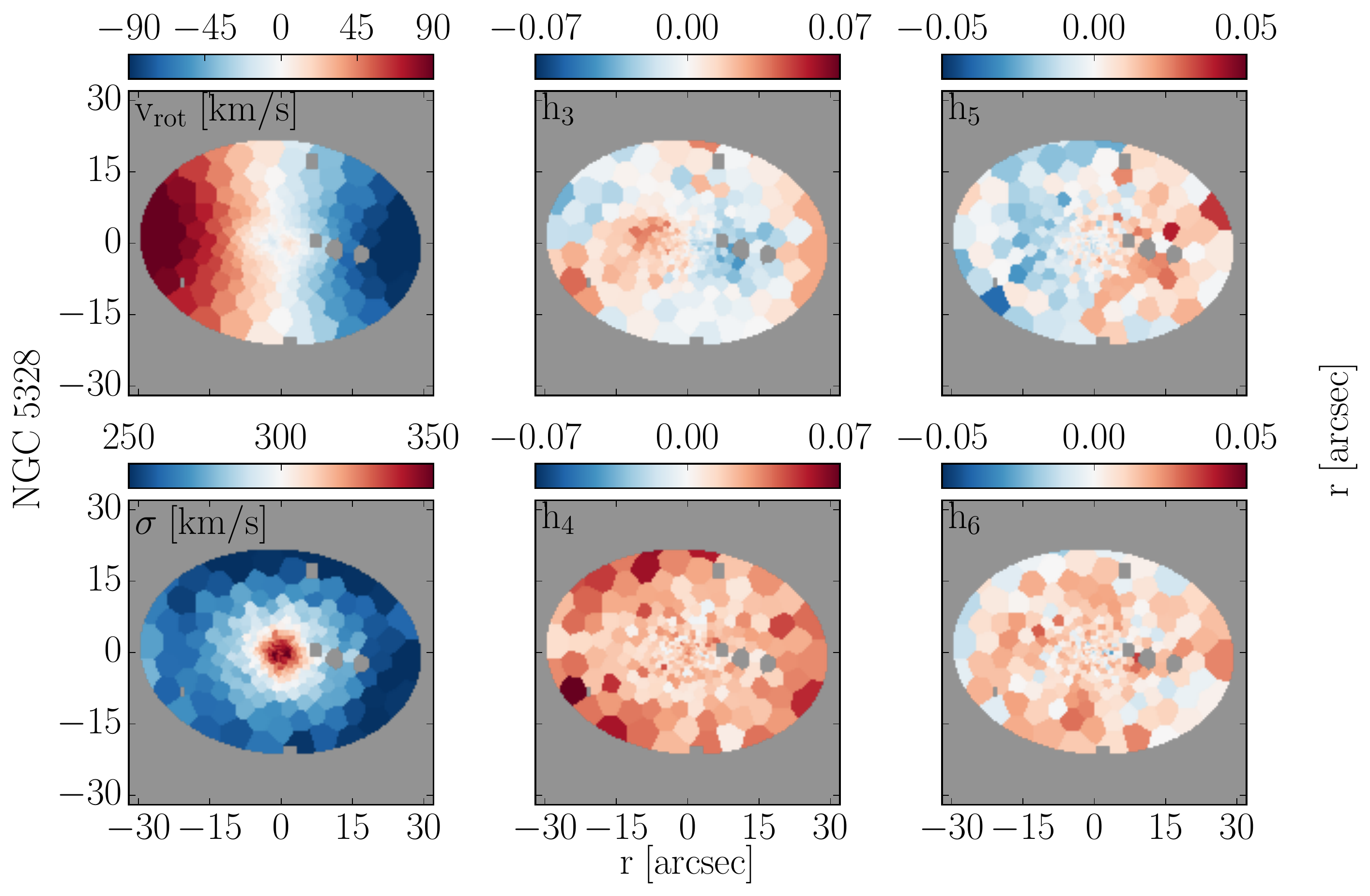}
 \includegraphics[width=1.7\columnwidth]{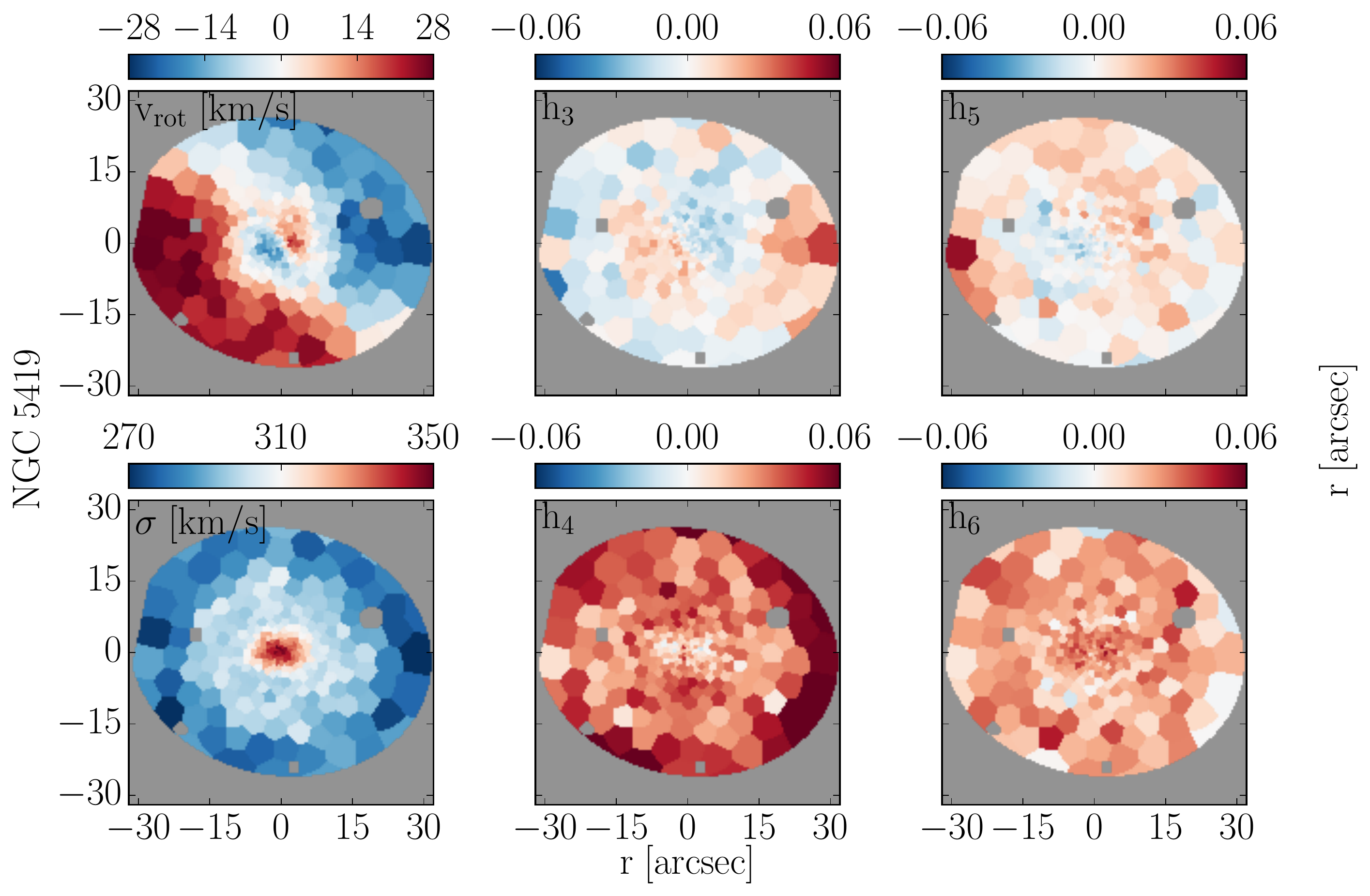}

 \caption{(continued)}
 \end{figure*}
 \addtocounter{figure}{-1}
 \begin{figure*}[t!]
\centering
 \includegraphics[width=1.7\columnwidth]{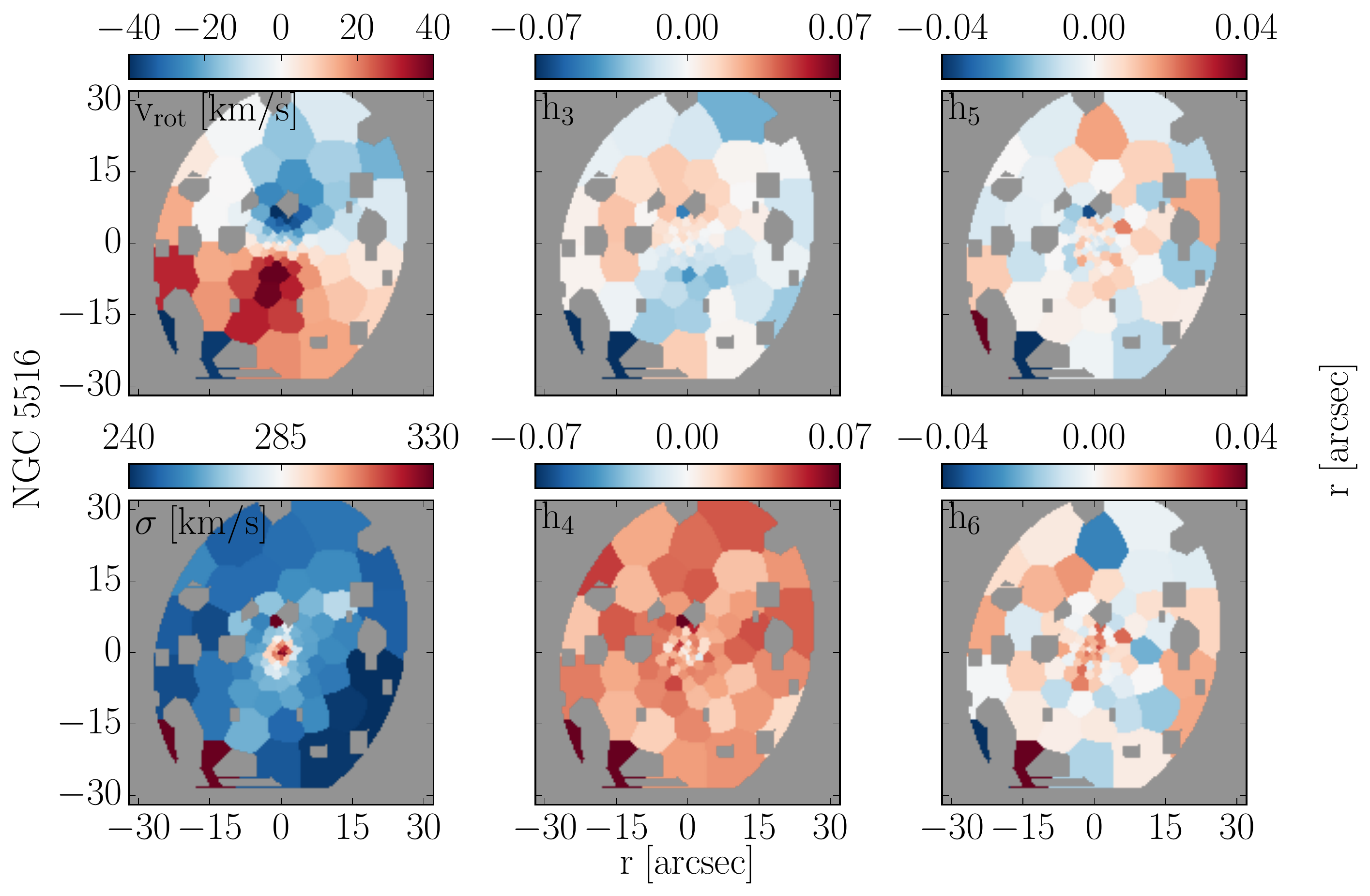}
 \includegraphics[width=1.7\columnwidth]{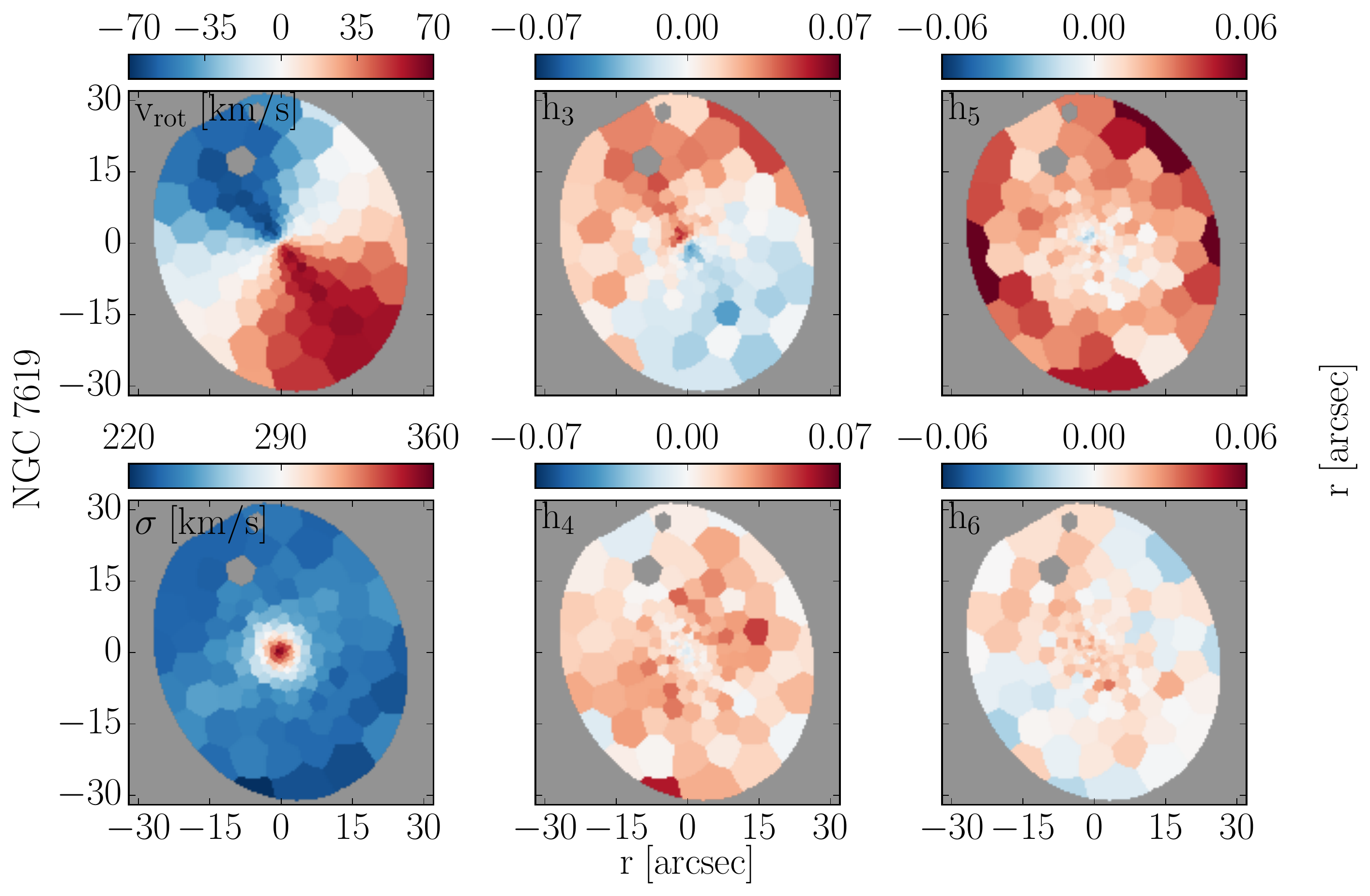}

 \caption{(continued)}
 \end{figure*}
\section{Discussion}
\label{sec:discussion}

\begin{figure}[!t]
\centering
 \includegraphics[width=0.8\columnwidth]{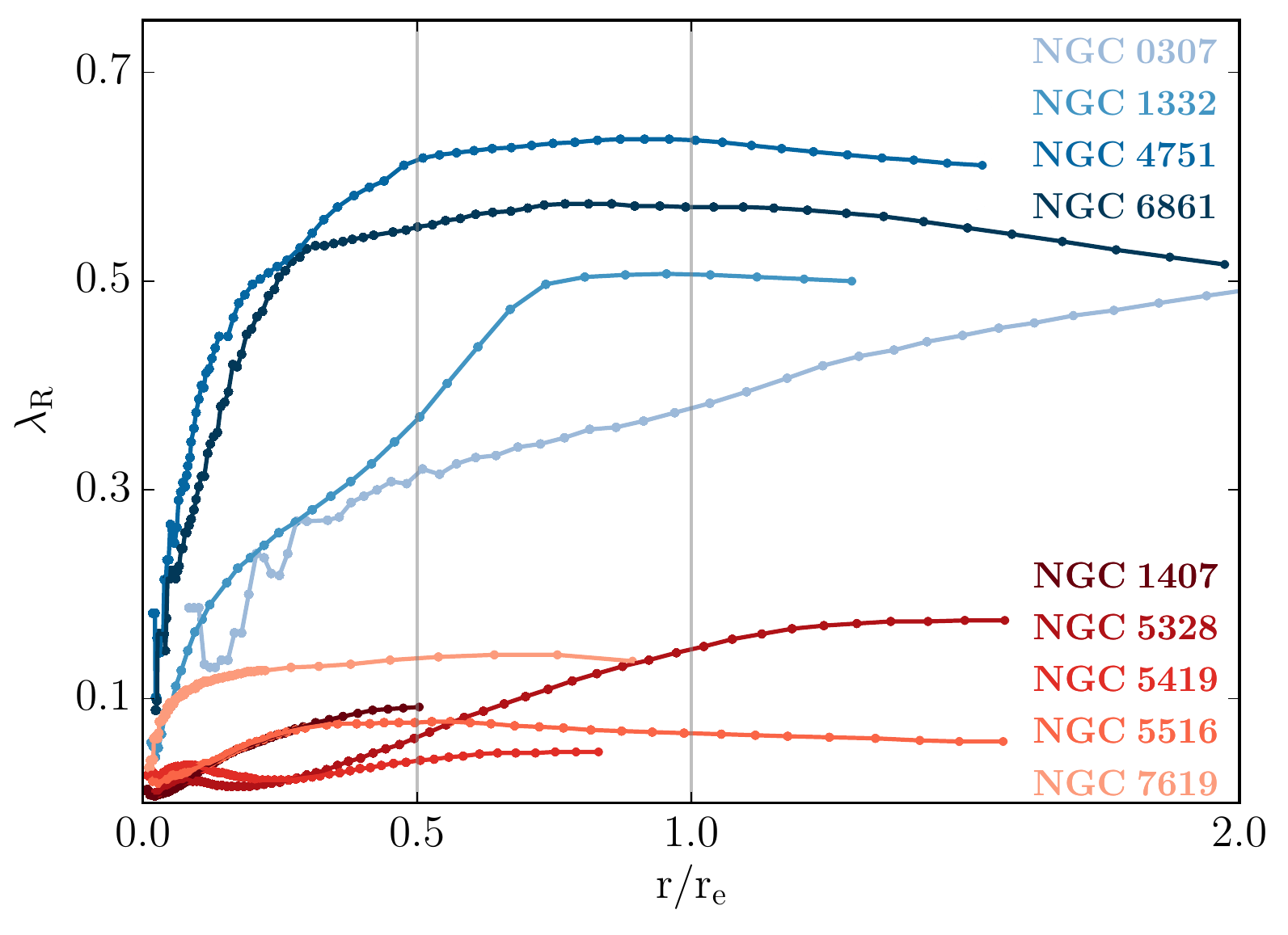}
 \caption{Angular momentum parameter $\lambda$ over radius, scaled by the effective radius of each galaxy, $r/r_e$. Galaxies are separated into fast rotators (shades of blue) and slow rotators (shades of red).}
   \label{fig:LambdaProfiles}
\end{figure}

\subsection{Robustness of the measurements}
\label{ref:wingauth}
We are here for the first time investigating the full non-parametric shape of the LOSVDs of massive ETGs up to the highest bound stellar velocities. Our setup was specifically designed with the goal of bringing new, previously inaccessible kinematic aspects of ETGs to light. We have paid a lot of attention to possible sources of template mismatch and ways to minimise its effect on the shape of the recovered LOSVDs. In Sec.~\ref{subsec:realTemp} we have discussed mock tests that have shown that our setup is robust against the expected forms of template mismatch. A posteriori, our measured LOSVDs for the observed galaxies confirm the robustness of our setup. Most significantly:

\begin{itemize}
    \item The Gauss-Hermite parameters derived from our LOSVDs give virtually no indication of template mismatch, given the almost complete lack of bias in $h_3$ and even, to a lesser extend, the other higher order odd moments.
    \item The high-velocity wings that we observe in many galaxies do not follow the velocity shifts of the LOSVD peaks in galaxies with rotation: While the main trunk of the LOSVD is centered on different $v_\mathrm{los}$ in different places of a rotating galaxy the wings remain stationary relative to $v_{los} \sim 0$ (e.g. NGC~4751 and NGC~6861 in Fig.~\ref{fig:FRellipses}). This makes template mismatch unlikely as the origin of the wings.
    \item Template mismatch is likely strongest in the centers of our galaxies due to increased metallicity and $\alpha$ elemental abundances there. However, we find LOSVD wings both at small and at large radii. 
\end{itemize}

We note that fitting the ETGs with the library of stars from \citet{MUSEtemplates}, which were observed with MUSE, produced the same winged LOSVDs as in our analysis. The wings are slightly less symmetric in this case, due to the smaller number of templates (only 35) which gives less flexibility to reduce the template mismatch in our pre-selection approach. This test excludes that the shape of the instrumental line-spread function (LSF) of MUSE could bias the LOSVDs towards a winged shape.

Finally, in Appendix \ref{ap:Litcompare}, we compare our kinematic measurements from the ETGs with previous published ones using different data and methods.


\subsection{Origins of the LOSVD wings}
\label{ref:wingdiscussion}
In this section we will speculate on the possible origin of extended, high-velocity wings for the non-parametric LOSVDs of ETGs. In addition to the LOSVDs presented here, the non-parametric kinematic study of the brightest cluster galaxy (BCG) Holm~15A with WINGFIT \citep{Mehrgan2019} had also presented evidence for wings.
Furthermore, there are examples of previous studies \citep[e.g.][]{vanDeSande2017,Veale2018} noting increasing $h_4$ towards larger radii in massive ETGs. \citet{Veale2018} in particular, noted positive values of $h_4$ for most ETGs from the MASSIVE sample.
While there is no one-to-one relation between $h_4$ and wings, wings always tend to increase $h_4$.




\subsubsection{LOSVD wings at small radii: PSF light?}
\label{ref:wingcluster}
The PSF redistributes light between different regions on scales on the order of a couple of times the FWHM. In the central regions, within the cusp or cuspy core of ETGs, especially approaching the central SMBH, the intrinsic stellar kinematics have very steep gradients with scales smaller than the PSF in this study \citep{Rusli2013a, Rusli2013b}. Therefore, it is very likely that most of the light at large projected velocities - i.e. the wings - is light from the very center of the galaxy, possibly including contributions from within the sphere of influence (SOI) of the central black hole. 

If this is true, fitting non-parametric LOSVDs with the advanced accuracy as we do here, might enable us to determine the masses of the central black holes from Schwarzschild dynamical modeling, even when the SOI is smaller than the PSF. We will investigate this possibility in our dynamical follow-up study of these ETGs (Mehrgan in prep.).

Alternatively, or in addition, there could also be an overlap of two kinematically distinct stellar populations in the central parts of the MUSE FOV. Wings would then be produced by the contrast between a compact high-$\sigma$ component and a brighter, but narrower component which becomes more dominant with distance from the center of the galaxy.

\subsubsection{LOSVD-wings at large radii: a faint stellar envelope?}

Wings outside the PSF region must have a different physical origin. Given the low signal but large stellar line-of-sight velocities in the wings $v_{los} \gtrsim \SI{1000}{km/s}$ it is likely that must suggest highly eccentric, loosely bound orbits which take the stars out to even larger radii. We therefore propose that these stars are to be associated with a faint, outer envelope of weakly bound stars around the ETGs, similar to the much larger and brighter envelopes around BCGs.

In the latter galaxies, stellar envelopes are usually readily apparent in the stellar kinematics as a rise in the stellar velocity dispersion towards larger radii \citep[e.g.][]{Carter81, Carter85, Ventimiglia2010,Arnaboldi2012, Spiniello2018, Loubser2020}. For NGC~6166, \citet{Bender2015} found that this rising velocity dispersion profile increased towards larger radii, until it meets the dispersion of the cluster, Abell~2199. They concluded that the stellar envelope of the galaxy was more dominated by the gravity of the cluster than the central galaxy to which it was only weakly bound, which was also evidenced by a growing bias in the rotation velocity towards larger radii. We propose a similar scenario for our ETGs, except for the stars being \textit{necessarily} bound to a galaxy cluster.

Naturally, light from the outer envelopes of our ETGs would be much fainter than for the BCGs, which are at large radii entirely dominated by their envelopes. In our ETGs, in particular given our comparatively small spatial coverage with respect to the scales that are relevant here ($r > r_e$), we are in a regime where we only expect a faint high-$\sigma$ LOSVD component superimposed on the dominant LOSVD associated to the inner parts of the galaxy -- similar to how we have constructed the winged mock-LOSVD (of the DEEP mock) in Sec.~\ref{subsec:verif}. 

While in BCGs the velocity dispersion at large radius rises because the broader envelope component overshadows the main galaxy component, for the ETGs, we would only expect to see an increase in the wing light with radius while the broad component remains still the minor component.
This \textit{could} lead to an increase in $h_4$ with radius.
By eye, we see increasing wing-light towards larger radii in NGC~6861, NGC~1407, NGC~5328, NGC~5419 (see Figs. \ref{fig:FRellipses} and \ref{fig:SRellipses}). In these same galaxies we also find an increase of $h_4$ (see Figs. \ref{fig:FRkinmap} and \ref{fig:SRkinmap}). 

However, it must be noted that we are here only observing these ETGs at relatively small radii, $\lesssim 2 r_e$ -- we would need observations at larger radii to properly confirm this trend in these and perhaps also the other galaxies.

This scenario could also help explain the fact that in some galaxies the wings are slightly asymmetric. If the wings were an artifact related to template mismatch we would expect them to be larger in the $\alpha$-enriched galaxy centers rather than in the outskirts. But asymmetry between the narrower and broader components could reflect distortions in the dynamical equilibrium between the inner and the outer parts of the galaxies. For example, massive BCGs are often not properly settled in their respective cluster/envelope (see e.g. the aforementioned study of NGC~6166 by \citealp{Bender2015}, where the BCG's central velocity is at odds with the cluster component on a scale of $50 - \SI{100}{km/s}$) -- we would expect this trend to be typically worse for our less massive ETGs. Hence the asymmetry of the wings in our ETGs could originated from the broad LOSVD component of the envelope being shifted relative to the narrower component of the galaxy.

Of course, the analogy between such faint stellar envelopes and those of BCGs can only go so far, as for the latter the envelope is mostly a halo of stars ripped free, stripped and accreted from other galaxies in the cluster \citep[e.g.][]{Bender2015, Kluge2020}, while for our much less massive ETGs, which do not sit at the bottom of the gravitational well of some massive cluster, this probably cannot account for all of the envelope stars. Instead, we suggest a scenario in which the faint envelopes are mostly remnants of past mergers, wherein stars were placed on loose, higher eccentricity orbits.

Here, we have introduced the idea of the faint stellar envelope purely on kinematic grounds, but this is in itself not remarkable, as for BCGs there is typically no conclusive way of confirming an envelope from photometry alone, as in NGC~6166, were the impetus for the photometric decomposition by \citet{Bender2015} was derived from the increase in the velocity dispersion towards larger radii. Yet, for several of the galaxies in our sample, photometric decompositions require an outer stellar envelope component: This was the case for NGC~1407, 5516 \citep{Rusli2013b} and NGC~6861 (Mehrgan et al. in prep.). The latter has been noted for being surrounded by an especially rich and dense globular cluster system \citep{Escudero2015}.


\subsection{An end-on hidden bar component in NGC~1332?}
One of the most striking kinematic patterns that we found for the kinematic maps which we produced for our sample are doubtlessly the ones
for NGC~1332 (See Fig. \ref{fig:FRkinmap}): The $h_3$ map shows x-shaped or butterfly-like regions of $v_{rot}-h_3$ correlation at roughly $45 \deg$ angles to the major axis of the galaxy, interrupting the usual $v_{rot}-h_3$ anti-correlation regions at smaller and larger radii. At the same time, we find two lobes of lowered $h_4$ for the central regions of the $h_4$ map, as well as extended handles or lobes of relatively heightened $\sigma$
along the major axis in the $\sigma$ map.

These patterns look surprisingly similar to the 2D kinematic signatures of boxy/peanut bulges in simulated disk galaxies from \citet{Iannuzzi2015}:
The x-shaped $v_{rot}-h_3$ correlation regions in particular seem to be a good match for maps of an end-on bar shown in Figure 29 of \citet{Iannuzzi2015} for a disk inclination of $i = 80 ^\circ$. Intriguingly, a similar inclination of $i \sim 85 ^\circ$ has been found by ALMA observations of the circumnuclear disk of NGC~1332 by \citet{Barth2015}. 

We therefore cautiously suggest the presence of an end on bar in NGC~1332. If this turns out to be true, this would likely lead to a 
bias in the determination of dynamical mass estimates of the galaxy if the bar component is not accounted for. This could \textit{potentially} account for the discrepancy between the black hole mass measurements from dynamical models of the stars in galaxy from \citet{Rusli2011} and those from emission from its circumnuclear disk observed with ALMA \citep{Barth2015}. 

We note that \citet{CrettonBosch1999} and \citet{Emsellem2011} have also suggested (but ultimately not favored) the possibility of end-on bars in the massive ETGs NGC~4342 and NGC~1277, respectively, based on similar kinematic signals as we have detected in NGC~1332.

\subsection{Decoupled cores in NGC~1407, NGC~5328 and NGC~5419?}

Similarly striking in the kinematics maps are the decoupled cores which we found particularly for NGC~5328 and NGC~5419, as well as to a lesser extend, NGC~1407 (see Fig. \ref{fig:SRkinmap}). They are central regions were the rotation patterns of the galaxies suddenly flip and change signs. As previously discussed in Section \ref{subsec:angmom}, the three galaxies also seem to have central local maxima of their angular momentum parameter $\lambda$ associated with the decoupled regions, which distinguish them from the other cored ETGs.

The counter-rotation in the center of NGC~5419 was already reported by \citet{Mazzalay2016}, who, using both HST and adaptive optics based SINFONI observations found a double nucleus in the galaxy.
For NGC~1407 \citet{Johnston2018} claimed a decoupled core from their analysis of the MUSE data from our proposal. It is not based on non-parametric LOSVDs though and the $h_3$ distribution is biased, likely indicating template mismatch. 

Numerical simulations of merging ETGs from \citet{Rantala2019, Frigo2021} produced rotation and dispersion patterns matching those we discuss here. Their simulations suggest that counter-rotating decoupled cores are the products of binary SMBH mergers which occur during the mergers of ETGs which produce cored ETGs, such as NGC~1407, 5328 and 5419. Indeed, \citet{Mazzalay2016} suggested that there are two SMBHs at a distance of $\sim \SI{70}{pc}$ in the center of NGC~5419, which are associated with the double nuclei. In our central rotation map, the inner counter-rotating regions appears to be connected to the larger velocity field in a way that suggest an inspiraling motion of the stars, which, in this scenario would have been caused by dynamical drag of the sinking SMBHs/nuclei of the merging ETGs \citep{Rantala2019, Frigo2021}.

\subsection{On the inclusion of NaD}
\label{sec:NaD}

The $5890, \SI{5896}{\angstrom}$ NaD absorption feature is commonly masked for stellar kinematics/populations due to the danger of excess NaD absorption (or emission, but this only applies to low mass galaxies unlike our ETGs \citealp{Concas2019}) from cold neutral gas in the interstellar medium (ISM). Such excess absorption would lead an artificial template mismatch, whereby the template would appear to be more deficient in [Na/H] relative to the galaxy than it actually is. The result of this negative template mismatch would be a suppression of wing light (see Fig. \ref{fig:NaDmismatch}.)

To investigate for the presence of cold, neutral gas we performed a number of tests on the MUSE data of the galaxies.
This included (i) inspecting the residuals of kinematic fits, which did not yield an excess beyond the local noise level and (ii) using an isolated absorption line doublet at $5890, \SI{5896}{\angstrom}$ as an additional kinematic component in the fit which, using WINGFIT's multi-component fitting capabilities, was allowed to have its own kinematics LOSVD independent of that of the stars. This additional kinematic component, however, always yielded an LOSVD similar to that of the stars. Hence, we could not find evidence for an absorption signal in the NaD lines that has a different kinematics than the stars. For galaxies with spatial regions with significant emission lines -- which in the first place meant ionized gas in the cold disks of NGC~6861 and NGC~4751 -- we attempted to at first fit their spectra with a stellar and an emission component while masking NaD. Then, in a second step we re-fitted the spectra with two components, but added the extra NaD absorption component to the templates of the ``emission'' component and fixed its kinematics to the kinematics of the emission lines from the first step.
Tying the extra absorption component to the kinematics of the cold disks in this way did not produce acceptable fits to the spectrum. 

Thus, we could find no tangible evidence for excess absorption within the ISM of any of the galaxies in our sample. However, there is evidence for such absorption from the ISM of the Milky Way (which can be easily masked; App.~\ref{ap:masking}). Therefore, we saw no reason to discard this strong spectral feature from our analysis.
 
Including this feature in our mock-tests (Section \ref{sec:mocktesting}) consistently provided better recoveries of the LOSVD. For instance, expanding the upper limit of the wavelength range for our the Mg and Fe mismatch tests (see Fig. \ref{fig:Wingmechanism}) to $\SI{6200}{\angstrom}$ to include NaD, reduced the dominant distortion effects of all tests. Most notably, for the test where our template was in excess of [Fe/H], the artificially produced wings decreased from $h_4 \sim 0.03$ to $0.01$. The positive effect of including NaD is hardly surprising, as the continuum of the NaD absorption feature is especially smooth and featureless on both sides and the feature itself is usually the strongest in this wavelength range. 


 Positive $h_4$ values have been observed in several massive ETGs. Our tests suggest that a good indicator for the robustness of these measurements is their invariance under inclusion of the NaD region in the fit.

\section{Summary and Conclusions}
\label{sec:conclusion}

We have presented the first systematic study of the detailed LOSVD shapes of massive ETGs using non-parametric spectral fitting. Our sample encompasses nine galaxies, four of which are highly flattened, fast-rotating power-law galaxies. The remaining galaxies are massive slowly rotating galaxies with a depleted stellar core. The high signal of the MUSE observations allows to extract 40-400 individual LOSVDs for each galaxy from binned spectra that
still have a SNR$>100$ per spectral bin. Our LOSVDs are determined with our recently developed spectral fitting code WINGFIT. It recovers the LOSVDs in a non-parametric fashion and uses a novel technique to adaptively optimise the smoothing \citep{ThomasLipka2022}.

We have extensively discussed how well the detailed shapes of the LOSVDs can be measured and identified various types of LOSVD distortions related to potential template mismatch.
Based on Monte-Carlo simulations with stellar-population models that allow to vary the abundances of individual elements we found an optimal strategy to avoid LOSVD distortions from template mismatch. Our setup combining the advanced LOSVD extraction method with very-high SNR data was designed to measure the shapes of LOSVD with unprecedented precision.

Our most important findings concerning template mismatch are
\begin{itemize}

    \item Only when the exact template is among the candidate template spectra, the LOSVD recovery is largely independent of the fitting setup, e.g. independent of the fitted wavelength range and of the applied additive or multiplicative polynomials
    \item Template mismatch which affects only individual stellar absorption features results in the long-known asymmetric LOSVD distortions, which are easily identifiable as template mismatch since they lead to a bias in $h_3$.
    \item Template mismatch which affects multiple features simultaneously or a particularly dominant feature, results in symmetric LOSVD distortions, most often in the form of excess wings which bias $h_4$ and $\sigma$ high.
    \item Such artificial wings can easily ``hide'' template mismatch since their presence cannot unambiguously be attributed to template mismatch.
    \item When the right template spectrum is not among the candidate spectra, liberal usage of polynomials makes this ``hidden'' mismatch more likely and amplifies its distorting effect on LOSVD shapes.
    \item Our simulations of fitting massive ETGs in the optical regions without NaD almost always led to an overprediction of light in the LOSVD wings.
\end{itemize}
    
    To minimise template mismatch we have developed a strategy that includes a proper preselection of template stars, minimal use of spectral masking and polynomials, and a wavelength basis that includes NaD. 
 We carefully checked the NaD region in our sample of observed galaxies for contamination by an absorbing component with kinematics different from that of the stars. We do not find evidence for such a component. Fitting all the galaxies with our fiducial best setup, our main findings are

\begin{itemize}

    \item The complexity of observed non-parametrically derived LOSVD shapes of our ETGs requires Gauss-Hermite polynomials of at least 6th or 8th order to be represented well.
    
    \item All galaxies show some kind of wings in their LOSVDs which are unlikely an artefact of hidden template mismatch.



    \item In the central regions of the galaxies the wings are always strongest and likely originate from light at the very centers of the galaxies, close to the central SMBH, which has been redistributed by the PSF. Additionally, a secondary, faint and compact central high-$\sigma$ stellar population, could also produce a wing component in the LOSVD.
    \item In some galaxies wings are also present at larger radii. We propose that these wings could be associated with faint stellar envelopes consisting of stars that are only loosely bound to the galaxies, similar to the cluster-bound stellar envelopes found in many BCGs.

\end{itemize}

In some of our ETGs photometric evidence for faint outer envelopes has been found \citep{Rusli2013b}. We are developing a decomposition method for LOSVDs to compare photometric and kinematic evidence for distinct stellar components in galaxies. In addition, deeper spectral observations, larger samples and detailed dynamical models are required to follow up on the outer envelope of massive ETGs in the future.

The latter in particular will be important here: If extended LOSVD wings are to some degree resultant from the overlap of kinematically distinct stellar populations, stellar dynamical models with multiple kinematic components might be necessary to properly represent the galaxies. Furthermore, dynamical models together with independent mass constraints are crucial to determine the existence or absence of further systematic issues which affect the shape of the LOSVDs as we have recovered them here.



\section{Acknowledgement}
We acknowledge project/application support by the Max Planck Computing and Data Facility. Most kinematic computations were performed on the HPC system Raven and Cobra at the Max Planck Computing and Data Facility. We also made use of the computing facilities of the Computational Center for Particle and Astrophysics (C2PAP) and we are grateful for the support by A. Krukau and F. Beaujean through the C2PAP.

This study is based on observations collected at the European organisation for Astronomical Research in the Southern Hemisphere under ESO program 095.B-0624(A), P.I. J. Thomas.

\appendix
\section{Creating Mock Spectra and Templates using \textit{alf}}
\label{ap:MockCreation}For the creation of mock-galaxy spectra and synthetic stellar templates we used the software $\textit{alf}$ (\href{https://github.com/cconroy20/alf}{v2.1}), which is an implementation of the stellar population models first presented in \citet{Conroy&vanDokkum2012}, adapted for fitting the spectra of old stellar populations ($\gtrsim \SI{1}{Gyr}$) in the optical to near infrared \citep{SanchezBlazquez2006, Choi2014, Conroy2014, Choi2016, Villaume2017, Conroy2018}. The ``spec\_from\_sum''-subroutine of $\textit{alf}$ allows one to create model-spectra by manually defining model-parameters (instead of deriving them from fitted data), such as the metallicity or individual elemental abundances such as [Fe/H], but also the spectral broadening from a LOSVD, parameterized by Gauss-Hermite polynomials. By manually setting  the LOSVD to a Gaussian with $v_{los} = \SI{0}{km/s}$ and $\sigma = \SI{1}{km/s}$ we created kinematically unbroadened template stellar spectra with an instrumental velocity-resolution of $\sim \SI{100}{km/s}$ for our mock-tests. 
 
For the creation of the Shallow features + Gaussian mock, specifically, from Section \ref{subsec:verif}, for which we fitted the Deep features + Wings mock with alf using a Gaussian LOSVD, we required some additional tweaking of the code: The stellar population models are usually fit in several, same-sized wavelength-sections of the full spectrum, as small as \SI{100}{\angstrom} in extend, each with its own multiplicative polynomial for the normalization of the continuum. This however, results in a model, which, even with the correct, intrinsic template, cannot be properly fit with WINGFIT, as the the polynomial of the final model, which is ``strung together'' from the polynomials of all intervals, is too complex. This would have diluted the purpose of the mock-tests based on this model, as the shape of the polynomial might then have interfered with the shape of the recovered LOSVD which would have been used to partially compensate for the mismatch in polynomials. Therefore, during this fit, we manually set up the fit such that the spectrum was not separated into intervals but fit as a whole. This resulted in a model with a lower-order multiplicative polynomial which could easily be reproduced with the multiplicative polynomial of our WINGFIT-analysis. It should be noted that the $\textit{alf}$-generated templates of our analysis do not include the polynomial, and that these had to be multiplied after broadening with the LOSVD to match original model-fit to the Deep features + Wings mock.

\begin{figure}
\centering
 \includegraphics[width=0.9\columnwidth]{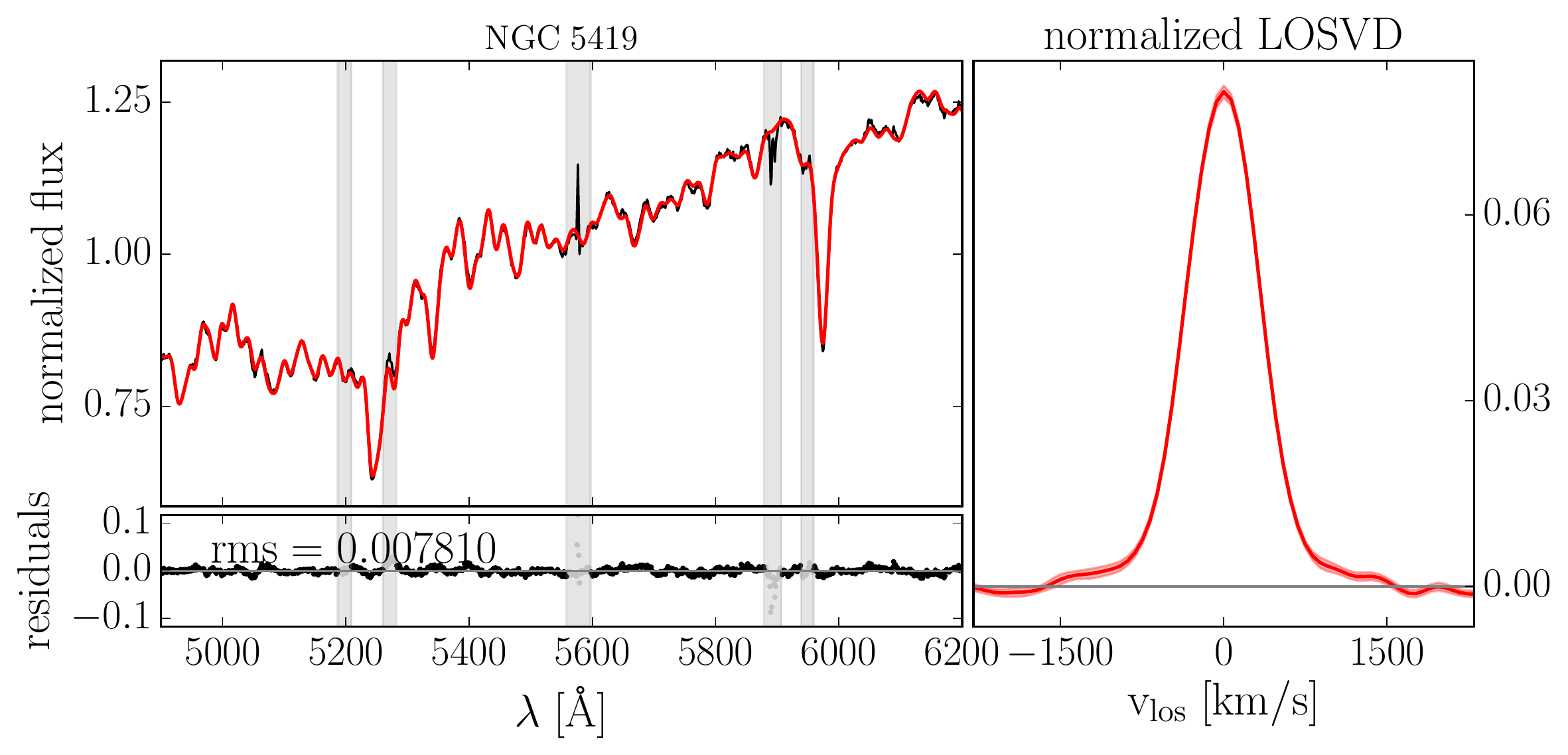}
 \caption{Left: Kinematic fit (red) to the spectrum (black) of a Voronoi-bin of NGC~5419 (bottom) using WINGFIT. At $\sim \SI{5890} {\angstrom}$ the NaD absorption feature of the Milky Way (z = 0) is clearly visible. Grey shaded areas indicate spectral regions which were masked during the fit -- including this excess absortption from the Milky Way. Right: Non-parametric LOSVDs (solid red) recovered from these fits. The shaded envelope indicates the statistical uncertainties of the LOSVD from 100 Monte-Carlo simulations. The line-of-sight velocities $v_{los}$ are relative to the systemic velocity of the galaxy.}
   \label{fig:exampleMWNaD}
\end{figure}

\section{spectral masking and treatment of emission}
\label{ap:masking}

Apart from template-mismatch, the true shape of the LOSVDs can also be distorted by {\textbf{a)}} including spectral regions in the fit affected by systemic issues such as under- or over-subtracted skylines or regions in which the detector is affected by other instrumental issues, {\textbf{b)}} excess absorption from neutral gas in the interstellar medium (ISM), or {\textbf{c)}} not properly treating emission lines from ionized gas.

Concerning issue a) we spectrally masked or excluded from the wavelength-interval of the fit all affected regions. This mostly meant spectrally masking the strong oxygen $\SI{5577}{\angstrom}$ sky emission line and excluding wavelength regions bluer than $\SI{4800}{\angstrom}$ which are strongly affected by a sharp upturn of systemic uncertainties in the case of MUSE. Also, we set the upper limit of the wavelength range at around $\SI{6200}{\angstrom}$, since, as we briefly described in Section \ref{sec:mocktesting}, there were strong residuals from the telluric correction in between $\sim 6200 - \SI{6350}{\angstrom}$. The mock-tests from that section showed that, for the recovery of the LOSVD, cutting off the wavelength-interval to the exclusion of contaminated spectral regions is preferable to including large spectrally masked regions in the fit, which discontinuously section the spectrum. The particular wavelength-interval and spectral masking we chose, varies from galaxy to galaxy and was determined on a case by case basis, but we generally aimed to maximize the number of (good) spectral pixel in the fit, as per our prescription from Section \ref{subsec:verif}.

For treating issue b), we consider the $5890, \SI{5896}{\angstrom}$ NaD absorption feature at the redshift of the galaxy for excess absorption from the ISM within the galaxy and in the rest frame $z = 0$ for foreground absorption from within the Milky Way. The latter is frequently ignored in stellar population $\&$ kinematic analyses of galaxies, but is readily apparent for most galaxies of our sample when inspecting the residuals of kinematic fits to the galactic spectra (see NGC~5419 in Fig. \ref{fig:exampleMWNaD}), varying in strength with the position of each galaxy on the night sky relative to the plane of the Milky Way. For the former, we could find no evidence of excess absorption within the ISM of any of the galaxies, see Section \ref{sec:NaD}.

As for issue c), we adopted a two stage fitting procedure:  first we fit the stellar spectrum over a range wide enough to cover $H\beta$, as well as the strong $H\alpha$ and [NII] $\SI{6583}{\angstrom}$ emission lines using the in-built emission line fitting capabilities of WINGFIT to fit both absorption and emission features simultaneously. If we measure a flux large enough such that the local SNR of the emission $> 3$ for these emission lines, as well as the [OIII] $\SI{5007}{\angstrom}$ emission line, we treat the presence of ionized emission for the spectrum in question as ``significant''. We include the [OIII] emission lines in this SNR-check to avoid an issue wherein the fit uses emission lines which, unlike [OIII] emission, nominally lie on top of stellar absorption, as is the case for example for $H\beta$, to better ``fill-out'' the absorption feature in the spectrum with an effective model consisting of an insufficiently deep stellar absorption + an emission component with an extremely broad LOSVD ($\sigma \gtrsim \SI{500}{km/s}$). This effect is in essence a compensation for template mismatch. In the second stage we narrow the fitted wavelength range to match that of the final WINGFIT analysis, which does not include the $H\alpha$-region, $\sim 4800 - \SI{6200}{\angstrom}$, mask all emission lines within this range where we judged the ionized emission of the concerned spectrum to be significant in the previous step and fit only a stellar component (no emission). Where this was not the case we did not apply any additional spectral masking in addition to the spectral masking from $a)$ and $b)$ to maximise the available constraints on the stellar abundances of the local template sets.

If, for fits from the first stage of the fitting approach of our template selection process (see Section \ref{subsec:TemplateSelection}), we detect emission with $SNR > 3$ for any elliptical ring-section of a galaxy, we apply the same spatial masking - the spatial masking accounting for emission-lines - to all spectra of that galaxy, even if some particular spectrum lies within an aperture for which we did not find any significant emission. If we did not detect any significant emission for any of the elliptical sections of a galaxy, we do not apply any masking of ionized gas for any bins of that galaxy. 

\section{Comparison with previous measurements}
\label{ap:Litcompare}

\begin{figure}
\centering
    \begin{tabular}{c c}
        \centering
        \textbf{{ \ \  \ \ \ \ NGC~0307}} & \textbf{{ \ \  \ \ \ \ NGC~0307}} \\
 \includegraphics[width=0.37\columnwidth]{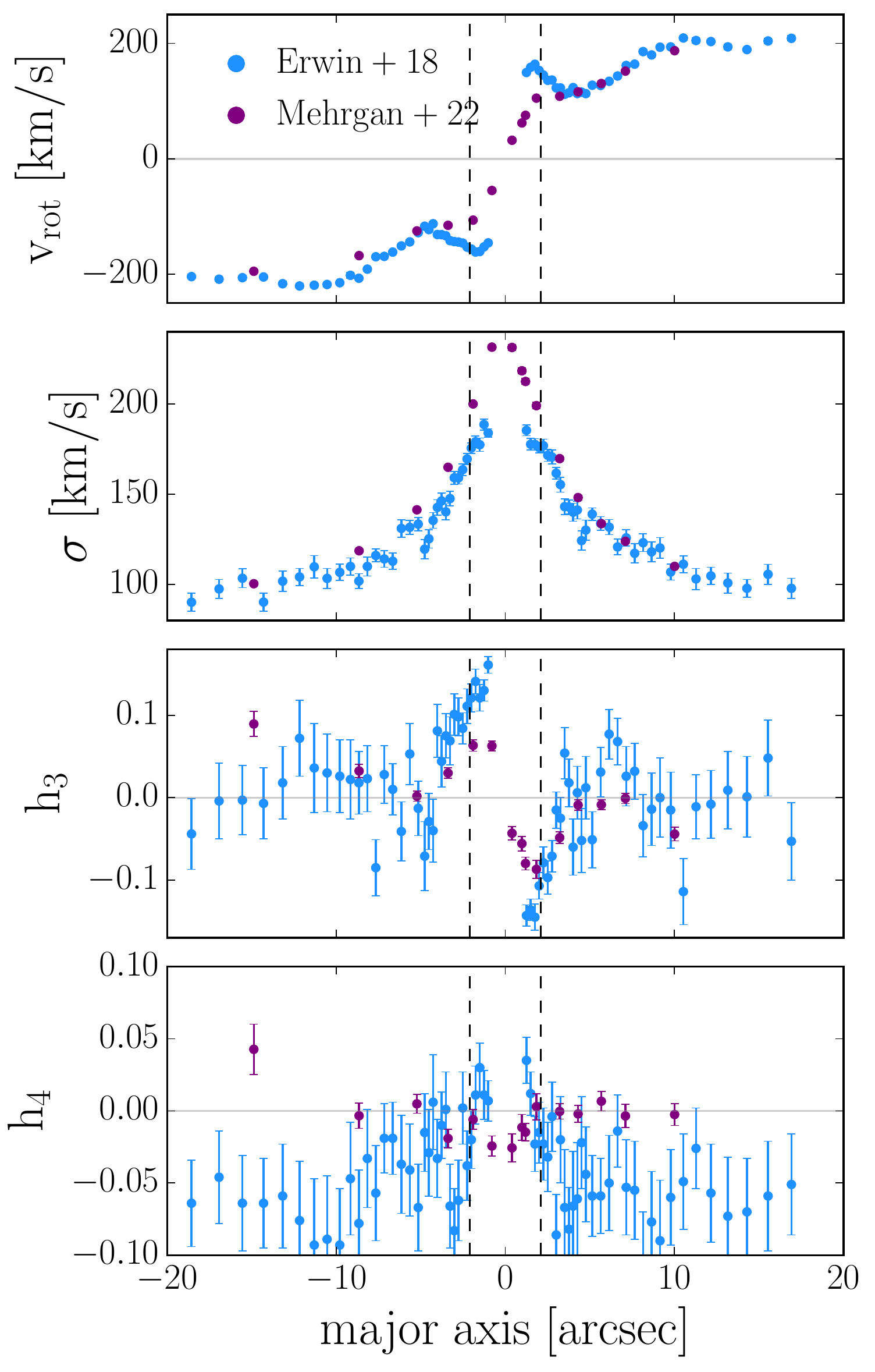} &
  \includegraphics[width=0.37\columnwidth]{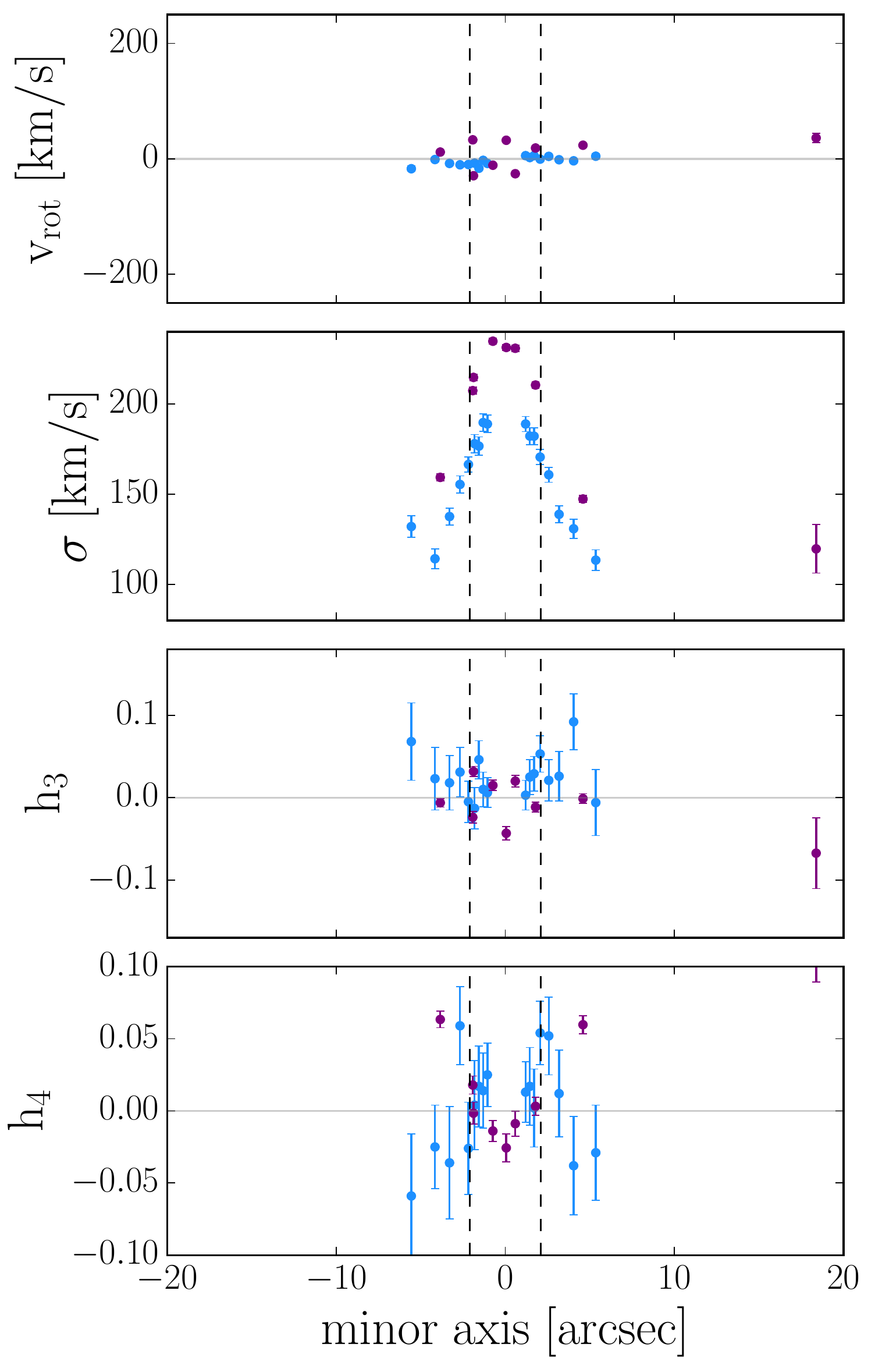}
    \end{tabular}
        \begin{tabular}{c c}
        \centering
        \textbf{{ \ \  \ \ \ \ NGC~1332}} & \textbf{{ \ \  \ \ \ \ NGC~6861}} \\
 \includegraphics[width=0.37\columnwidth]{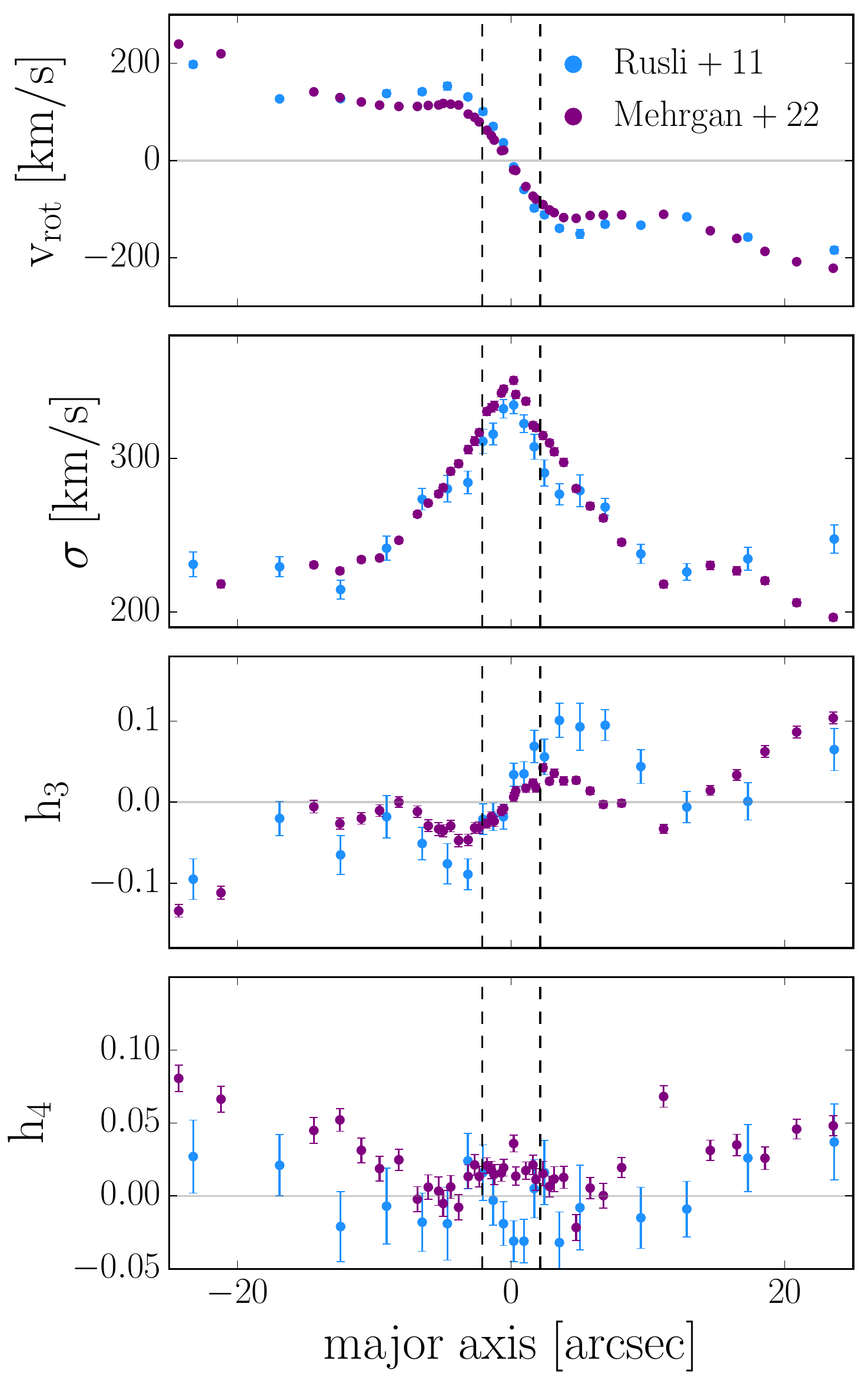} &
  \includegraphics[width=0.37\columnwidth]{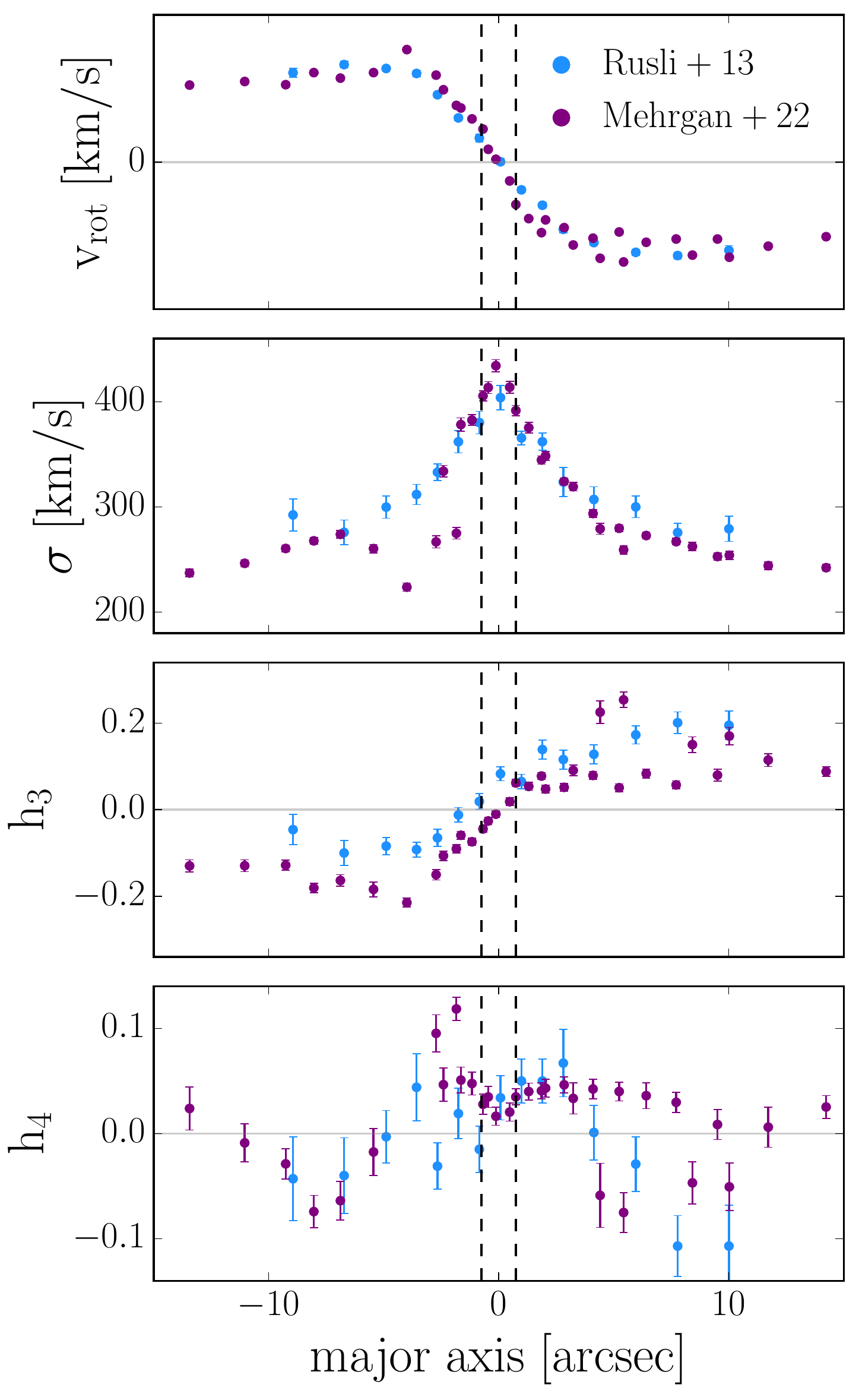}
    \end{tabular}

 \caption{Stellar kinematics of NGC~0307, NGC~1332 and NGC~6861 parameterized by 4th order Gauss-Hermite polynomials along the galaxies' major and, in the case of NGC~0307, minor axis for non-parametric measurements of the LOSVDs from this study (purple points), and \citet{Rusli2011, Rusli2013b, Erwin2018} (blue points). We show statistical uncertainties for all measurements. Vertical dashed lines show the FWHM of the PSF from our MUSE measurements.}
    \label{fig:ngc0307Profile}
\end{figure}

\begin{figure}
  \centering

    \begin{tabular}{c c}
        \centering
        \textbf{{ \ \  \ \ \ \ NGC~5328}} & \textbf{{ \ \  \ \ \ \ NGC~5328}} \\
 \includegraphics[width=0.37\columnwidth]{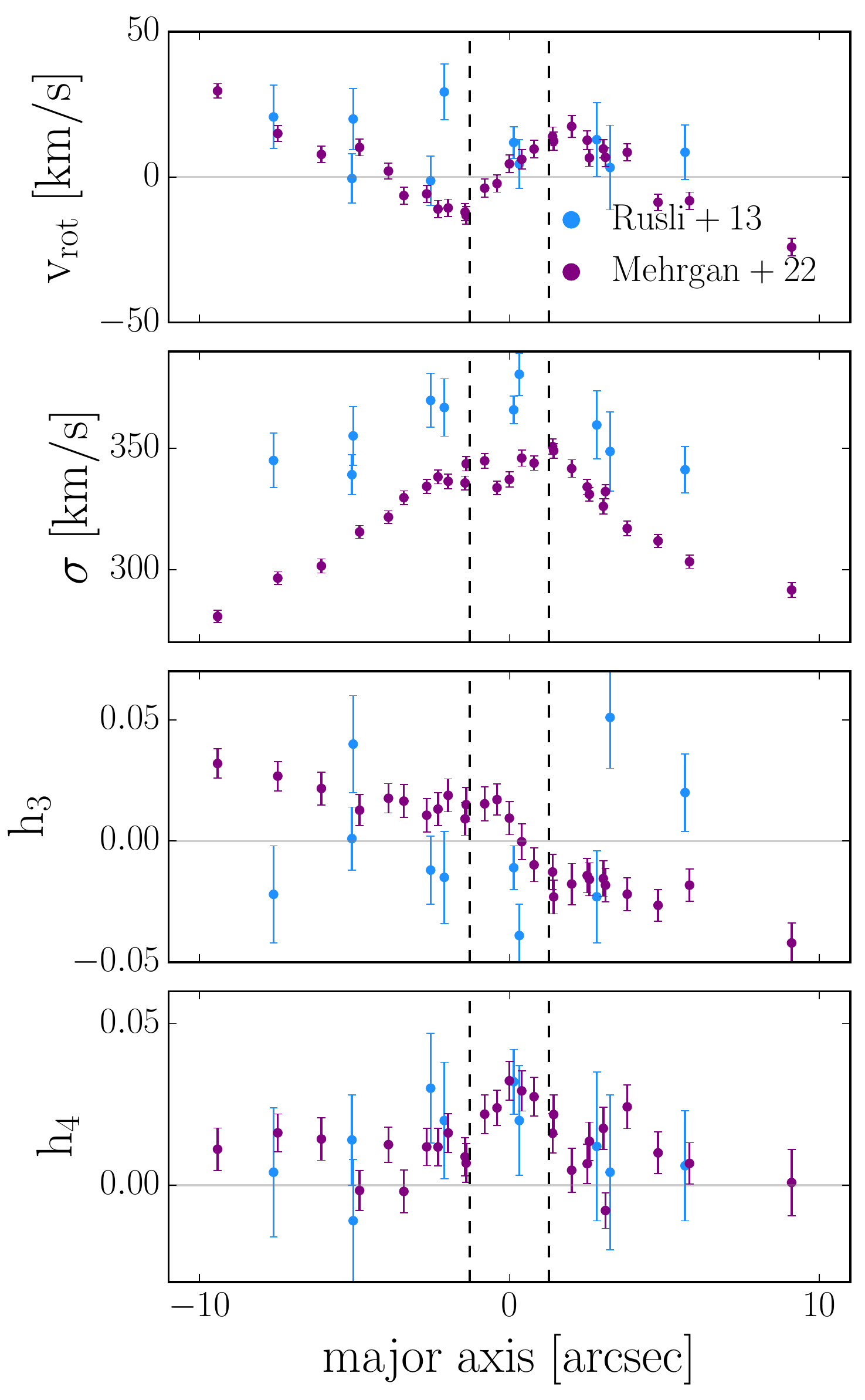} &
  \includegraphics[width=0.37\columnwidth]{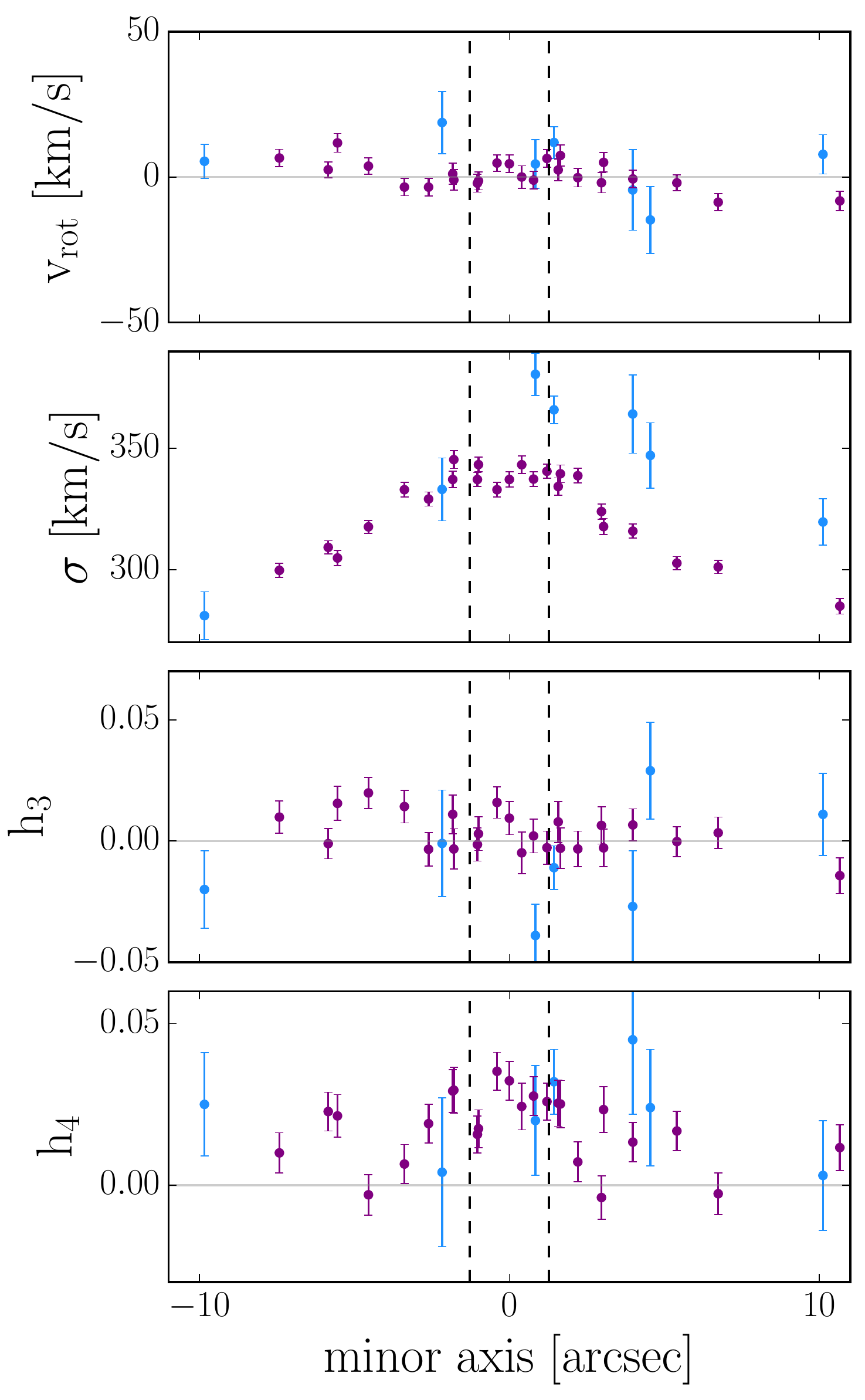}
    \end{tabular}
    
        \begin{tabular}{c c}
        \centering
        \textbf{{ \ \  \ \ \ \ NGC~7619}} & \textbf{{ \ \  \ \ \ \ NGC~7619}} \\
 \includegraphics[width=0.37\columnwidth]{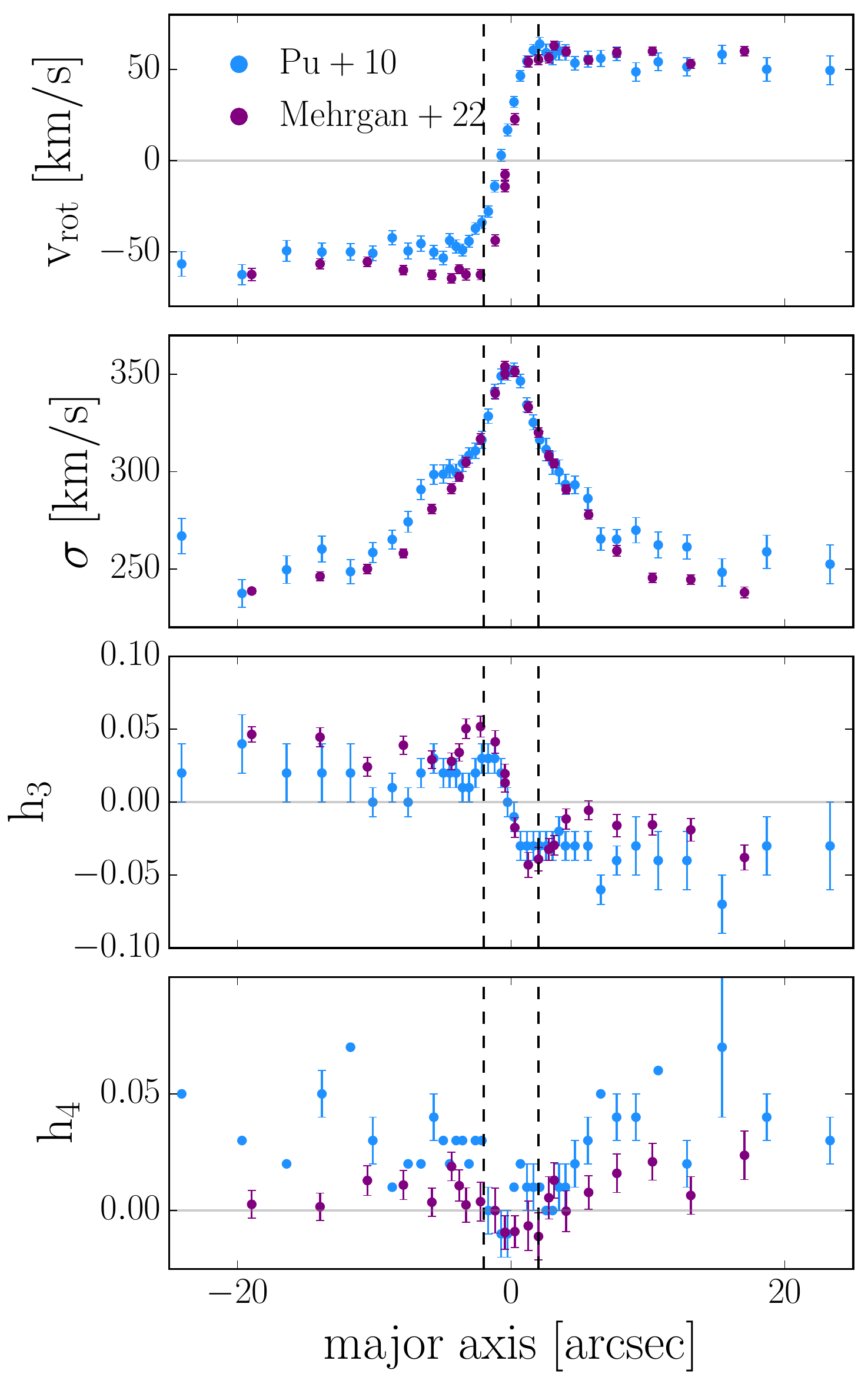} &
  \includegraphics[width=0.37\columnwidth]{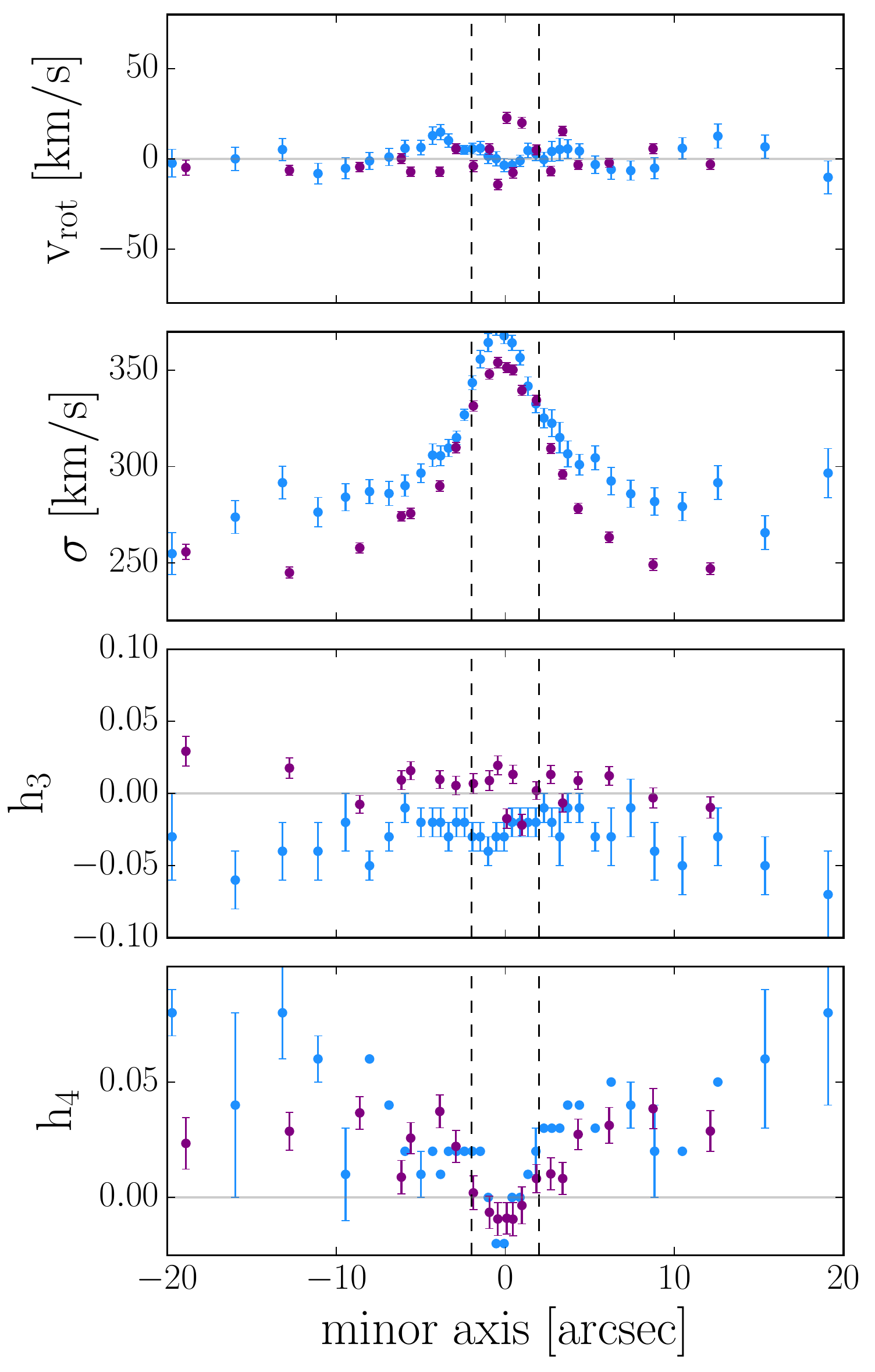}
    \end{tabular}

 \caption{Stellar kinematics of NGC~5328 and NGC~7619 parameterized by 4th order Gauss-Hermite polynomials along the galaxies' major and minor axes for measurements from this study (purple points), and \citet{Rusli2013b, Pu2010} (blue points). We show statistical uncertainties for all measurements. Vertical dashed lines show the the PSF from our measurements.}
   \label{fig:ngc5328Profile}
\end{figure}

We will here briefly compare our new kinematic measurements with those of previously published measurements, which were on similar spatial scales as the MUSE FOV, using different data and methods. For notable cases, we show the full kinematic measurements of the galaxies along their major axis and -- where such data was available -- the minor axis in comparison to ours. The observations with which we here compare our own were mostly long-slit data. In that case, for either axis, we plot our values of $v_{rot}$, $\sigma$, $h_{3}$, $h_{4}$ of all those bins which a) included spaxel lying on the axis in question and, which b) encompassed more spaxel within the slit-width of the comparison data than outside (see Figs. \ref{fig:ngc0307Profile} \& \ref{fig:ngc5328Profile}).

All comparison measurements were performed using the FCQ method, which, like for our WINGFIT code, produced non-parametric LOSVDs which were fit a posteriori with Gauss-Hermite Polynomials, in this case of fourth order. For better comparison we refitted all of our non-parametric LOSVDs with only 4th order Gauss-Hermite Polynomials, as well.

It should be kept in mind that for all these measurements the comparison data was binned for a SNR that was a factor $3-5$ lower than our MUSE data.

\begin{itemize}
    \item {\textbf{NGC~0307}}: Our measurements for $v_{rot}$, $\sigma$, $h_{3}$, $h_{4}$ matched the measurements of \citet{Erwin2018} which were based on VLT-FORS1 data (see top panels of Fig. \ref{fig:ngc0307Profile})
    \item {\textbf{NGC~1332}}: The measured kinematics
    from \citet{Rusli2011} seem to overall agree with our own (see bottom left panel of Fig. \ref{fig:ngc0307Profile}). Their measurements were based on observations with the RCS at the Multiple Mirror telescope. The small differences between the measurements can be explained by differences in the PSF which for \citet{Rusli2011} was likely smaller than our very large PSF $=2.12 \arcsec$, resulting in a more smeared out profile in our case (Their PSF was unknown).
    \item {\textbf{NGC~4751 \& NGC~5516}}:  for these two galaxies the available WiFes-based kinematics were based on observations with very poor signal \citep{Rusli2013b}. Therefore both $h_{3}$ and $h_{4}$ were essentially not resolved. Nonetheless, within the uncertainties and for $v_{rot}$ and $\sigma$, the kinematics match our own measurements.
    \item {\textbf{NGC~5328}}: Most notably for the kinematic measurements of NGC~5328 from \citet{Rusli2013a}, $\sigma$ seems to be offset by $10 - 15 \%$ from our own measurements (see top panels of Fig. \ref{fig:ngc5328Profile}). The measurement from \citet{Rusli2013b} was based on spectra from the VIRUSW IFU spectograph. Their kinematic fits were performed using only a single template star in contrast to our own pre-selected $\sim 30$ template stars for this galaxy. Refitting a central bin from our own MUSE data using only one template star from MILES of the same class and spectral type as the one used in \citet{Rusli2013b}, we were able to reproduce the $10 - 15 \%$ offset in $\sigma$. 
     \item {\textbf{NGC~5419}}: With the exception of some outliers in $\sigma$ at large radii, the kinematic measurements from \citet{Mazzalay2016}, which were based on long-slit spectroscopy from the Southern African Large Telescope, agree well with our own measurements. Their $h_3$, however, is much more strongly biased compared to our own -- particularly for the aforementioned $\sigma$-outliers -- such that the diffences can be explained by template mismatch.
     \item {\textbf{NGC~6861}}:  for the most part, the stellar kinematics from \citet{Rusli2013b}, based on observations from the EMMI spectrograph at the ESO New Technology Telescope, agree with our measurements (see bottom right panel of \ref{fig:ngc0307Profile}). However, there is a notable difference around $\pm \SI{5}{\arcsec}$ along the major axis:  here our values of $\sigma$ seem to ``drop'' by $20 - 30 \%$, relative to the values from \citet{Rusli2013b}. Our study of the ionized emissions from the cold disk of the galaxy, which we will publish in a subsequent publication, reveal star forming regions here, which bias $\sigma$ low for our measurements. The measurements of \citet{Rusli2013b} on the other hand, were performed in a much redder region of the spectrum, namely around die CaT triplet. As a result, their measurements are more robust for the affected bins. These are in any case only a handful of bins out of the $\sim 300$ bins of our MUSE FOV. Furthermore, while for both measurements of $h_3$ there appears to be a similar amount of template mismatch, our own measurements of $h_4$ appear to be more asymmetric about the center of the galaxy. 
     
     \item {\textbf{NGC~7619}}:
      Along the major axis, within $\sim \SI{5}{\arcsec}$ the kinematics 
      of \citet{Pu2010} match ours  well (see bottom left panel of Fig. \ref{fig:ngc5328Profile}). At larger radii, and along the minor axis, the $\sigma$ and $h_4$ measured by \citet{Pu2010} (see bottom right panel of Fig. \ref{fig:ngc5328Profile})
    are larger than our measurements. Furthermore, their $v_{rot}$ and $\sigma$ profiles appear less symmetric about the center of the galaxy. All together, this could be symptomatic of template mismatch, as all forms of template mismatch we have encountered in this study (see Section \ref{sec:mocktesting}) have invariably produced excess wing light. This, by necessity biases $\sigma$ and $h_4$ high. Our setup by comparison had specific precautions put in place to avoid this (see Sections \ref{subsub:realrecov} and \ref{ref:wingauth}). Furthermore, along the minor axis there is a significant amount of bias in $h_3 \sim -0.03$, larger than for our measurements, a dead giveaway for likely template mismatch. The data of \citet{Pu2010} were obtained with the Low Resolution Spectograph (LRS) of the Hobby-Eberly Telescope (HET) and only a single template from the \citet{Vazdekis1999} library of \textit{synthetic} stellar spectra was used.

\end{itemize}

\bibliography{main} 

\begin{thebibliography}{}
\expandafter\ifx\csname natexlab\endcsname\relax\def\natexlab#1{#1}\fi
\providecommand{\url}[1]{\href{#1}{#1}}
\providecommand{\dodoi}[1]{doi:~\href{http://doi.org/#1}{\nolinkurl{#1}}}
\providecommand{\doeprint}[1]{\href{http://ascl.net/#1}{\nolinkurl{http://ascl.net/#1}}}
\providecommand{\doarXiv}[1]{\href{https://arxiv.org/abs/#1}{\nolinkurl{https://arxiv.org/abs/#1}}}

\bibitem[{{Arnaboldi} {et~al.}(2012){Arnaboldi}, {Ventimiglia}, {Iodice},
  {Gerhard}, \& {Coccato}}]{Arnaboldi2012}
{Arnaboldi}, M., {Ventimiglia}, G., {Iodice}, E., {Gerhard}, O., \& {Coccato},
  L. 2012, \aap, 545, A37, \dodoi{10.1051/0004-6361/201116752}

\bibitem[{{Barth} {et~al.}(2016){Barth}, {Boizelle}, {Darling}, {Baker},
  {Buote}, {Ho}, \& {Walsh}}]{Barth2015}
{Barth}, A.~J., {Boizelle}, B.~D., {Darling}, J., {et~al.} 2016, \apjl, 822,
  L28, \dodoi{10.3847/2041-8205/822/2/L28}

\bibitem[{{Barth} {et~al.}(2002){Barth}, {Ho}, \& {Sargent}}]{Barth2002}
{Barth}, A.~J., {Ho}, L.~C., \& {Sargent}, W. L.~W. 2002, \aj, 124, 2607,
  \dodoi{10.1086/343840}

\bibitem[{{Bender}(1990)}]{Bender1990}
{Bender}, R. 1990, \aap, 229, 441

\bibitem[{{Bender} {et~al.}(2015){Bender}, {Kormendy}, {Cornell}, \&
  {Fisher}}]{Bender2015}
{Bender}, R., {Kormendy}, J., {Cornell}, M.~E., \& {Fisher}, D.~B. 2015, \apj,
  807, 56, \dodoi{10.1088/0004-637X/807/1/56}

\bibitem[{{Bender} {et~al.}(1994){Bender}, {Saglia}, \& {Gerhard}}]{Bender1994}
{Bender}, R., {Saglia}, R.~P., \& {Gerhard}, O.~E. 1994, \mnras, 269, 785,
  \dodoi{10.1093/mnras/269.3.785}

\bibitem[{{Binney} \& {Mamon}(1982)}]{Binney1982}
{Binney}, J., \& {Mamon}, G.~A. 1982, \mnras, 200, 361,
  \dodoi{10.1093/mnras/200.2.361}

\bibitem[{{Cappellari}(2008)}]{Cappellari2008}
{Cappellari}, M. 2008, \mnras, 390, 71,
  \dodoi{10.1111/j.1365-2966.2008.13754.x}

\bibitem[{{Cappellari}(2016)}]{Cappellari2016}
---. 2016, \araa, 54, 597, \dodoi{10.1146/annurev-astro-082214-122432}

\bibitem[{{Cappellari}(2020)}]{Cappellari2020}
---. 2020, \mnras, 494, 4819, \dodoi{10.1093/mnras/staa959}

\bibitem[{{Cappellari} \& {Copin}(2003)}]{Voronoi2003}
{Cappellari}, M., \& {Copin}, Y. 2003, \mnras, 342, 345,
  \dodoi{10.1046/j.1365-8711.2003.06541.x}

\bibitem[{{Cappellari} {et~al.}(2012){Cappellari}, {McDermid}, {Alatalo},
  {Blitz}, {Bois}, {Bournaud}, {Bureau}, {Crocker}, {Davies}, {Davis}, {de
  Zeeuw}, {Duc}, {Emsellem}, {Khochfar}, {Krajnovi{\'c}}, {Kuntschner},
  {Lablanche}, {Morganti}, {Naab}, {Oosterloo}, {Sarzi}, {Scott}, {Serra},
  {Weijmans}, \& {Young}}]{Cappellari2012}
{Cappellari}, M., {McDermid}, R.~M., {Alatalo}, K., {et~al.} 2012, \nat, 484,
  485, \dodoi{10.1038/nature10972}

\bibitem[{{Carter} {et~al.}(1981){Carter}, {Efstathiou}, {Ellis}, {Inglis}, \&
  {Godwin}}]{Carter81}
{Carter}, D., {Efstathiou}, G., {Ellis}, R.~S., {Inglis}, I., \& {Godwin}, J.
  1981, \mnras, 195, 15P, \dodoi{10.1093/mnras/195.1.15P}

\bibitem[{{Carter} {et~al.}(1985){Carter}, {Inglis}, {Ellis}, {Efstathiou}, \&
  {Godwin}}]{Carter85}
{Carter}, D., {Inglis}, I., {Ellis}, R.~S., {Efstathiou}, G., \& {Godwin},
  J.~G. 1985, \mnras, 212, 471, \dodoi{10.1093/mnras/212.2.471}

\bibitem[{{Choi} {et~al.}(2014){Choi}, {Conroy}, {Moustakas}, {Graves},
  {Holden}, {Brodwin}, {Brown}, \& {van Dokkum}}]{Choi2014}
{Choi}, J., {Conroy}, C., {Moustakas}, J., {et~al.} 2014, \apj, 792, 95,
  \dodoi{10.1088/0004-637X/792/2/95}

\bibitem[{{Choi} {et~al.}(2016){Choi}, {Dotter}, {Conroy}, {Cantiello},
  {Paxton}, \& {Johnson}}]{Choi2016}
{Choi}, J., {Dotter}, A., {Conroy}, C., {et~al.} 2016, \apj, 823, 102,
  \dodoi{10.3847/0004-637X/823/2/102}

\bibitem[{{Concas} {et~al.}(2019){Concas}, {Popesso}, {Brusa}, {Mainieri}, \&
  {Thomas}}]{Concas2019}
{Concas}, A., {Popesso}, P., {Brusa}, M., {Mainieri}, V., \& {Thomas}, D. 2019,
  \aap, 622, A188, \dodoi{10.1051/0004-6361/201732152}

\bibitem[{{Conroy} {et~al.}(2014){Conroy}, {Graves}, \& {van
  Dokkum}}]{Conroy2014}
{Conroy}, C., {Graves}, G.~J., \& {van Dokkum}, P.~G. 2014, \apj, 780, 33,
  \dodoi{10.1088/0004-637X/780/1/33}

\bibitem[{{Conroy} \& {van Dokkum}(2012)}]{Conroy&vanDokkum2012}
{Conroy}, C., \& {van Dokkum}, P.~G. 2012, \apj, 760, 71,
  \dodoi{10.1088/0004-637X/760/1/71}

\bibitem[{{Conroy} {et~al.}(2018){Conroy}, {Villaume}, {van Dokkum}, \&
  {Lind}}]{Conroy2018}
{Conroy}, C., {Villaume}, A., {van Dokkum}, P.~G., \& {Lind}, K. 2018, \apj,
  854, 139, \dodoi{10.3847/1538-4357/aaab49}

\bibitem[{{Cretton} {et~al.}(1999){Cretton}, {de Zeeuw}, {van der Marel}, \&
  {Rix}}]{Cretton1999}
{Cretton}, N., {de Zeeuw}, P.~T., {van der Marel}, R.~P., \& {Rix}, H.-W. 1999,
  \apjs, 124, 383, \dodoi{10.1086/313264}

\bibitem[{{Cretton} \& {van den Bosch}(1999)}]{CrettonBosch1999}
{Cretton}, N., \& {van den Bosch}, F.~C. 1999, \apj, 514, 704,
  \dodoi{10.1086/306971}

\bibitem[{{de Lorenzi} {et~al.}(2007){de Lorenzi}, {Debattista}, {Gerhard}, \&
  {Sambhus}}]{delorenzi2007}
{de Lorenzi}, F., {Debattista}, V.~P., {Gerhard}, O., \& {Sambhus}, N. 2007,
  \mnras, 376, 71, \dodoi{10.1111/j.1365-2966.2007.11434.x}

\bibitem[{{de Nicola} {et~al.}(2020){de Nicola}, {Saglia}, {Thomas}, {Dehnen},
  \& {Bender}}]{denicola2020}
{de Nicola}, S., {Saglia}, R.~P., {Thomas}, J., {Dehnen}, W., \& {Bender}, R.
  2020, \mnras, 496, 3076, \dodoi{10.1093/mnras/staa1703}

\bibitem[{{Dejonghe} \& {Merritt}(1992)}]{Dejonghe1992}
{Dejonghe}, H., \& {Merritt}, D. 1992, \apj, 391, 531, \dodoi{10.1086/171368}

\bibitem[{{Emsellem} {et~al.}(2007){Emsellem}, {Cappellari}, {Krajnovi{\'c}},
  {van de Ven}, {Bacon}, {Bureau}, {Davies}, {de Zeeuw}, {Falc{\'o}n-Barroso},
  {Kuntschner}, {McDermid}, {Peletier}, \& {Sarzi}}]{Emsellem2007}
{Emsellem}, E., {Cappellari}, M., {Krajnovi{\'c}}, D., {et~al.} 2007, \mnras,
  379, 401, \dodoi{10.1111/j.1365-2966.2007.11752.x}

\bibitem[{{Emsellem} {et~al.}(2011){Emsellem}, {Cappellari}, {Krajnovi{\'c}},
  {Alatalo}, {Blitz}, {Bois}, {Bournaud}, {Bureau}, {Davies}, {Davis}, {de
  Zeeuw}, {Khochfar}, {Kuntschner}, {Lablanche}, {McDermid}, {Morganti},
  {Naab}, {Oosterloo}, {Sarzi}, {Scott}, {Serra}, {van de Ven}, {Weijmans}, \&
  {Young}}]{Emsellem2011}
---. 2011, \mnras, 414, 888, \dodoi{10.1111/j.1365-2966.2011.18496.x}

\bibitem[{{Erwin} {et~al.}(2018){Erwin}, {Thomas}, {Saglia}, {Fabricius},
  {Rusli}, {Seitz}, \& {Bender}}]{Erwin2018}
{Erwin}, P., {Thomas}, J., {Saglia}, R.~P., {et~al.} 2018, \mnras, 473, 2251,
  \dodoi{10.1093/mnras/stx2499}

\bibitem[{{Erwin} {et~al.}(2015){Erwin}, {Saglia}, {Fabricius}, {Thomas},
  {Nowak}, {Rusli}, {Bender}, {Vega Beltr{\'a}n}, \& {Beckman}}]{Erwin2015}
{Erwin}, P., {Saglia}, R.~P., {Fabricius}, M., {et~al.} 2015, \mnras, 446,
  4039, \dodoi{10.1093/mnras/stu2376}

\bibitem[{{Escudero} {et~al.}(2015){Escudero}, {Faifer}, {Bassino},
  {Calder{\'o}n}, \& {Caso}}]{Escudero2015}
{Escudero}, C.~G., {Faifer}, F.~R., {Bassino}, L.~P., {Calder{\'o}n}, J.~P., \&
  {Caso}, J.~P. 2015, \mnras, 449, 612, \dodoi{10.1093/mnras/stv283}

\bibitem[{{Faber} {et~al.}(1997){Faber}, {Tremaine}, {Ajhar}, {Byun},
  {Dressler}, {Gebhardt}, {Grillmair}, {Kormendy}, {Lauer}, \&
  {Richstone}}]{Faber1997}
{Faber}, S.~M., {Tremaine}, S., {Ajhar}, E.~A., {et~al.} 1997, \aj, 114, 1771,
  \dodoi{10.1086/118606}

\bibitem[{{Fabricius} {et~al.}(2014){Fabricius}, {Coccato}, {Bender}, {Drory},
  {G{\"o}ssl}, {Landriau}, {Saglia}, {Thomas}, \& {Williams}}]{Fabricius2014}
{Fabricius}, M.~H., {Coccato}, L., {Bender}, R., {et~al.} 2014, \mnras, 441,
  2212, \dodoi{10.1093/mnras/stu694}

\bibitem[{{Falc{\'o}n-Barroso} \& {Martig}(2021)}]{FalconBarroso2021}
{Falc{\'o}n-Barroso}, J., \& {Martig}, M. 2021, \aap, 646, A31,
  \dodoi{10.1051/0004-6361/202039624}

\bibitem[{{Freudling} {et~al.}(2013){Freudling}, {Romaniello}, {Bramich},
  {Ballester}, {Forchi}, {Garc{\'{\i}}a-Dabl{\'o}}, {Moehler}, \&
  {Neeser}}]{Esoreflex2013}
{Freudling}, W., {Romaniello}, M., {Bramich}, D.~M., {et~al.} 2013, \aap, 559,
  A96, \dodoi{10.1051/0004-6361/201322494}

\bibitem[{{Frigo} {et~al.}(2021){Frigo}, {Naab}, {Rantala}, {Johansson},
  {Neureiter}, {Thomas}, \& {Rizzuto}}]{Frigo2021}
{Frigo}, M., {Naab}, T., {Rantala}, A., {et~al.} 2021, \mnras, 508, 4610,
  \dodoi{10.1093/mnras/stab2754}

\bibitem[{{Gebhardt} {et~al.}(2002){Gebhardt}, {Rich}, \& {Ho}}]{Gebhardt2002}
{Gebhardt}, K., {Rich}, R.~M., \& {Ho}, L.~C. 2002, \apjl, 578, L41,
  \dodoi{10.1086/342980}

\bibitem[{{Gebhardt} {et~al.}(2000){Gebhardt}, {Richstone}, {Kormendy},
  {Lauer}, {Ajhar}, {Bender}, {Dressler}, {Faber}, {Grillmair}, {Magorrian}, \&
  {Tremaine}}]{Gebhardt2000}
{Gebhardt}, K., {Richstone}, D., {Kormendy}, J., {et~al.} 2000, \aj, 119, 1157,
  \dodoi{10.1086/301240}

\bibitem[{{Gerhard}(1993)}]{Gerhard1993}
{Gerhard}, O.~E. 1993, \mnras, 265, 213, \dodoi{10.1093/mnras/265.1.213}

\bibitem[{{Houghton} {et~al.}(2006){Houghton}, {Magorrian}, {Sarzi}, {Thatte},
  {Davies}, \& {Krajnovi{\'c}}}]{Houghton2006}
{Houghton}, R.~C.~W., {Magorrian}, J., {Sarzi}, M., {et~al.} 2006, \mnras, 367,
  2, \dodoi{10.1111/j.1365-2966.2005.09713.x}

\bibitem[{{Iannuzzi} \& {Athanassoula}(2015)}]{Iannuzzi2015}
{Iannuzzi}, F., \& {Athanassoula}, E. 2015, \mnras, 450, 2514,
  \dodoi{10.1093/mnras/stv764}

\bibitem[{{Ivanov} {et~al.}(2019){Ivanov}, {Coccato}, {Neeser}, {Selman},
  {Pizzella}, {Dalla Bont{\`a}}, {Corsini}, \& {Morelli}}]{MUSEtemplates}
{Ivanov}, V.~D., {Coccato}, L., {Neeser}, M.~J., {et~al.} 2019, \aap, 629,
  A100, \dodoi{10.1051/0004-6361/201936178}

\bibitem[{{Jethwa} {et~al.}(2020){Jethwa}, {Thater}, {Maindl}, \& {Van de
  Ven}}]{Jethwa2020}
{Jethwa}, P., {Thater}, S., {Maindl}, T., \& {Van de Ven}, G. 2020, {DYNAMITE:
  DYnamics, Age and Metallicity Indicators Tracing Evolution}, Astrophysics
  Source Code Library, record ascl:2011.007.
\newblock \doeprint{2011.007}

\bibitem[{{Johnston} {et~al.}(2018){Johnston}, {Hau}, {Coccato}, \&
  {Herrera}}]{Johnston2018}
{Johnston}, E.~J., {Hau}, G. K.~T., {Coccato}, L., \& {Herrera}, C. 2018,
  \mnras, 480, 3215, \dodoi{10.1093/mnras/sty2048}

\bibitem[{{Kausch} {et~al.}(2015){Kausch}, {Noll}, {Smette}, {Kimeswenger},
  {Barden}, {Szyszka}, {Jones}, {Sana}, {Horst}, \& {Kerber}}]{Molecfit2015b}
{Kausch}, W., {Noll}, S., {Smette}, A., {et~al.} 2015, \aap, 576, A78,
  \dodoi{10.1051/0004-6361/201423909}

\bibitem[{{Kluge} {et~al.}(2020){Kluge}, {Neureiter}, {Riffeser}, {Bender},
  {Goessl}, {Hopp}, {Schmidt}, {Ries}, \& {Brosch}}]{Kluge2020}
{Kluge}, M., {Neureiter}, B., {Riffeser}, A., {et~al.} 2020, \apjs, 247, 43,
  \dodoi{10.3847/1538-4365/ab733b}

\bibitem[{{Kormendy} \& {Bender}(1996)}]{KormendyBender1997}
{Kormendy}, J., \& {Bender}, R. 1996, \apjl, 464, L119, \dodoi{10.1086/310095}

\bibitem[{{Krajnovi{\'c}} {et~al.}(2015){Krajnovi{\'c}}, {Weilbacher},
  {Urrutia}, {Emsellem}, {Carollo}, {Shirazi}, {Bacon}, {Contini}, {Epinat},
  {Kamann}, {Martinsson}, \& {Steinmetz}}]{Krajnovic2015}
{Krajnovi{\'c}}, D., {Weilbacher}, P.~M., {Urrutia}, T., {et~al.} 2015, \mnras,
  452, 2, \dodoi{10.1093/mnras/stv958}

\bibitem[{{Lauer}(2012)}]{Lauer2012}
{Lauer}, T.~R. 2012, \apj, 759, 64, \dodoi{10.1088/0004-637X/759/1/64}

\bibitem[{{Liepold} {et~al.}(2020){Liepold}, {Quenneville}, {Ma}, {Walsh},
  {McConnell}, {Greene}, \& {Blakeslee}}]{Liepold2020ApJ}
{Liepold}, C.~M., {Quenneville}, M.~E., {Ma}, C.-P., {et~al.} 2020, \apj, 891,
  4, \dodoi{10.3847/1538-4357/ab6f71}

\bibitem[{{Lipka} \& {Thomas}(2021)}]{Lipka2021}
{Lipka}, M., \& {Thomas}, J. 2021, \mnras, 504, 4599,
  \dodoi{10.1093/mnras/stab1092}

\bibitem[{{Loubser} {et~al.}(2020){Loubser}, {Babul}, {Hoekstra}, {Bah{\'e}},
  {O'Sullivan}, \& {Donahue}}]{Loubser2020}
{Loubser}, S.~I., {Babul}, A., {Hoekstra}, H., {et~al.} 2020, \mnras, 496,
  1857, \dodoi{10.1093/mnras/staa1682}

\bibitem[{{Mazzalay} {et~al.}(2016){Mazzalay}, {Thomas}, {Saglia}, {Wegner},
  {Bender}, {Erwin}, {Fabricius}, \& {Rusli}}]{Mazzalay2016}
{Mazzalay}, X., {Thomas}, J., {Saglia}, R.~P., {et~al.} 2016, \mnras, 462,
  2847, \dodoi{10.1093/mnras/stw1802}

\bibitem[{{Mehrgan} {et~al.}(2019){Mehrgan}, {Thomas}, {Saglia}, {Mazzalay},
  {Erwin}, {Bender}, {Kluge}, \& {Fabricius}}]{Mehrgan2019}
{Mehrgan}, K., {Thomas}, J., {Saglia}, R., {et~al.} 2019, \apj, 887, 195,
  \dodoi{10.3847/1538-4357/ab5856}

\bibitem[{{Mei} {et~al.}(2005){Mei}, {Blakeslee}, {Tonry}, {Jord{\'a}n},
  {Peng}, {C{\^o}t{\'e}}, {Ferrarese}, {West}, {Merritt}, \&
  {Milosavljevi{\'c}}}]{Mei2005}
{Mei}, S., {Blakeslee}, J.~P., {Tonry}, J.~L., {et~al.} 2005, \apj, 625, 121,
  \dodoi{10.1086/429554}

\bibitem[{{Merritt} \& {Saha}(1993)}]{Merritt1993}
{Merritt}, D., \& {Saha}, P. 1993, \apj, 409, 75, \dodoi{10.1086/172643}

\bibitem[{{Napolitano} {et~al.}(2011){Napolitano}, {Romanowsky}, {Capaccioli},
  {Douglas}, {Arnaboldi}, {Coccato}, {Gerhard}, {Kuijken}, {Merrifield},
  {Bamford}, {Cortesi}, {Das}, \& {Freeman}}]{Napolitano2011}
{Napolitano}, N.~R., {Romanowsky}, A.~J., {Capaccioli}, M., {et~al.} 2011,
  \mnras, 411, 2035, \dodoi{10.1111/j.1365-2966.2010.17833.x}

\bibitem[{{Neureiter} {et~al.}(2021){Neureiter}, {Thomas}, {Saglia}, {Bender},
  {Finozzi}, {Krukau}, {Naab}, {Rantala}, \& {Frigo}}]{Neureiter2021}
{Neureiter}, B., {Thomas}, J., {Saglia}, R., {et~al.} 2021, \mnras, 500, 1437,
  \dodoi{10.1093/mnras/staa3014}

\bibitem[{{Parikh} {et~al.}(2018){Parikh}, {Thomas}, {Maraston}, {Westfall},
  {Goddard}, {Lian}, {Meneses-Goytia}, {Jones}, {Vaughan}, {Andrews},
  {Bershady}, {Bizyaev}, {Brinkmann}, {Brownstein}, {Bundy}, {Drory},
  {Emsellem}, {Law}, {Newman}, {Roman-Lopes}, {Wake}, {Yan}, \&
  {Zheng}}]{Parikh2018}
{Parikh}, T., {Thomas}, D., {Maraston}, C., {et~al.} 2018, \mnras, 477, 3954,
  \dodoi{10.1093/mnras/sty785}

\bibitem[{{Pinkney} {et~al.}(2003){Pinkney}, {Gebhardt}, {Bender}, {Bower},
  {Dressler}, {Faber}, {Filippenko}, {Green}, {Ho}, {Kormendy}, {Lauer},
  {Magorrian}, {Richstone}, \& {Tremaine}}]{Pinkney2003}
{Pinkney}, J., {Gebhardt}, K., {Bender}, R., {et~al.} 2003, \apj, 596, 903,
  \dodoi{10.1086/378118}

\bibitem[{{Pu} {et~al.}(2010){Pu}, {Saglia}, {Thomas}, {Fabricius}, {Bender},
  \& {Han}}]{Pu2010}
{Pu}, S.~B., {Saglia}, R.~P., {Thomas}, J., {et~al.} 2010, VizieR Online Data
  Catalog, J/A+A/516/A4

\bibitem[{{Quenneville} {et~al.}(2021){Quenneville}, {Liepold}, \&
  {Ma}}]{Quenneville2021}
{Quenneville}, M.~E., {Liepold}, C.~M., \& {Ma}, C.-P. 2021, \apjs, 254, 25,
  \dodoi{10.3847/1538-4365/abe6a0}

\bibitem[{{Quenneville} {et~al.}(2022){Quenneville}, {Liepold}, \&
  {Ma}}]{Quenneville2022}
---. 2022, \apj, 926, 30, \dodoi{10.3847/1538-4357/ac3e68}

\bibitem[{{Rantala} {et~al.}(2019){Rantala}, {Johansson}, {Naab}, {Thomas}, \&
  {Frigo}}]{Rantala2019}
{Rantala}, A., {Johansson}, P.~H., {Naab}, T., {Thomas}, J., \& {Frigo}, M.
  2019, \apjl, 872, L17, \dodoi{10.3847/2041-8213/ab04b1}

\bibitem[{{Rusli} {et~al.}(2013{\natexlab{a}}){Rusli}, {Erwin}, {Saglia},
  {Thomas}, {Fabricius}, {Bender}, \& {Nowak}}]{Rusli2013b}
{Rusli}, S.~P., {Erwin}, P., {Saglia}, R.~P., {et~al.} 2013{\natexlab{a}}, \aj,
  146, 160, \dodoi{10.1088/0004-6256/146/6/160}

\bibitem[{{Rusli} {et~al.}(2011{\natexlab{a}}){Rusli}, {Thomas}, {Erwin},
  {Saglia}, {Nowak}, \& {Bender}}]{Thomas2011}
{Rusli}, S.~P., {Thomas}, J., {Erwin}, P., {et~al.} 2011{\natexlab{a}}, \mnras,
  410, 1223, \dodoi{10.1111/j.1365-2966.2010.17610.x}

\bibitem[{{Rusli} {et~al.}(2011{\natexlab{b}}){Rusli}, {Thomas}, {Erwin},
  {Saglia}, {Nowak}, \& {Bender}}]{Rusli2011}
---. 2011{\natexlab{b}}, \mnras, 410, 1223,
  \dodoi{10.1111/j.1365-2966.2010.17610.x}

\bibitem[{{Rusli} {et~al.}(2013{\natexlab{b}}){Rusli}, {Thomas}, {Saglia},
  {Fabricius}, {Erwin}, {Bender}, {Nowak}, {Lee}, {Riffeser}, \&
  {Sharp}}]{Rusli2013a}
{Rusli}, S.~P., {Thomas}, J., {Saglia}, R.~P., {et~al.} 2013{\natexlab{b}},
  \aj, 146, 45, \dodoi{10.1088/0004-6256/146/3/45}

\bibitem[{{S{\'a}nchez-Bl{\'a}zquez} {et~al.}(2006){S{\'a}nchez-Bl{\'a}zquez},
  {Peletier}, {Jim{\'e}nez-Vicente}, {Cardiel}, {Cenarro},
  {Falc{\'o}n-Barroso}, {Gorgas}, {Selam}, \& {Vazdekis}}]{SanchezBlazquez2006}
{S{\'a}nchez-Bl{\'a}zquez}, P., {Peletier}, R.~F., {Jim{\'e}nez-Vicente}, J.,
  {et~al.} 2006, \mnras, 371, 703, \dodoi{10.1111/j.1365-2966.2006.10699.x}

\bibitem[{{Sarazin} \& {Roddier}(1990)}]{ESODIMM1990}
{Sarazin}, M., \& {Roddier}, F. 1990, \aap, 227, 294

\bibitem[{{Schwarzschild}(1979)}]{Schwarzschild1979}
{Schwarzschild}, M. 1979, \apj, 232, 236, \dodoi{10.1086/157282}

\bibitem[{{Siopis} \& {Kandrup}(2000)}]{Siopis2000}
{Siopis}, C., \& {Kandrup}, H.~E. 2000, \mnras, 319, 43,
  \dodoi{10.1046/j.1365-8711.2000.03740.x}

\bibitem[{{Smette} {et~al.}(2015){Smette}, {Sana}, {Noll}, {Horst}, {Kausch},
  {Kimeswenger}, {Barden}, {Szyszka}, {Jones}, {Gallenne}, {Vinther},
  {Ballester}, \& {Taylor}}]{Molecfit2015a}
{Smette}, A., {Sana}, H., {Noll}, S., {et~al.} 2015, \aap, 576, A77,
  \dodoi{10.1051/0004-6361/201423932}

\bibitem[{{Spiniello} {et~al.}(2011){Spiniello}, {Koopmans}, {Trager},
  {Czoske}, \& {Treu}}]{Spiniello2011}
{Spiniello}, C., {Koopmans}, L.~V.~E., {Trager}, S.~C., {Czoske}, O., \&
  {Treu}, T. 2011, \mnras, 417, 3000, \dodoi{10.1111/j.1365-2966.2011.19458.x}

\bibitem[{{Spiniello} {et~al.}(2012){Spiniello}, {Trager}, {Koopmans}, \&
  {Chen}}]{Spiniello2012}
{Spiniello}, C., {Trager}, S.~C., {Koopmans}, L.~V.~E., \& {Chen}, Y.~P. 2012,
  \apjl, 753, L32, \dodoi{10.1088/2041-8205/753/2/L32}

\bibitem[{{Spiniello} {et~al.}(2018){Spiniello}, {Napolitano}, {Arnaboldi},
  {Tortora}, {Coccato}, {Capaccioli}, {Gerhard}, {Iodice}, {Spavone},
  {Cantiello}, {Peletier}, {Paolillo}, \& {Schipani}}]{Spiniello2018}
{Spiniello}, C., {Napolitano}, N.~R., {Arnaboldi}, M., {et~al.} 2018, \mnras,
  477, 1880, \dodoi{10.1093/mnras/sty663}

\bibitem[{{Syer} \& {Tremaine}(1996)}]{Syer1996}
{Syer}, D., \& {Tremaine}, S. 1996, \mnras, 282, 223,
  \dodoi{10.1093/mnras/282.1.223}

\bibitem[{{Thater} {et~al.}(2022){Thater}, {Krajnovi{\'c}}, {Weilbacher},
  {Nguyen}, {Bureau}, {Cappellari}, {Davis}, {Iguchi}, {McDermid}, {Onishi},
  {Sarzi}, \& {van de Ven}}]{Thater2022}
{Thater}, S., {Krajnovi{\'c}}, D., {Weilbacher}, P.~M., {et~al.} 2022, \mnras,
  509, 5416, \dodoi{10.1093/mnras/stab3210}

\bibitem[{{Thomas} {et~al.}(1999){Thomas}, {Greggio}, \&
  {Bender}}]{DThomas1999}
{Thomas}, D., {Greggio}, L., \& {Bender}, R. 1999, \mnras, 302, 537,
  \dodoi{10.1046/j.1365-8711.1999.02138.x}

\bibitem[{{Thomas} {et~al.}(2005){Thomas}, {Maraston}, {Bender}, \& {Mendes de
  Oliveira}}]{DThomas2005}
{Thomas}, D., {Maraston}, C., {Bender}, R., \& {Mendes de Oliveira}, C. 2005,
  \apj, 621, 673, \dodoi{10.1086/426932}

\bibitem[{{Thomas} \& {Lipka}(2022)}]{ThomasLipka2022}
{Thomas}, J., \& {Lipka}, M. 2022, \mnras, \dodoi{10.1093/mnras/stac1581}

\bibitem[{{Thomas} {et~al.}(2014){Thomas}, {Saglia}, {Bender}, {Erwin}, \&
  {Fabricius}}]{Thomas2014}
{Thomas}, J., {Saglia}, R.~P., {Bender}, R., {Erwin}, P., \& {Fabricius}, M.
  2014, \apj, 782, 39, \dodoi{10.1088/0004-637X/782/1/39}

\bibitem[{{Thomas} {et~al.}(2004){Thomas}, {Saglia}, {Bender}, {Thomas},
  {Gebhardt}, {Magorrian}, \& {Richstone}}]{Thomas2004}
{Thomas}, J., {Saglia}, R.~P., {Bender}, R., {et~al.} 2004, \mnras, 353, 391,
  \dodoi{10.1111/j.1365-2966.2004.08072.x}

\bibitem[{{Tonry} {et~al.}(2001){Tonry}, {Dressler}, {Blakeslee}, {Ajhar},
  {Fletcher}, {Luppino}, {Metzger}, \& {Moore}}]{Tonry2001}
{Tonry}, J.~L., {Dressler}, A., {Blakeslee}, J.~P., {et~al.} 2001, \apj, 546,
  681, \dodoi{10.1086/318301}

\bibitem[{{van de Sande} {et~al.}(2017){van de Sande}, {Bland-Hawthorn},
  {Fogarty}, {Cortese}, {d'Eugenio}, {Croom}, {Scott}, {Allen}, {Brough},
  {Bryant}, {Cecil}, {Colless}, {Couch}, {Davies}, {Elahi}, {Foster},
  {Goldstein}, {Goodwin}, {Groves}, {Ho}, {Jeong}, {Jones}, {Konstantopoulos},
  {Lawrence}, {Leslie}, {L{\'o}pez-S{\'a}nchez}, {McDermid}, {McElroy},
  {Medling}, {Oh}, {Owers}, {Richards}, {Schaefer}, {Sharp}, {Sweet}, {Taranu},
  {Tonini}, {Walcher}, \& {Yi}}]{vanDeSande2017}
{van de Sande}, J., {Bland-Hawthorn}, J., {Fogarty}, L. M.~R., {et~al.} 2017,
  \apj, 835, 104, \dodoi{10.3847/1538-4357/835/1/104}

\bibitem[{{van den Bosch} {et~al.}(2008){van den Bosch}, {van de Ven},
  {Verolme}, {Cappellari}, \& {de Zeeuw}}]{vanDenBosch2008}
{van den Bosch}, R.~C.~E., {van de Ven}, G., {Verolme}, E.~K., {Cappellari},
  M., \& {de Zeeuw}, P.~T. 2008, \mnras, 385, 647,
  \dodoi{10.1111/j.1365-2966.2008.12874.x}

\bibitem[{{van der Marel} \& {Franx}(1993)}]{vanDerMarel1993}
{van der Marel}, R.~P., \& {Franx}, M. 1993, \apj, 407, 525,
  \dodoi{10.1086/172534}

\bibitem[{{van Dokkum} \& {Conroy}(2012)}]{vanDokkum2012}
{van Dokkum}, P.~G., \& {Conroy}, C. 2012, \apj, 760, 70,
  \dodoi{10.1088/0004-637X/760/1/70}

\bibitem[{{Vasiliev} \& {Valluri}(2020)}]{Vasilev2020}
{Vasiliev}, E., \& {Valluri}, M. 2020, \apj, 889, 39,
  \dodoi{10.3847/1538-4357/ab5fe0}

\bibitem[{{Vazdekis}(1999)}]{Vazdekis1999}
{Vazdekis}, A. 1999, \apj, 513, 224, \dodoi{10.1086/306843}

\bibitem[{{Veale} {et~al.}(2018){Veale}, {Ma}, {Greene}, {Thomas}, {Blakeslee},
  {Walsh}, \& {Ito}}]{Veale2018}
{Veale}, M., {Ma}, C.-P., {Greene}, J.~E., {et~al.} 2018, \mnras, 473, 5446,
  \dodoi{10.1093/mnras/stx2717}

\bibitem[{{Ventimiglia} {et~al.}(2010){Ventimiglia}, {Gerhard}, {Arnaboldi}, \&
  {Coccato}}]{Ventimiglia2010}
{Ventimiglia}, G., {Gerhard}, O., {Arnaboldi}, M., \& {Coccato}, L. 2010, \aap,
  520, L9, \dodoi{10.1051/0004-6361/201015485}

\bibitem[{{Villaume} {et~al.}(2017){Villaume}, {Conroy}, {Johnson}, {Rayner},
  {Mann}, \& {van Dokkum}}]{Villaume2017}
{Villaume}, A., {Conroy}, C., {Johnson}, B., {et~al.} 2017, \apjs, 230, 23,
  \dodoi{10.3847/1538-4365/aa72ed}

\end{thebibliography}
\end{document}